\documentclass[12pt]{article}

\usepackage{graphics,amssymb,float}
\usepackage{graphicx}
\usepackage{rotating}
\usepackage{dcolumn}
\usepackage{bm}
\usepackage{cite}
\usepackage{amssymb,amsmath,amsbsy}
\usepackage{mathrsfs}
\usepackage{caption}
\usepackage{indentfirst}

\usepackage[colorlinks=true
,urlcolor=blue 
,anchorcolor=blue
,citecolor=blue
,filecolor=blue
,linkcolor=blue
,menucolor=blue
,pagecolor=blue
,linktocpage=true
]{hyperref}
\usepackage{epstopdf}
\usepackage{slashed,xcolor}

\DeclareMathOperator\sgn{sgn}
\newenvironment{Eqnarray}
         {\arraycolsep 0.14em\begin{eqnarray}}{\end{eqnarray}}
\def\br{{\mathcal B}}

\def\sina{\sin\alpha}
\def\sa{s_\alpha}
\def\ca{c_\alpha}
\def\cosa{\cos\alpha}
\def\tanb{\tan\beta}
\def\sinb{\sin\beta}
\def\cosb{\cos\beta}
\def\cotb{\cot\beta}
\def\sbma{s_{\beta-\alpha}}
\def\cbma{c_{\beta-\alpha}}
\def\sbmaii{s^2_{\beta-\alpha}}
\def\cbmaii{c^2_{\beta-\alpha}}
\def\sbmaiii{s^3_{\beta-\alpha}}
\def\cbmaiii{c^3_{\beta-\alpha}}

\def\gam{\gamma}
\def\ie{{\it i.e.}}
\def\eg{{\it e.g.}}

\def\gev{~{\rm GeV}}

\def\tev{~{\rm TeV}}
\def\bit{\begin{itemize}}
\def\eit{\end{itemize}}
\def\ben{\begin{enumerate}}
\def\een{\end{enumerate}}
\def\beq{\begin{equation}}
\def\eeq{\end{equation}}
\def\mw{m_W}

\def\Eq#1{Eq.~(\ref{#1})}

\def\phm{\phantom{-}}

\def\lam{\lambda}
\def\bea{\begin{Eqnarray}}
\def\eea{\end{Eqnarray}}

\def\cbma{\cos(\beta-\alpha)}

\def\beq{\begin{equation}}
\def\eeq{\end{equation}}
\def\beqa{\begin{Eqnarray}}
\def\eeqa{\end{Eqnarray}}

\def\calo{\mathcal{O}}

\def\cbma{c_{\beta-\alpha}}

\def\sbma{s_{\beta-\alpha}}
\def\ctwob{c_{2\beta}}
\def\stwob{s_{2\beta}}

\def\lamtil{\lam\ls{345}}

\def\phm{\phantom{-}}

\def\beq{\begin{equation}}
\def\eeq{\end{equation}}
\def\ifmath#1{\relax\ifmmode #1\else $#1$\fi}
\def\calm{\mathcal{M}}

\def\tb{t_{\beta}}

\def\sb  {s_{\beta}}
\def\cb  {c_{\beta}}
\def\stwob  {s_{2\beta}}
\def\ctwob  {c_{2\beta}}
\def\sa  {s_{\alpha}}
\def\ca  {c_{\alpha}}

\def\sba  {s_{\beta-\alpha}}
\def\cba  {c_{\beta-\alpha}}
\def\tanb{\tan\beta}
\def\cotb{\cot\beta}
\def\sinb{\sin\beta}
\def\cosb{\cos\beta}
\def\sina{\sin\alpha}
\def\cosa{\cos\alpha}

\def\sbmaii{s^2_{\beta-\alpha}}
\def\cbmaii{c^2_{\beta-\alpha}}

\def\hl{h}
\def\ha{A}
\def\hh{H}
\def\hpm{{H^\pm}}

\def\hp{{H^+}}
\def\hm{{H^-}}
\def\go{G}

\def\lamtil{\lam_{345}}

\def\mha{m_{\ha}}
\def\mhl{m_{\hl}}
\def\mhh{m_{\hh}}
\def\mhpm{m_{\hpm}}

\def\mw{m_W}

\def\ls#1{\ifmath{_{\lower1.5pt\hbox{$\scriptstyle #1$}}}}
\def\lss#1{\ifmath{^{\,\lower2.5pt\hbox{$\scriptstyle #1$}}}}

\def\half{\tfrac{1}{2}}

\def\ie{{\it i.e.}}
\def\eg{{\it e.g.}}

\def\lsim{\mathrel{\raise.3ex\hbox{$<$\kern-.75em\lower1ex\hbox{$\sim$}}}}
\def\gsim{\mathrel{\raise.3ex\hbox{$>$\kern-.75em\lower1ex\hbox{$\sim$}}}}
\def\ifmath#1{\relax\ifmmode #1\else $#1$\fi}

\def\Ref#1{Ref.~\cite{#1}}

\def\eq#1{Eq.~(\ref{#1})}
\def\Eq#1{Eq.~(\ref{#1})}

\def\eqs#1#2{Eqs.~(\ref{#1}) and (\ref{#2})}
\def\eqss#1#2#3{Eqs.~(\ref{#1}), (\ref{#2}) and (\ref{#3})}
\def\eqst#1#2{Eqs.~(\ref{#1})--(\ref{#2})}

\begin{document}
\begin{titlepage}
\begin{center}

\vspace*{-1.5cm}
\begin{flushright}
LPSC15154,\\
UCD-2015-001,\\  
SCIPP-15/08\\
\end{flushright}

\vspace*{1.5cm}
{\Large\bf Scrutinizing the Alignment Limit\\[2mm] in Two-Higgs-Doublet Models} 

\vspace*{0.8cm}

{\large\bf\boldmath Part~1: $m_h=125$~GeV} 

\vspace*{0.8cm}

\renewcommand{\thefootnote}{\fnsymbol{footnote}}

{\large
J\'er\'emy~Bernon$^{1}$\footnote[1]{Email: bernon@lpsc.in2p3.fr},
John~F.~Gunion$^{2}$\footnote[2]{Email: jfgunion@ucdavis.edu},
Howard~E.~Haber$^{3}$\footnote[3]{Email: haber@scipp.ucsc.edu},
Yun~Jiang$^{2}$\footnote[4]{Email: yunjiang@ucdavis.edu},
Sabine~Kraml$^{1}$\footnote[5]{Email: sabine.kraml@lpsc.in2p3.fr} 
}

\renewcommand{\thefootnote}{\arabic{footnote}}

\vspace*{0.8cm}
{\normalsize \it
$^1\,$Laboratoire de Physique Subatomique et de Cosmologie, Universit\'e Grenoble-Alpes,
CNRS/IN2P3, 53 Avenue des Martyrs, F-38026 Grenoble, France\\[2mm]
$^2\,$Department of Physics, University of California, Davis, CA 95616, USA\\[2mm]
$^3\,$Santa Cruz Institute for Particle Physics, Santa Cruz, CA 95064, USA}
\vspace{0.8cm}

\begin{abstract}
In the alignment limit of a multi-doublet Higgs sector, one of the Higgs mass eigenstates aligns with the direction of the scalar field vacuum expectation values, and its couplings approach those of the Standard Model (SM) Higgs boson.  We consider CP-conserving Two-Higgs-Doublet Models (2HDMs) of Type~I and Type~II near the alignment limit in which the lighter of the two CP-even Higgs bosons,
$h$, is the SM-like state observed at  $125\gev$.  In particular, we focus on the 2HDM parameter regime where the coupling of $h$ to gauge bosons approaches that of the SM.
We review the theoretical structure and analyze the phenomenological implications of the regime of alignment limit without decoupling, in which the other Higgs scalar masses are not significantly larger than $m_h$ and thus do not decouple from the effective theory at the electroweak scale. 
For the numerical analysis, we perform scans of the 2HDM parameter space employing the software packages \texttt{2HDMC} and \texttt{Lilith}, taking into account all relevant pre-LHC constraints, the latest constraints from the measurements of the $125\gev$ Higgs signal at the LHC,  as well as the most recent limits coming from searches for heavy Higgs-like states.  We contrast these results with the alignment limit achieved via the decoupling of heavier scalar states, where $h$ is the only light Higgs scalar.
Implications for Run~2 at the LHC, including expectations for observing the other scalar states, are also discussed.
\end{abstract}

\end{center}
\end{titlepage}

\section{Introduction}\label{sec:intro}

The minimal version of the Standard Model (SM) contains one
complex Higgs doublet, resulting in one physical neutral CP-even Higgs
boson after electroweak symmetry breaking.
The discovery \cite{Aad:2012tfa,Chatrchyan:2012ufa} of a new particle with mass of about $125\gev$~\cite{Aad:2015zhl}
and properties that match very well those expected for a SM Higgs boson was a real triumph of Run~1 of the LHC. 
Fits of the Higgs couplings performed by ATLAS~\cite{ATLAS-CONF-2015-007} and CMS~\cite{Khachatryan:2014jba} 
show no significant deviations from SM expectations. 
(A combined global fit of the Higgs couplings based on the Run~1 results was performed by some of us in \cite{Bernon:2014vta}.)  
However, one has to keep in mind that the present precisions on the Higgs couplings are, roughly, of the order of tens of percent, so substantial deviations are still possible. 
Indeed, the SM is not necessarily the ultimate theoretical structure responsible for electroweak symmetry breaking, 
and theories that go beyond the SM, such as supersymmetry, typically require an extended Higgs sector~\cite{Fayet:1974pd,Inoue:1982ej,Flores:1982pr,Gunion:1984yn}. 
Hence, the challenge for Run~2 of the LHC, and other future collider programs, is to
determine whether the observed state is \textit{the} SM Higgs boson, or whether it is part of a non-minimal Higgs sector of a more fundamental theory. 

In this paper, we take Two-Higgs-Doublet Models (2HDMs) of Type~I and Type~II~\cite{Hall:1981bc} as the prototypes for studying 
the effects of an extended Higgs sector.  Our focus will be on a particularly interesting limit of these models, 
namely the case in which one of the neutral Higgs mass eigenstates is approximately aligned with the direction of the scalar field vacuum expectation values. 
In this case, the coupling to gauge bosons of the Higgs boson observed at the LHC tends towards the SM limit, $C_V\to 1$.\footnote{We use the notation of coupling scale factors, or {\em reduced couplings},  employed in \cite{Bernon:2014vta}: $C_V$ ($V=W,Z$) for the coupling to gauge bosons, 
$C_{U,D}$ for the couplings to up-type and down-type fermions and $C_{\gamma,g}$ for the loop-induced couplings to photons and gluons.} 
This so-called \textit{alignment limit} is most easily attained in the decoupling limit~\cite{Gunion:2002zf}, where all the other non-SM-like Higgs scalars of the model are heavy.  However, the alignment limit of the 2HDM can also be achieved in a parameter regime in which one or more of the non-SM-like Higgs scalars are light (and in some cases very light).  This region of \textit{alignment without decoupling} is a primary focus of this paper.

An extensive review of the status of 2HDMs of Type~I and Type~II was given in \cite{Dumont:2014wha,Dumont:2014kna}.
Interpretations of the recently discovered Higgs boson at 125 GeV in the context of the 2HDMs were also studied in~\cite{Coleppa:2013dya,Eberhardt:2013uba,Chang:2013ona,Cheung:2013rva,Baglio:2014nea,Craig:2015jba,Chowdhury:2015yja}.
The possibility of alignment without decoupling was first noted in \cite{Gunion:2002zf} and further clarified in
\cite{Haber:2013mia,Asner:2013psa}.
Previous studies of alignment without decoupling scenarios in the light of the LHC Higgs results were conducted in  \cite{Craig:2013hca,Carena:2013ooa,Carena:2014nza}.
The specific case of additional light Higgs states in 2HDMs with mass below $\sim 125/2$~GeV was studied in \cite{Bernon:2014nxa}.

Considering experimental as well as theoretical uncertainties, the expected precision for coupling measurements at the LHC after collecting 300~fb$^{-1}$ of data is about 4--6\% for the coupling to gauge bosons, and of the level of 6--13\% for the couplings to fermions \cite{Dawson:2013bba}. The precision improves by roughly a factor of 2 for at the high-luminosity run of the LHC with 3000~fb$^{-1}$. At a future $e^+e^-$ international linear collider (ILC) with $\sqrt{s}=250$~GeV to $1$~TeV, one may measure the couplings to fermions at the percent level, and the coupling to gauge bosons 
at the sub-percent level.
A detailed discussion of the prospects of various future colliders can be found in \cite{Dawson:2013bba}.

We take this envisaged $\sim 1\%$ accuracy on $C_V$ as the starting point for the numerical analysis of the alignment case. Concretely, we investigate the parameter spaces of the 2HDMs of Type~I and Type~II assuming
that the observed $125\gev$ state is the $h$, the lighter of the two CP-even Higgs bosons in these models,
and imposing that $C_V^h>0.99$ (note that $|C_V|\le 1$ in any model whose Higgs sector consists of only 
doublets and/or singlets). The case of the heavier CP-even $H$ being the state at $125\gev$ is discussed in a separate paper \cite{part2}. 

Taking into account all relevant theoretical and phenomenological constraints, including the signal strengths of the observed Higgs boson, as well as the most recent limits from the non-observation of any other Higgs-like states, we then analyse the phenomenological consequences of this scenario. In particular, we study the variations in the couplings to fermions and in the triple-Higgs couplings that are possible as a function of the amount of alignment when the other Higgs states are light, and contrast this to what happens in the decoupling regime. Moreover, we study the prospects to discover the additional Higgs states when they are light. 

The public tools used in this study include 
\texttt{2HDMC}~\cite{Eriksson:2009ws} for computing couplings and decay widths and for testing theoretical constraints within the 2HDM context, \texttt{Lilith 1.1.2}~\cite{Bernon:2015hsa} for evaluating the Higgs signal strength constraints, and  \texttt{SusHi-1.3.0}~\cite{Harlander:2012pb} and \texttt{VBFNLO-2.6.3}~\cite{Arnold:2008rz} for computing production cross sections at the LHC. 

The paper is organised as follows. In Section~\ref{modelreview} we first review the theoretical structure of the 2HDM.   A softly-broken discrete $\mathbb{Z}_2$-symmetric scalar potential is introduced using a basis of scalar doublet fields (called the $\mathbb{Z}_2$-basis) in which a the symmetry is manifest.  The Higgs basis is then introduced, which provides an elegant framework for exhibiting the alignment limit.  We then provide a comprehensive discussion of the Higgs couplings in the alignment regime. In Section~\ref{setup}, we explain the setup of the numerical analysis and the tools used. The results are presented in Section~\ref{results-lighth}. 
Section~\ref{conclusions} contains our conclusions.   In Appendix~\ref{AppInverse}, detailed formulae relating the quartic coefficients of the Higgs potential in the $\mathbb{Z}_2$-basis to those of the Higgs basis are given.
Some useful analytical expressions regarding the trilinear Higgs self-couplings in terms of physical Higgs masses are collected in Appendix~\ref{AppTrilinear}. 

{\bf\em NB: This version of the paper has been updated to include exactly the same constraints as Part~2 with \boldmath $m_H=125$~GeV~\cite{part2}.}

\section{CP-conserving 2HDM of Types~I and II}
\label{modelreview}

In this section, we review the theoretical structure of the two-Higgs doublet model. 
Comprehensive reviews of the model can also be found in, 
\eg,~\cite{Gunion:1989we,Asner:2013psa,Gunion:2002zf,Branco:2011iw}.  
In order to avoid tree-level Higgs-mediated flavor changing neutral currents (FCNCs), we shall
impose a Type-I or II structure on the Higgs-fermion interactions.  This structure can
be naturally implemented~\cite{Glashow:1976nt,Paschos:1976ay} by imposing a discrete $\mathbb{Z}_2$ symmetry on the
dimension-four terms of the Higgs Lagrangian. This discrete symmetry is softly-broken by
mass terms that appear in the Higgs scalar potential.  Nevertheless, the absence of
tree-level Higgs-mediated FCNCs is maintained, and FCNC effects generated at one loop
are all small enough to be consistent with phenomenological constraints 
over a significant fraction of the 2HDM parameter space~\cite{Haisch:2008ar,Mahmoudi:2009zx,Gupta:2009wn,Jung:2010ik}.

Even with the imposition of the softly-broken discrete $\mathbb{Z}_2$ symmetry mentioned above,
new CP-violating phenomena in the Higgs sector are still possible, either
explicitly due to a physical complex phase that cannot be removed from the scalar
potential parameters or spontaneously due to a CP-violating vacuum state.   To simplify
the analysis in this paper, we shall assume that these CP-violating effects
are absent, in which case one can choose a basis of scalar doublet Higgs fields such that
all scalar potential parameters and the two neutral Higgs field vacuum expectation values
are simultaneously real.  Moreover, we assume that only the neutral Higgs fields acquire
non-zero vacuum expectation values, i.e.~the scalar potential does not admit the possibility
of stable charge-breaking minima~\cite{Barroso:2005sm,Ivanov:2006yq}.

We first exhibit the Higgs scalar potential, the corresponding Higgs scalar spectrum and the
Higgs-fermion interactions subject to the restrictions discussed above.
Motivated by the Higgs data, we then examine
the conditions that yield an approximately  SM-like Higgs boson.

\subsection{Higgs scalar potential}
\label{higgsscalarpotential}

Let $\Phi_1$ and $\Phi_2$ denote two complex $Y=1$, SU(2)$\ls{L}$ doublet scalar fields.
The most general gauge invariant renormalizable scalar potential is given by
\beqa
\label{lambdapotential}
\mathcal{V}&=& m_{11}^2\Phi_1^\dagger\Phi_1+m_{22}^2\Phi_2^\dagger\Phi_2
-[m_{12}^2\Phi_1^\dagger\Phi_2+{\rm h.c.}]
 +\half\lambda_1(\Phi_1^\dagger\Phi_1)^2
+\half\lambda_2(\Phi_2^\dagger\Phi_2)^2
+\lambda_3(\Phi_1^\dagger\Phi_1)(\Phi_2^\dagger\Phi_2)\nonumber\\[8pt]
&&\qquad\qquad\,\,
+\lambda_4(\Phi_1^\dagger\Phi_2)(\Phi_2^\dagger\Phi_1)
+\left\{\half\lambda_5(\Phi_1^\dagger\Phi_2)^2
+\big[\lambda_6(\Phi_1^\dagger\Phi_1)
+\lambda_7(\Phi_2^\dagger\Phi_2)\big]
\Phi_1^\dagger\Phi_2+{\rm h.c.}\right\}\,. \label{pot}
\eeqa
In general, $m_{12}^2$, $\lambda_5$, $\lambda_6$ and $\lambda_7$ can be complex.
As noted above, to avoid tree-level Higgs-mediated FCNCs, we impose a softly-broken discrete
$\mathbb{Z}_2$ symmetry, $\Phi_1\to +\Phi_1$ and $\Phi_2\to -\Phi_2$
on the quartic terms of \eq{pot}, which implies that $\lambda_6=\lambda_7=0$, whereas $m_{12}^2\neq 0$ is allowed.  In this basis of scalar doublet fields (denoted as the $\mathbb{Z}_2$-basis),
the discrete $\mathbb{Z}_2$ symmetry of the quartic terms of \eq{lambdapotential} is manifest.
Furthermore, we assume that the scalar fields can be rephased such that
$m_{12}^2$ and $\lambda_5$ are both real.  The resulting scalar potential
is then explicitly CP-conserving.

The scalar fields will
develop non-zero vacuum expectation values if the Higgs mass matrix
$m_{ij}^2$ has at least one negative eigenvalue.
We assume that the
parameters of the scalar potential are chosen such that
the minimum of the scalar potential respects the
U(1)$\ls{\rm EM}$ gauge symmetry.  Then, the scalar field
vacuum expectations values are of the form
\beq
\langle \Phi_1 \rangle={1\over\sqrt{2}} \left(
\begin{array}{c} 0\\ v_1\end{array}\right), \qquad \langle
\Phi_2\rangle=
{1\over\sqrt{2}}\left(\begin{array}{c}0\\ v_2
\end{array}\right)\,.\label{potmin}
\eeq
As noted in Appendix B of Ref.~\cite{Gunion:2002zf},
if $|m_{12}^2|\geq \lambda_5 |v_1| |v_2|$, then the vacuum is
CP-conserving and the vacuum expectation values $v_1$ and $v_2$
can be chosen to be non-negative without loss of generality.
In this case, the
corresponding potential minimum conditions are:\footnote{Here and in the following, we use the
shorthand notation $\cb\equiv\cos\beta$, $\sb\equiv\sin\beta$,
$\ca\equiv\cos\alpha$, $\sa\equiv\sin\alpha$,
$\ctwob\equiv\cos2\beta$, $\stwob\equiv\sin2\beta$,
$\cbma\equiv\cos(\beta-\alpha)$, $\sbma\equiv\sin(\beta-\alpha)$, etc.}
\beqa
m_{11}^2 &=& m_{12}^2\tb -\half
v^2\left(\lam_1\cb^2+\lamtil\sb^2\right)
\,,\label{minconditionsa} \\
m_{22}^2 &=& m_{12}^2\tb^{-1}-\half v^2
\left(\lam_2\sb^2+\lamtil\cb^2\right)\,,
\label{minconditionsb}
\eeqa
where we have defined:
\beq
\lamtil\equiv\lam_3+\lam_4+\lam_5\,,\qquad\qquad\tb\equiv\tanb\equiv{v_2\over
  v_1}\,,
\label{tanbdef}
\eeq
where $0\leq\beta\leq\half\pi$, and
\beq
v^2\equiv v_1^2+v_2^2={4\mw^2\over g^2}=(246~{\rm GeV})^2\,.
\label{v246}
\eeq

Of the original eight scalar degrees of freedom, three Goldstone
bosons ($G^\pm$ and~$\go$) are absorbed (``eaten'') by the $W^\pm$ and
$Z$.  The remaining five physical Higgs particles are: two CP-even
scalars ($\hl$ and $\hh$, with $\mhl\leq \mhh$), one CP-odd scalar
($\ha$) and a charged Higgs pair ($\hpm$).
The resulting squared-masses for the CP-odd and charged
Higgs states are\beqa
\mha^2 &=&\overline{m}^{\,2}-\lambda_5 v^2\,,\label{massha}\\[6pt]
m_{H^{\pm}}^2 &=& m_{A}^2+\half v^2(\lambda_5-\lambda_4)\,,
\label{mamthree}
\eeqa
where
\beq \label{mbar}
\overline{m}^{\,2}\equiv\frac{2m_{12}^2}{s_{2\beta}}\,.
\eeq

The two neutral CP-even Higgs states mix according to the following squared-mass
matrix:
\beq \label{massmhh}
\calm^2\equiv
\left( \begin{array}{cc}
  \lambda_1 v^2\cb^2+(\mha^2+\lambda_5 v^2)\sb^2
    &\bigl[\lamtil v^2-(\mha^2+\lambda_5 v^2)]\sb\cb \\[3pt]
 \bigl[\lamtil v^2-(\mha^2+\lambda_5 v^2)\bigr]\sb\cb
    &\lambda_2 v^2 \sb^2 +(\mha^2+\lambda_5 v^2)\cb^2
\end{array}\right) \,.
\eeq
Defining the physical mass eigenstates
\beqa
\hh &=&(\sqrt{2}{\rm Re\,}\Phi_1^0-v_1)\ca+
(\sqrt{2}{\rm Re\,}\Phi_2^0-v_2)\sa\,,\label{HZ2scalareigenstates}\\
\hl &=&-(\sqrt{2}{\rm Re\,}\Phi_1^0-v_1)\sa+
(\sqrt{2}{\rm Re\,}\Phi_2^0-v_2)\ca\,,
\label{hZ2scalareigenstates}
\eeqa
the masses and mixing angle $\alpha$ are found from the diagonalization
process
\beqa
&& \!\!\!\!\!\!\!\!\!\!\!
\left(\begin{array}{cc} \mhh^2 & 0 \cr 0 &\mhl^2\end{array}\right)=
\left(\begin{array}{cc} \phm\ca & \sa \cr -\sa & \ca \end{array}\right)
\left(\begin{array}{cc} \calm_{11}^2 & \calm_{12}^2 \\
\calm_{12}^2 &\calm_{22}^2\end{array}\right)
\left(\begin{array}{cc} \ca & -\sa \cr \sa & \phm\ca \end{array}\right)
\nonumber\\[8pt]
&& \!\!\!
=\left(\begin{array}{cc} \calm_{11}^2\ca^2+2\calm_{12}^2\ca\sa
+\calm_{22}^2\sa^2 & \quad\calm_{12}^2(\ca^2-\sa^2)+(\calm_{22}^2-
\calm_{11}^2)\sa\ca \\[4pt]
\calm_{12}^2(\ca^2-\sa^2)+(\calm_{22}^2-\calm_{11}^2)\sa\ca
& \quad\calm_{11}^2\sa^2-2\calm_{12}^2\ca\sa
+\calm_{22}^2\ca^2 \end{array}\right).
\label{diagn}
\eeqa
Note that the two equations, ${\rm Tr}~\mathcal{M}^2=m_H^2+m_h^2$ and
${\rm det}~\mathcal{M}^2=m_H^2 m_h^2$,
yield the following result:
\beq
|\mathcal{M}^2_{12}|
=\sqrt{(m_H^2-\mathcal{M}^2_{11})(\mathcal{M}^2_{11}-m_h^2)}
=\sqrt{(\mathcal{M}^2_{22}-m_h^2)(\mathcal{M}^2_{11}-m_h^2)}
\,.\label{relation}
\eeq
Explicitly, the squared-masses of the neutral CP-even Higgs bosons are given by
\beq \label{higgsmasses}
m^2_{H,h}=\half\bigl[\mathcal{M}^2_{11}+\mathcal{M}^2_{22}\pm\Delta\bigr]\,,
\eeq
where $m_h\leq m_H$ and the non-negative quantity $\Delta$ is defined by
\beq \label{deltadef}
\Delta\equiv\sqrt{(\mathcal{M}^2_{11}-\mathcal{M}^2_{22})^2+4(\mathcal{M}^2_{12})^2}\,.
\eeq

The mixing angle $\alpha$, which is defined modulo $\pi$, is evaluated by setting the off-diagonal
elements of the CP-even scalar squared-mass matrix given in \eq{diagn}
to zero.  It is often convenient to restrict the range of the mixing angle to $|\alpha|\leq\half\pi$.  In this case, $c_\alpha$ is non-negative and is given by
\beq \label{ca}
c_\alpha=\sqrt{\frac{\Delta+\mathcal{M}^2_{11}-\mathcal{M}^2_{22}}{2\Delta}}=\sqrt{\frac{\mathcal{M}^2_{11}-m_h^2}{m_H^2-m_h^2}}\,,
\eeq
and the sign of $s_\alpha$ is given by the sign of $\mathcal{M}_{12}^2$.  Explicitly, we have
\beq \label{sa}
s_\alpha=\frac{\sqrt{2}\,\mathcal{M}_{12}^2}{\sqrt{\Delta(\Delta+\mathcal{M}^2_{11}-\mathcal{M}^2_{22})}}={\rm sgn}(\mathcal{M}^2_{12})\sqrt{\frac{m_H^2-\mathcal{M}^2_{11}}{m_H^2-m_h^2}}\,.
\eeq
In deriving \eqs{ca}{sa}, we have assumed that $\mhl\neq\mhh$.   The case of $\mhl=\mhh$ is singular; in this case, the angle $\alpha$ is undefined since any two linearly independent combinations
of $h$ and $H$ can serve as the physical states.   In the rest of this paper, we shall not consider this mass-degenerate case further.

\subsection{SM-limit in the Higgs basis} 

The scalar potential given in \eq{lambdapotential} is expressed in the $\mathbb{Z}_2$-basis of scalar doublet fields in which the $\mathbb{Z}_2$ discrete symmetry of the quartic terms is manifest.  
It will prove convenient to re-express the scalar doublet fields in the Higgs basis~\cite{Branco:1999fs,Davidson:2005cw}, defined by
\beq \label{cphiggsbasisfields}
H_1=\begin{pmatrix}H_1^+\\ H_1^0\end{pmatrix}\equiv \Phi_1\cb+\Phi_2\sb\,,
\qquad\quad H_2=\begin{pmatrix} H_2^+\\ H_2^0\end{pmatrix}\equiv -\Phi_1\sb+\Phi_2\cb\,,
\eeq
so that $\langle{H_1^0}\rangle=v/\sqrt{2}$ and $\langle{H_2^0}\rangle=0$.  The scalar doublet $H_1$ possesses SM tree-level couplings to all the SM particles.
Therefore, if one of the CP-even neutral Higgs mass eigenstates is SM-like, then it must be approximately aligned with the real part of the neutral field $H_1^0$.

The scalar potential, when expressed in terms of the doublet fields, $H_1$ and $H_2$, has the same form as \eq{pot},
\beqa
\mathcal{V}&=& Y_1 H_1^\dagger H_1+Y_2 H_2^\dagger H_2
+Y_3[H_1^\dagger H_2+{\rm h.c.}]
+\half Z_1(H_1^\dagger H_1)^2
+\half Z_2(H_2^\dagger H_2)^2
+Z_3(H_1^\dagger H_1)(H_2^\dagger H_2)
\nonumber\\[8pt]
&&\quad\quad 
+Z_4(H_1^\dagger H_2)(H_2^\dagger H_1)
+\left\{\half Z_5(H_1^\dagger H_2)^2
+\big[Z_6(H_1^\dagger H_1)
+Z_7(H_2^\dagger H_2)\big]
H_1^\dagger H_2+{\rm h.c.}\right\}, \label{potZ}
\eeqa
where the $Y_i$ are real linear combinations of the $m_{ij}^2$ and the $Z_i$ are real linear combinations of the $\lambda_i$.
In particular, since $\lambda_6=\lambda_7=0$, we have~\cite{Davidson:2005cw,Haber:2015pua}\footnote{To make contact with the notation
of \Ref{Gunion:2002zf},
$\lambda\equiv Z_1$, $\lambda_V\equiv Z_2$,
$\lambda_T\equiv Z_3+Z_4-Z_5$, $\lambda_F\equiv Z_5-Z_4$,
$\lambda_A\equiv Z_1-Z_5$, $\hat{\lambda}\equiv -Z_6$ and
$\lambda_{U}\equiv -Z_7$.}
\beqa
Z_1 & \equiv & \lambda_1 c^4_\beta+\lambda_2 s^4_\beta+\half\lamtil s^2_{2\beta}\,,\label{zeeone}\\
Z_2 & \equiv & \lambda_1 s^4_\beta+\lambda_2 c^4_\beta+\half\lamtil s^2_{2\beta}\,,\label{zeetwo}\\
Z_i & \equiv & \tfrac{1}{4} s^2_{2\beta}\bigl[\lambda_1+\lambda_2-2\lamtil\bigr]+\lambda_i\,,\quad \text{(for $i=3,4$ or 5)}\,,\label{zeefive}\\
Z_6 & \equiv & -\half s_{2\beta}\bigl[\lambda_1 c^2_\beta-\lambda_2 s^2_\beta-\lamtil c_{2\beta}\bigr]\,,\label{zeesix} \\
Z_7 & \equiv & -\half s_{2\beta}\bigl[\lambda_1 s^2_\beta-\lambda_2 c^2_\beta+\lamtil c_{2\beta}\bigr]\,.\label{zeeseven}
\eeqa
Since there are five nonzero $\lambda_i$ and seven nonzero $Z_i$, there must be two relations.  The following two identities are satisfied if $\beta\neq 0$, $\tfrac{1}{4}\pi$, $\half\pi$~\cite{Haber:2015pua}:\footnote{For $\beta=0$, $\half\pi$, the $\mathbb{Z}_2$-basis and the Higgs basis coincide, in which case $Z_6=Z_7=0$ and $Z_1$, $Z_2$, $Z_{345}$ are independent quantities.   
For $\beta=\tfrac{1}{4}\pi$, the two relations are $Z_1=Z_2$ and $Z_6=Z_7$, and $Z_{345}$ is an independent quantity.}  
\beqa \label{ids}
Z_2&=& Z_1+2(Z_6+Z_7)\cot 2\beta\,,\label{ztwo}\\
Z_{345}&=&Z_1+2Z_6\cot 2\beta-(Z_6-Z_7)\tan 2\beta\,,\label{z345}
\eeqa
where $Z_{345}\equiv Z_3+Z_4+Z_5$.  One can invert the expressions given in \eqst{zeeone}{zeeseven}, subject to the relations given by \eqs{ztwo}{z345}.  The results are given in Appendix~\ref{AppInverse}.

The squared mass parameters $Y_i$ are given by
\beqa
Y_1&=& m_{11}^2 c^2_\beta+m_{22}^2 s^2_\beta-m_{12}^2 s_{2\beta}\,,\\
Y_2&=& m_{11}^2 s^2_\beta+m_{22}^2 c^2_\beta+m_{12}^2 s_{2\beta}\,,\\
Y_3&=& \half(m_{22}^2-m_{11}^2)s_{2\beta}-m_{12}^2 c_{2\beta}\,.\label{whythree}
\eeqa
$Y_1$ and $Y_3$ are fixed by the scalar potential minimum conditions,
\beq \label{sminconds}
Y_1=-\half Z_1 v^2\,,\qquad\quad Y_3=-\half Z_6 v^2\,.
\eeq
Using \eqs{mbar}{sminconds}, we can express $\overline{m}^{\,2}$ in terms of $Y_2$, $Z_1$ and $Z_6$,
\beq \label{mbarid}
\overline{m}^{\,2}=Y_2+\half Z_1 v^2+Z_6 v^2\cot 2\beta\,.
\eeq

\noindent The masses of $H^\pm$ and $A$ are given by
\beqa
\mhpm^2&=&Y_2+\half Z_3 v^2\,,\label{chhiggs}\\
\mha^2&=& Y_2+\half(Z_3+Z_4-Z_5)v^2\label{cpodd}\,.
\eeqa

It is straightforward to compute the  CP-even Higgs squared-mass matrix in the Higgs basis~\cite{Haber:2006ue,Branco:1999fs}, 
 \beq \label{Hmm}
\mathcal{M}_H^2=\begin{pmatrix} Z_1 v^2 & \quad Z_6 v^2 \\  Z_6 v^2 & \quad m_A^2+Z_5 v^2\end{pmatrix}\,.
\eeq
From \eq{Hmm}, one can immediately derive the conditions that yield a SM-like Higgs boson.  Since
$\langle{H_1^0}\rangle=v/\sqrt{2}$ and $\langle{H_2^0}\rangle=0$, the couplings of $H_1$
are precisely those of the Standard Model.  Thus a SM-like Higgs boson exists if $\sqrt{2}\,{\rm Re}~H_1^0-v$
is an approximate mass eigenstate.
That is, the mixing of $H_1^0$ and $H_2^0$ is subdominant, which implies that either
$|Z_6|\ll 1$ and/or $m_A^2+Z_5 v^2\gg Z_1 v^2$, $Z_6 v^2$.   Moreover, if in addition $Z_1 v^2<\mha^2+Z_5 v^2$, then $h$ is SM-like,
whereas if $Z_1 v^2>\mha^2+Z_5 v^2$, then $H$ is SM-like.  In both cases, the squared-mass of the SM-like Higgs
boson is approximately equal to $Z_1 v^2$.  

The physical mass eigenstates are identified from \eq{HZ2scalareigenstates},~\eqref{hZ2scalareigenstates} and~\eqref{cphiggsbasisfields} as
\beqa
\hh &=&(\sqrt{2}{\rm Re\,}H_1^0-v)\cbma-
\sqrt{2}{\rm Re\,}H_2^0\sbma\,,\label{Hscalareigenstates}\\
\hl &=&(\sqrt{2}{\rm Re\,}H_1^0-v)\sbma+
\sqrt{2}{\rm Re\,}H_2^0\cbma\,.
\label{hscalareigenstates}
\eeqa
Then, \eqs{higgsmasses}{deltadef} yield
\beq
m^2_{H,h}=\half\bigl[\mha^2+(Z_1+Z_5)v^2\pm\Delta_H\bigr]\,,
\eeq
where
\beq
\Delta_H\equiv\sqrt{\bigl[\mha^2+(Z_5-Z_1)v^2\bigr]^2+4Z_6^2 v^4}\,.
\eeq
In addition, \eq{relation} yields
\beq \label{Hrelation}
|Z_6| v^2=\sqrt{\bigl(\mhh^2-Z_1 v^2)(Z_1 v^2-\mhl^2\bigr)}\,.
\eeq

Comparing \eqs{HZ2scalareigenstates}{hZ2scalareigenstates} with \eqs{Hscalareigenstates}{hscalareigenstates}, we identify the corresponding mixing angle by
$\alpha-\beta$, which is defined modulo $\pi$.  Diagonalizing the squared mass matrix, \eq{Hmm},
it is straightforward to derive the following expressions:
\beqa
Z_1 v^2&=&\mhl^2 s^2_{\beta-\alpha}+\mhh^2 c^2_{\beta-\alpha}\,,\label{z1v}\\
Z_6 v^2&=&(\mhl^2-\mhh^2)\sbma\cbma\,,\label{z6v} \\
\mha^2+Z_5 v^2&=&\mhh^2 s^2_{\beta-\alpha}+\mhl^2 c^2_{\beta-\alpha}\,.\label{z5v}
\eeqa
It follows that
\beqa
\mhl^2&=&\left(Z_1 +Z_6 \frac{\cbma}{\sbma}\right)v^2\,,\label{hl2mass} \\
\mhh^2&=&m_A^2+\left(Z_5-Z_6 \frac{\cbma}{\sbma}\right)v^2\,.\label{hh2mass} 
\eeqa
Note that \eq{z6v} implies that\footnote{Having established a convention where $0\leq\beta\leq\half\pi$, we are no longer free to redefine the Higgs basis field $H_2\to -H_2$.  Consequently, the sign of $Z_6$ is meaningful in this convention.}
\beq \label{sgnz6}
Z_6\sbma\cbma\leq 0\,.
\eeq

One can also derive expressions for $\cbma$ and $\sbma$ either directly from \eqs{z1v}{z6v} or by using
\eqs{ca}{sa} with $\alpha$ replaced by $\alpha-\beta$.  Using \eq{sgnz6}, the sign of the product $\sbma\cbma$ is fixed by
the sign of $Z_6$.  However, since $\beta-\alpha$ is defined only modulo $\pi$, we are free to choose
a convention where either $\cbma$ or $\sbma$ is always non-negative.\footnote{Such a convention, if adopted, would replace the convention employed in \eq{ca} in which $c_\alpha$ is taken to be non-negative.} 
In a convention where
$\sbma$ is non-negative (this is a convenient choice when the $h$ is SM-like),
\beq \label{cbmaeq}
\cbma=-\sgn(Z_6)\sqrt{\frac{Z_1 v^2-\mhl^2}{\mhh^2-\mhl^2}}=\frac{-Z_6 v^2}{\sqrt{(\mhh^2-\mhl^2)(\mhh^2-Z_1 v^2)}}\,,
\eeq
where we have used \eq{Hrelation} to obtain the second form for $\cbma$ in \eq{cbmaeq}.

Finally, we record the following useful formula that is easily obtained from 
\eqs{massha}{Z345},\footnote{In \eq{useful}, the term in the expression for $\overline{m}^{\,2}$ that is
proportional to $(Z_6-Z_7)v^2\tan 2\beta$ is never greater than $\mathcal{O}(v^2)$ for \textit{all} values of
$\beta$, since \eqs{zeesix}{zeeseven} imply that $(Z_6-Z_7)\tan 2\beta=-\half s^2_{2\beta}(\lambda_1-\lambda_2-2\lamtil)\lsim\mathcal{O}(1)$. \label{fn}}
\beq \label{useful}
\overline{m}^{\,2}=\mha^2+Z_5 v^2+\half(Z_6-Z_7)v^2\tan 2\beta\,.
\eeq
Combining \eq{useful} with \eqs{z6v}{z5v} yields
\beq \label{z7v}
Z_7 v^2=(\mhl^2-\mhh^2)\sbma\cbma+2\cot 2\beta\bigl[\mhh^2 s^2_{\beta-\alpha}+\mhl^2 c^2_{\beta-\alpha}-\overline{m}^{\,2}\bigr]\,.
\eeq
Using \eqs{ztwo}{z345}, one can likewise obtain expressions for $Z_2 v^2$ and $Z_{345}v^2$ in terms of $m_h^2$, $m_H^2$,
and $\overline{m}^{\,2}$.  However, these expressions are not particularly illuminating, so we do not write them out explicitly here.

\subsection{Higgs couplings and the alignment limit}

As noted in the previous subsection, the Higgs basis field $H_1$ behaves precisely as the Standard Model Higgs
boson.  Thus, if one of the neutral CP-even Higgs mass eigenstates is approximately aligned with $\sqrt{2}\,{\rm Re\,}H_1^0-v$,
then its properties will approximately coincide with those of the SM Higgs boson.  Thus, we shall define the
\textit{alignment limit} as the limit in which the one of the two neutral CP-even Higgs mass eigenstates
aligns with the direction of the scalar field vacuum expectation values.   Defined in this way, it is clear that
the alignment limit is independent of the choice of basis for the two Higgs doublet fields.  Nevertheless,
the alignment limit is most clearly exhibited in the Higgs basis.   In light of \eqs{Hscalareigenstates}{hscalareigenstates},
the alignment limit corresponds either to the limit of $\cbma\to 0$ if $h$ is identified as the SM-like Higgs boson,
or to the limit of $\sbma\to 0$ if $H$ is identified as the SM-like Higgs boson.

Consider first the case of a SM-like $h$, with $m_h\approx 125$~GeV.  In this case, $Z_1 v^2<m_A^2+Z_5 v^2$, $|\cbma|\ll 1$, and $m_h^2\approx Z_1 v^2$.  It follows from
\eq{cbmaeq} that the alignment limit can be achieved in two ways: (i) $Z_6\to 0$ or (ii) $\mhh\gg v$.
The case of $\mhh\gg v$ (or equivalently $Y_2\gg v$) is called the \textit{decoupling limit} in the literature.\footnote{More precisely, we are
assuming that $\mhh^2\gg |Z_6|v^2$.  Since $Z_6$ is a dimensionless coefficient in the Higgs basis scalar potential,
we are implicitly assuming that $Z_6$ cannot get too large without spoiling perturbativity and/or unitarity.
One might roughly expect $|Z_6|\lsim 4\pi$, in which case $\mhh\gg v$ provides a reasonable indication of the domain
of the decoupling limit.}
In this case, one
finds that $\mhh\sim\mha\sim\mhpm$, so one can integrate out the heavy scalar states below the scale of $\mhh$.
The effective Higgs theory below the scale $\mhh$ is a theory with one Higgs doublet and corresponds to the Higgs
sector of the Standard Model.  Thus not surprisingly, $h$ is a SM-like Higgs boson.  However, it is possible
to achieve the alignment limit even if the masses of all scalar states are similar in magnitude in the limit of
$Z_6\to 0$.  This is the case of \textit{alignment without decoupling} and the main focus of this study. 
Finally, if both $|Z_6|\ll 1$ and $\mhh\gg\mhl$ are satisfied, the alignment is even more pronounced; 
when relevant we shall denote this case as the \textit{double decoupling limit}. 

For completeness we note that in the case of a SM-like $H$ we have $Z_1 v^2>m_A^2+Z_5 v^2$, $|\sbma|\ll 1$ and $\mhh^2\approx Z_1 v^2$.
Here, it is more convenient to employ a convention where $\cbma$ is non-negative.  One can then
use \eqss{Hrelation}{sgnz6}{cbmaeq} to obtain an expression for $\sbma$.  In a convention where
$\cbma$ is non-negative,
\beq
\sbma=-\sgn(Z_6)\sqrt{\frac{\mhh^2-Z_1 v^2}{\mhh^2-\mhl^2}}=\frac{-Z_6 v^2}{\sqrt{(\mhh^2-\mhl^2)(Z_1 v^2-\mhl^2)}}\,.
\eeq
Taking $m_H\approx 125$~GeV, there is no decoupling limit as in the case of a SM-like $h$.
However, the alignment limit without decoupling can be achieved in the limit of $Z_6\to 0$. 
This case will be discussed in detail in \cite{part2}. 

We now turn to the tree-level Higgs couplings.  Denoting the SM Higgs boson by $h_{\rm SM}$, the coupling of the
CP-even Higgs bosons to $VV$ (where $V=W^\pm$ or $Z$) normalized to the $h\ls{\rm SM}VV$ coupling is given by
\beq
C_V^{h}=\sbma\,,\qquad\quad C_V^H=\cbma\,.
\eeq
As expected, if $h$ is a SM-like Higgs boson then $C_V^{h}\approx 1$ in the alignment limit, whereas if
$H$ is a SM-like Higgs boson then $C_V^{H}\approx 1$ in the alignment limit.

Next, we consider the Higgs boson couplings to fermions.  The most general renormalizable
Yukawa couplings of the two Higgs doublets to a single generation of up and down-type quarks and leptons (using third generation notation) is given by
\beq \label{thdmyuk}
-\mathscr{L}_{\rm Yuk}=\mathcal{Y}^1_b\overline{b}_R\Phi_1^{i\,*}Q^i_L
+\mathcal{Y}^2_b\overline{b}_R\Phi_2^{i\,*}Q^i_L+\mathcal{Y}^1_\tau\overline{\tau}_R\Phi_1^{i\,*}L^i_L
+\mathcal{Y}^2_\tau\overline{\tau}_R\Phi_2^{i\,*}L^i_L
+\epsilon_{ij}\bigl[\mathcal{Y}^1_t\overline{t}_R Q_L^i\Phi^j_1+\mathcal{Y}^2_t\overline{t}_R Q_L^i\Phi^j_2\bigr]
+{\rm h.c.}\,,
\eeq
where $\epsilon_{12}=-\epsilon_{21}=1$, $\epsilon_{11}=\epsilon_{22}=0$, $Q_L=(t_L\,,\,b_L)$
and $L_L=(\nu_L\,,\,e_L)$ are the doublet left handed quark and lepton fields and
$t_R$, $b_R$ and $e_R$ are the singlet right-handed quark and lepton fields.  However, if all terms in \eq{thdmyuk} are
present, then tree-level Higgs-mediated FCNCs would be present, in conflict with experimental constraints.
To avoid tree-level Higgs-mediated FCNCs, we
extend the discrete $\mathbb{Z}_2$ symmetry to the Higgs-fermion Lagrangian.   There are four possible choices for
the transformation properties of the fermions with respect to $\mathbb{Z}_2$, which we exhibit in Table~\ref{Tab:type}.

\begin{table}[t!]
 \begin{center}
 \caption{Four possible $\mathbb{Z}_2$ charge assignments that forbid
tree-level Higgs-mediated FCNC effects in the 2HDM~\cite{Aoki:2009ha}.}
\label{Tab:type}
\begin{tabular}{|cl||c|c|c|c|c|c|}
\hline && $\Phi_1$ & $\Phi_2$ & $t_R^{}$ & $b_R^{}$ & $\tau_R^{}$ &
 $t_L$, $b_L$, $\nu_L$, $e_L$ \\  \hline
Type I  && $+$ & $-$ & $-$ & $-$ & $-$ & $+$ \\
Type II && $+$ & $-$ & $-$ & $+$ & $+$ & $+$ \\
Type X  &(lepton specific) & $+$ & $-$ & $-$ & $-$ & $+$ & $+$ \\
Type Y  &(flipped) & $+$ & $-$ & $-$ & $+$ & $-$ & $+$ \\
\hline
\end{tabular}
\end{center}
\end{table}

For simplicity, we shall assume in this paper that the pattern of the Higgs couplings to down-type quarks and leptons is the same.
This leaves two possible choices for the Higgs-fermion couplings~\cite{Hall:1981bc}:
\beqa
&&\text{Type I:}\quad\phantom{I} \mathcal{Y}_t^1=\mathcal{Y}_b^1=\mathcal{Y}_\tau^1=0\,,\\
&&\text{Type II:}\quad \mathcal{Y}_t^1=\mathcal{Y}_b^2=\mathcal{Y}_\tau^2=0\,.
\eeqa
In particular, the pattern of fermion couplings to the neutral Higgs bosons in the Type~I and Type~II models
is exhibited in Table~\ref{tab:2hdm-couplings}.

In the strict alignment limit, the fermion couplings to the SM-like Higgs boson should approach their Standard Model values.  To see this explicitly, we note
the identities,
\beqa
\frac{\cos\alpha}{\sin\beta}&=&\sbma+\cot\beta\cbma\,,\\
-\frac{\sin\alpha}{\cos\beta}&=& \sbma-\tan\beta\cbma\,,\\
\frac{\sin\alpha}{\sin\beta}&=& \cbma-\cot\beta\sbma\,,\\
\frac{\cos\alpha}{\cos\beta}&=&\cbma+\tan\beta\sbma\,.\label{yukid4}
\eeqa
If $h$ is the SM-like Higgs boson, then in the limit of $\cbma\to 0$, the fermion couplings of $h$ approach their Standard Model values.
However, if $\tan\beta\gg 1$, then the alignment limit is realized in the Type-II Yukawa couplings to down-type fermions only if
$|\cbma|\tan\beta\ll 1$.   That is, if $|\cbma|\ll 1$ but $|\cbma|\tan\beta\sim\mathcal{O}(1)$, then
the $hVV$ couplings and the $ht\bar{t}$ couplings are SM-like whereas the $hb\bar{b}$ and $h\tau^+\tau^-$ couplings deviate from their Standard Model values.   Thus the approach to the alignment limit is \textit{delayed} when $\tan\beta\gg 1$.  We denote this phenomenon as the \textit{delayed alignment limit}.  Similar considerations apply if $\cot\beta\gg 1$; however, this region of parameter space is disfavored as the corresponding $ht\bar{t}$ coupling quickly becomes non-perturbative if $\cot\beta$ is too large.

\begin{table}[t!]
\begin{center}
\caption{Tree-level vector boson couplings $C_V$ ($V=W,Z$) and fermionic couplings $C_{F}$
normalized to their SM values for the two scalars $h,H$ and the pseudoscalar $A$
in Type I and Type II Two-Higgs-doublet models.}
\label{tab:2hdm-couplings}
\begin{tabular}{|c|c|c|c|c|c|}
\hline
\ & Type I and II  & \multicolumn{2}{c|}  {Type I} & \multicolumn{2}{c|}{Type II} \\
\hline
Higgs & $VV$ & up quarks & down quarks  & up quarks & down quarks  \\
&  &  &  and leptons &  &  and leptons \cr
\hline
 $h$ & $\sin(\beta-\alpha)$ & $\cosa/ \sinb$ & $\cosa/ \sinb$  &  $\cosa/\sinb$ & $-{\sina/\cosb}$   \\
\hline
 $H$ & $\cos(\beta-\alpha)$ & $\sina/ \sinb$ &  $\sina/ \sinb$ &  $\sina/ \sinb$ & $\phm\cosa/\cosb$ \\
\hline
 $A$ & 0 & $\cotb$ & $-\cotb$ & $\cotb$  & $\tanb$ \\
\hline
\end{tabular}
\end{center}
\end{table}

Finally, we examine the trilinear Higgs self-couplings. Using the results of \Ref{Gunion:2002zf} (see also \Ref{Haber:2006ue}), 
the three-Higgs vertex Feynman rules
(including the corresponding symmetry factor for identical particles but excluding an overall factor of $i$) are given by:
\beqa
\!\!\!\!\!\! g\ls{\hl\ha\ha} &=&
   {-v}\bigl[(Z_3+Z_4-Z_5)\sbma+Z_7\cbma\bigr]\,, \label{hlhaha} \\[4pt]
\!\!\!\!\!\! g\ls{\hh\ha\ha} &=&
   {-v}\bigl[(Z_3+Z_4-Z_5)\cbma-Z_7\sbma\bigr]\,, \\[4pt]
\!\!\!\!\!\! g\ls{\hl\hh\hh} &=& -{3v}\bigl[
   Z_1\sbma\cbmaii
   +Z_{345}\sbma\left(\tfrac{1}{3}-\cbmaii\right)+Z_6\cbma(1-3\sbmaii)
    +Z_7\sbmaii\cbma\bigr]\,, \\[4pt]
\!\!\!\!\!\! g\ls{\hh\hl\hl} &=& -{3v}\bigl[
   Z_1\cbma\sbmaii
   +Z_{345}\cbma\left(\tfrac{1}{3}-\sbmaii\right)-Z_6\sbma(1-3\cbmaii)
    -Z_7\cbmaii\sbma\bigr]\,, \label{hhhlhl} \\[4pt]
\!\!\!\!\!\! g\ls{\hl\hl\hl} &=& {-3v}\bigl[
    Z_1\sbmaiii+Z_{345}\sbma\cbmaii+3Z_6\cbma\sbmaii
    +Z_7 c^3_{\beta-\alpha}\bigr]\,, \label{hlhlhl}\\[4pt]
\!\!\!\!\!\! g\ls{\hh\hh\hh} &=& {-3v}\bigl[
    Z_1\cbmaiii+Z_{345}\cbma\sbmaii-3Z_6\sbma\cbmaii
    -Z_7 s^3_{\beta-\alpha}\bigr]\,,\label{hhhhhh} \\[4pt]
\!\!\!\!\!\! g\ls{\hl\hp\hm} &=& \label{hhhphz}
   {-v}\bigl[Z_3\sbma+Z_7\cbma\bigr]\,,\\[4pt]
\!\!\!\!\!\! g\ls{\hh\hp\hm} &=&
   {-v}\bigl[Z_3\cbma-Z_7\sbma\bigr]\,. \label{hhhphm}
\eeqa
The trilinear Higgs couplings expressed in terms of the physical Higgs masses are given in Appendix~\ref{AppTrilinear}.

Consider the alignment limit, $\cbma\to 0$, where $h$ is SM-like.  Then \eqs{hl2mass}{hlhlhl} yield,\footnote{\Eq{hhhalign} is obtained in the convention
where $\sbma$ is non-negative, i.e.~$\sbma$ is close to 1.}
\beq \label{hhhalign}
g_{hhh}= g^{\rm SM}_{hhh}\biggl[1+\frac{2Z_6}{Z_1}\cbma+\left(\frac{Z_{345}}{Z_1}-\frac{2Z_6^2}{Z_1^2}-\frac{3}{2}\right)
c_{\beta-\alpha}^2+
\mathcal{O}(c_{\beta-\alpha}^3)\biggr]\,, 
\eeq
where the self-coupling of the SM Higgs boson is given by
\beq
g^{\rm SM}_{hhh}=-\frac{3\mhl^2}{v}\,.
\eeq
Note that in the alignment limit, $m_h^2\approx Z_1 v^2$ [cf. \eq{z1v}], which implies that $Z_1\approx 0.26$.

It is convenient to make use of \eq{cbmaeq} [in a convention where $\sbma\geq 0$] to write
\beq \label{cz}
\cbma=-\eta Z_6\,,
\eeq
where
\beq
\label{etastuff}
\eta\equiv \frac{v^2}{\sqrt{(\mhh^2-\mhl^2)(\mhh^2-Z_1 v^2)}}=\begin{cases}  \mathcal{O}(1)\,, & \text{for $\mhh^2\sim\mathcal{O}(v^2)$,} \\[10pt] \mathcal{O}\left(\displaystyle\frac{v^2}{\mhh^2}\right)\ll 1\,, & \text{in the decoupling limit.}\end{cases}
\eeq
Inserting \eq{cz} in \eq{hhhalign} yields
\beq \label{previous}
g_{hhh}= g^{\rm SM}_{hhh}\biggl\{1+\biggl[\bigl(Z_{345}-\tfrac{3}{2}Z_1\bigr)\eta^2-2\eta\biggr]\frac{Z_6^2}{Z_1}
+\mathcal{O}(\eta^3 Z_6^3)+\mathcal{O}(\eta^2 Z_6^4)\biggr\}\,.
\eeq

In the decoupling limit (where $\eta\ll 1$), 
\beq
g_{hhh}= g^{\rm SM}_{hhh}\biggl\{1-\frac{2\eta Z_6^2}{Z_1}+\mathcal{O}(\eta^2 Z_6^2)\biggr\}\,.
\label{hhhdecoup}
\eeq
It follows that $g_{hhh}$ is always suppressed with respect to the SM in the decoupling limit.\footnote{In the double decoupling limit where $\eta\ll 1$ \textit{and} $|Z_6|\ll 1$, \eq{hhhdecoup} shows that the deviation of $g_{hhh}$ from the corresponding SM value is highly suppressed.}
This behavior is confirmed in our numerical analysis.
In contrast, in the alignment limit without decoupling, $|Z_6|$ is significantly smaller than 1 and $\eta\sim\mathcal{O}(1)$.   It is now convenient to use
\eq{z345} to eliminate $Z_{345}$,
\beq \label{ghhh}
g_{hhh}= g^{\rm SM}_{hhh}\biggl\{1+\biggl[\bigl(Z_7\tan 2\beta-\tfrac{1}{2}Z_1\bigr)\eta^2-2\eta\biggr]\frac{Z_6^2}{Z_1}
+(2\cot 2\beta-\tan 2\beta)\eta^2\frac{Z_6^3}{Z_1}+\mathcal{O}(Z_6^3)\biggr\}\,,
\eeq
where the term above designated by $\mathcal{O}(Z_6^3)$ contains no potential enhancements in the limit of $s_{2\beta}\to 0$ or $c_{2\beta}\to 0$.  Given that $\eta\sim\calo( 1)$ in the alignment limit without decoupling, the form of \eq{ghhh} suggests two ways in which $g_{hhh}$ can be enhanced with respect to the SM.   For example if $\tan\beta\sim 1$, then one must satisfy
$(Z_7-Z_6)\eta\tan 2\beta\gsim 2+\half Z_1\eta$. 
Alternatively, if $\tan\beta\gg 1$, then one must satisfy
$Z_6\eta\cot 2\beta\gsim 1+\frac{1}{4}Z_1\eta$ (the latter inequality requires $Z_6<0$, since $\cot 2\beta<0$ when $\tfrac{1}{4}\pi<\beta<\half\pi$).
In both cases, $g_{hhh}>g_{hhh}^{\rm SM}$ is possible even when  $|Z_6|$ and $|Z_7|$ are significantly smaller than 1.  Indeed, both of the above alternatives correspond to $Z_{345}\gg Z_1$ and $\eta Z_{345}\gg 1$ in \eq{previous}.

As a second example, consider the $hAA$ coupling given in \eq{hlhaha} [or \eq{haa}]. Using \eq{z345},
we find that in the alignment limit,
\beqa \label{haaalign}
g_{hAA}&=&-\frac{1}{v}\biggl\{\mhl^2-2Z_5 v^2-(Z_6-Z_7)v^2\tan 2\beta+2Z_6 v^2\cot 2\beta+\mathcal{O}(\cbma)\biggr\}\nonumber \\
&=&-\frac{1}{v}\biggl\{\mhl^2-2\lambda_5 v^2+2Z_6 v^2\cot 2\beta+\mathcal{O}(\cbma)\biggr\}\,,
\eeqa
A similar computation yields the $Hhh$ coupling given in \eq{hhhlhl} [or \eq{Hhh}],
\beq \label{Hhhalign}
g_{Hhh}=\frac{1}{v}\biggl\{3Z_6 v^2-\bigl[m_h^2-4Z_6v^2\cot 2\beta+2(Z_6-Z_7)v^2\tan 2\beta\bigr]\cbma+\mathcal{O}(\cbma^2)\biggr\}\,.
\eeq
In the alignment limit without decoupling, the $\mathcal{O}(1)$ terms in \eqs{haaalign}{Hhhalign} that are proportional to $Z_6$
should be regarded as terms of $\mathcal{O}(\cbma)$ [cf.~\eqs{cz}{etastuff}].   That is, the decoupling limit [with $Z_6\sim\mathcal{O}(1)$] and the alignment limit without decoupling can be distinguished in the trilinear Higgs couplings.  Indeed, the $Hhh$ coupling is suppressed in the alignment limit without decoupling, whereas it can be of $\mathcal{O}(v)$ in the decoupling limit.
All the other trilinear Higgs self-couplings can be analyzed in the alignment limit following the procedure outlined above.

Last but not least, it is noteworthy that 
\beq \label{HHHlim}
g_{hH^+ H^-}=-v\bigl[Z_3+\mathcal{O}(\cbma)\bigr]\,,
\eeq
approaches a finite nonzero value in the alignment limit, with or without decoupling.  This is relevant for the analysis of the one-loop process $h\to\gamma\gamma$, which has a contribution that is mediated by a $H^\pm$ loop.  In the decoupling limit, the charged Higgs loop amplitude is suppressed by a factor of $\mathcal{O}(v^2/m_{H^\pm}^2)$ relative to the $W^\pm$ and the top quark loop contributions.   But, in the alignment limit without decoupling, the charged Higgs loop is parametrically of the same order as the corresponding SM loop contributions, thereby leading to a shift of the $h\to\gamma\gamma$ decay rate from its SM value.  This is in stark contrast to the behavior of tree-level Higgs couplings, which approach their SM values in the alignment limit with or without decoupling.
That is, the loop-corrected Higgs couplings to SM particles approach their SM values 
in the decoupling limit, but can yield deviations in the alignment limit without decoupling due to internal loops involving light non-SM-like Higgs states.

Before concluding this section, we examine a second theoretical distinction between the decoupling limit and alignment limit without decoupling.  
The SM Higgs sector is famously unnatural~\cite{Weisskopf:1939zz,Altarelli:2013lla}.  In particular, a fine tuning of the Higgs sector squared-mass parameter is required in order to explain the observed value of the vacuum expectation value (vev), $v\approx 246$~GeV.    
The 2HDM generically requires two separate and independent fine tunings.  In addition to identifying $v\approx 246$~GeV, which fixes the values of $Y_1$ and $Y_3$ in \eq{sminconds}, one must also perform a second fine-tuning to fix the squared-mass parameter $Y_2$ to be of $\mathcal{O}(v^2)$.
Thus, the regime of the decoupling limit (where $Y_2\gg v^2$) is less fine-tuned than the general 2HDM, since the natural value for $Y_2$ is the ultraviolet cutoff of the theory beyond which new physics presumably enters.  As long as the heavier Higgs scalars (whose squared masses are of order $Y_2$) are sufficiently massive,
then $h$ will be SM-like.\footnote{In general, $\mhh^2\gg |Z_6|v^2$ is sufficient to guarantee SM-like $h$ couplings.
However, in the 2HDM with Type-II Yukawa coupling and $\tan\beta>1$, a SM-like $h$ coupling to down-type quarks and leptons requires
$\mhh^2\gg |Z_6|v^2\tan\beta$, leading to the phenomenon of delayed decoupling~\cite{Haber:2000kq,Gunion:2002zf,Carena:2001bg,Ferreira:2014naa} at large $\tan\beta$.  This is a special case of delayed alignment introduced below \eq{yukid4}.}

In contrast, in the case of alignment without decoupling (or in the double decoupling limit), we have $|Z_6|\ll 1$, which is a finely-tuned region of the 2HDM parameter space (beyond the two tunings discussed above) unless we can demonstrate that $Z_6=0$ is a consequence of an enhanced symmetry of the theory.  The possibility of a natural implementation of alignment has been previously treated in~\cite{Dev:2014yca}. In the absence of Higgs--fermion Yukawa couplings, it is sufficient to consider the symmetry properties of the scalar potential.  Note that we have already imposed a softly-broken $\mathbb{Z}_2$ symmetry, which yields $\lambda_6=\lambda_7=0$ in the original basis.   In addition,
we observe that
$Z_6=Z_7=0$ [which also implies that $Y_3=0$ in light of \eq{sminconds}] corresponds to an exact
$\mathbb{Z}_2$ symmetry in the Higgs basis.   

The conditions $Z_6=Z_7=0$ can be implemented in three ways.  If $s_{2\beta}=0$, then only one of the two Higgs fields acquires a non-zero vev.  This means that our original basis and the Higgs basis coincide (in a convention where $H_1$ denotes the Higgs field with the non-zero vev), in which case the original $\mathbb{Z}_2$ symmetry is unbroken.  If $\lambda_6=\lambda_7=0$ and $s_{2\beta}c_{2\beta}\neq 0$, then setting $Z_6=Z_7=0$ in
\eqs{zeesix}{zeeseven} yields $\lambda_1=\lambda_2=\lambda_{345}$.   Such a scalar potential
exhibits a softly-broken CP3 symmetry, one of the three possible generalized CP symmetries
that can be imposed on the 2HDM~\cite{Ferreira:2009wh}.\footnote{If $m_{12}^2=0$ in \eq{lambdapotential} in addition to $\lambda_6=\lambda_7=0$, then the $\mathbb{Z}_2$ discrete symmetry ($\Phi_1\to +\Phi_1$, $\Phi_2\to -\Phi_2$) is exact.  In this case, $Z_6=Z_7=0$ implies that $\lambda_1=\lambda_2=\lambda_{345}$ \textit{and} $m_{11}^2=m_{22}^2$ [the latter via \eq{whythree}], and corresponds to an \textit{exact} CP3 symmetry of the scalar potential.  This restriction of scalar potential parameters has also been obtained in~\cite{Dev:2014yca}.}
Finally, if the scalar potential exhibits an exact CP2 symmetry, or equivalently there is a basis in which the $\mathbb{Z}_2$ discrete symmetry ($\Phi_1\to +\Phi_1$, $\Phi_2\to -\Phi_2$) and a second $\mathbb{Z}_2$ interchange symmetry ($\Phi_1\longleftrightarrow\Phi_2$) coexist~\cite{Davidson:2005cw,Ferreira:2009wh}, then it follows that
$\lambda_6=\lambda_7=0$, $\lambda_1=\lambda_2$ (with $\lambda_5$ real), $m_{11}^2=m_{22}^2$ and $m_{12}^2=0$.   In this case, \eqs{minconditionsa}{minconditionsb} yield $\tan\beta=1$.\footnote{Here we assume that $\lambda_1\neq \lambda_{345}$; otherwise, the CP2 symmetry is promoted to the CP3 symmetry previously considered.}
The latter can be maintained when the CP2 symmetry is softly broken such that $m_{12}^2\neq 0$.  Using \eqs{zeesix}{zeeseven} then yields $Z_6=Z_7=0$.
Thus, in the absence of the Higgs-fermion Yukawa couplings, $Z_6=0$ is a consequence of an enhanced symmetry of the scalar potential, in which case the regime of alignment without decoupling and the double decoupling regime
are both natural in the sense of 't Hooft\cite{'tHooft:1979bh}.

If we now include the Higgs-fermion Yukawa coupling, we can still maintain the symmetry of the scalar potential in special cases.  If the $\mathbb{Z}_2$ symmetry transformation is defined in the Higgs basis such that $H_2$ is odd  (i.e.~$H_2\to -H_2$) and $H_1$ and all fermion and vector fields are even, then the resulting model corresponds a Type-I 2HDM with $s_{2\beta}=0$, which we recognize as
the inert 2HDM~\cite{Deshpande:1977rw,Barbieri:2006dq}.     Indeed,
if we perturb the inert 2HDM by taking $Z_6$ and $Z_7$ small, then either $h$ or $H$ will be approximately SM-like. 
In the case of $s_{2\beta}\neq 0$, we would need to extend the (softly-broken) CP3 or CP2 symmetry of the scalar potential to the Higgs-fermion Yukawa sector.   As shown in \cite{Ferreira:2010bm}, 
no phenomenologically acceptable CP2-symmetric model exists.
A unique softly-broken CP3-symmetric 2HDM does exist with an acceptable fermion mass spectrum; however this model does not appear to be phenomenologically viable due to insufficient CP-violation and potentially large FCNC effects~\cite{Ferreira:2010bm}. 
Hence, for generic choices of the 2HDM parameters,  the regime of alignment without decoupling and the double decoupling regime must be regarded as more finely tuned than the generic 2HDM.

\section{\boldmath Setup of the numerical analysis}\label{setup}

In this section, we give details on the numerical procedure.  
In particular, we describe the scan of the 2HDM parameter space and the different constraints coming from theoretical requirements, signal strengths of the observed 125 GeV Higgs state, flavor physics and direct searches for extra Higgs states. 

Imposing a  softly broken $\mathbb{Z}_2$ symmetry ($\Phi_1\to +\Phi_1$, $\Phi_2\to -\Phi_2$) on the scalar potential given in \eq{lambdapotential} which sets $\lambda_6=\lambda_7=0$, the free parameters of the 2HDM scalar potential can be chosen to be the four physical Higgs masses $m_h, m_H, m_{H^\pm}, m_A$, the mass term $m_{12}^2$, the ratio of the two Higgs vacuum expectation values $\tan\beta$ and the mixing angle $\alpha$ of the CP-even Higgs squared-mass matrix. In this study, we choose the following ranges for the scan,
\beqa
\alpha\in[&-\pi/2,\pi/2],\ \ \  \tan\beta\in[0.5, 60], \ \ \ m_{12}^2\in[-(2000 \text{ GeV})^2,(2000\text{ GeV})^2], \nonumber\\[8pt]
&m_{H^\pm}\in[m^{*\!},2000\text{ GeV}], \ \ \ m_A\in[5\text{ GeV}, 2000\text{ GeV}],
\eeqa
where $m^*$ is a lower bound on the charged Higgs mass originating either from the LEP direct searches~\cite{Abbiendi:2013hk} or constraints from $B$-physics; mainly from the $Z\to b\bar{b}$ ($R_b$),  $\epsilon_K, \Delta m_{B_s}$, $B\to X_s\gamma$ and $B\to \tau\nu$ constraints~\cite{Haisch:2008ar,Mahmoudi:2009zx,Gupta:2009wn,Jung:2010ik,Misiak:2015xwa}.
In principle both $h$ and $H$ can have the same properties as the SM Higgs and thus serve as possible candidates for the observed SM-like Higgs state. In this paper, we consider $m_h\equiv 125.5\gev$\footnote{Having performed 
the parameter scans before the publication of \cite{Aad:2015zhl} which reports a central value of the Higgs mass of 125.09 GeV, we use $125.5\gev$ as the observed Higgs mass in this analysis.}, taking 
\beq
	m_H\in[129.5\text{ GeV},2000\text{ GeV}],
\eeq

As mentioned in Section~\ref{higgsscalarpotential}, the degenerate case $m_h\approx m_H$ is not considered in this study. Instead, we  require a 4 GeV mass splitting between $h$ and $H$ in order to avoid $H$ contamination of the $h$ signal.
Since we are primarily interested in the case that the electroweak gauge bosons acquire most of their masses from only one of the Higgs basis doublet fields, we impose $\sba\geq 0.99$, which translates into $|\cba|\lesssim 0.14$. This implies that we are allowing at most a 1\% deviation from $C^h_V=1$.
This should be compared with the expected ultimate precision for $C_V$ of about $2$--$4\%$ at the high-luminosity LHC, and about $0.2$--$0.5\%$ at the ILC~\cite{Dawson:2013bba,Asner:2013psa}.

We perform a flat random scan over this parameter space using the public code \texttt{2HDMC}~\cite{Eriksson:2009ws} for a precise state-of-the-art computation of the couplings and decay widths of the various Higgs states. 
Only points satisfying stability of the scalar potential [cf.~\eq{stability}], coupling perturbativity and tree-level S-matrix unitarity are retained. We also require the $S$, $T$, and $U$ Peskin-Takeuchi parameters~\cite{Peskin:1991sw} to be compatible with their corresponding values derived from electroweak precision observables~\cite{Baak:2014ora}. These constraints are also checked by means of \texttt{2HDMC}. 

Next we impose constraints from the non-observation of Higgs states other than the one at 125~GeV.   
From the LEP direct searches for light Higgs states, we consider the cross-section upper limits on $e^+e^-\to Zh/H$ and $e^+e^-\to Ah/H$ from~\cite{Barate:2003sz} and~\cite{Abbiendi:2004gn} respectively.  
For very light $A$ below $9.5\gev$, the limits from Upsilon decays \cite{Domingo:2008rr} are important, 
for which we follow the implementation in \texttt{NMSSMTools\,4.6.0}~\cite{nmssmtools}. 
Moreover, we consider the limits from CMS on light pseudo scalars decaying into $\mu^+\mu^-$~\cite{Chatrchyan:2012am}  in the mass range $m_A=5.5$--$9$ and $11.5$--$14$~GeV, which are relevant in particular in Type~II models.  
The limits from LHC searches for additional heavy Higgs states are also taken into account. 
These include the model-independent limits from the searches 
for $H\to ZZ^{(*)}\to 4\ell$ from ATLAS~\cite{ATLAS-CONF-2013-013} and CMS~\cite{Chatrchyan:2013mxa} and 
for $H\to ZZ^{(*)}\to 2\ell 2\nu$ from CMS~\cite{CMS-PAS-HIG-13-014}. However,
these limits are easily evaded in our study where it is the $h$ that has $C_V^h=\sbma>0.99$, while $HVV$ couplings behave as $\cbma$ and $|\cbma|\leq 0.14$.
(This also holds true in view of the Moriond 2015 update of the Higgs data~\cite{Khachatryan:2015cwa}.) 
More important are the limits from $H,A\to\tau\tau$ searches in gluon-fusion or associated production with a pair of $b$ quarks from ATLAS~\cite{Aad:2014vgg} and CMS~\cite{Khachatryan:2014wca}. These are particularly relevant in the large $\tanb$ region of the Type~II models where a significant enhancement of the down-type fermion coupling to the neutral Higgs states occurs. 
Finally, the limits derived from the pseudoscalar search $A\to Zh, h\to b\bar{b}$ from ATLAS~\cite{Aad:2015wra} and CMS~\cite{CMS-PAS-HIG-14-011} are imposed. 
(Limits from other searches, like for $A\to Z\gamma$ \cite{CMS-PAS-HIG-14-031} or $hh\to b\bar{b}b\bar{b}$ \cite{Khachatryan:2015yea}, have no effect on the results.)
To evaluate all these constraints, production of the $H$ and $A$  via gluon-gluon fusion (ggF) and  via associated production with a pair of bottom quarks (bbH,bbA) are computed at NNLO QCD\footnote{The NNLO corrections for ggF are only computed for the top quark loop, as those for the bottom quark loop are very small.}  accuracy using \texttt{SusHi-1.3.0}~\cite{Harlander:2012pb}, while the vector-boson fusion (VBF) mode for the $H$ is computed at NLO with \texttt{VBFNLO-2.6.3}~\cite{Arnold:2008rz}.

Signal strengths constraints coming from the precise measurements of the properties of the $125\gev$ state are taken into account by means of \texttt{Lilith 1.1.2}~\cite{Bernon:2015hsa}. We require each point of the analysis to be allowed at the 95\% confidence level (CL). 
The CL is derived from the log-likelihood ratio
\beq
\Delta(-2\ln L) (\mathcal{P})=-2\ln L(\mathcal{P}/\widehat{\text{2HDM}}),
\eeq
where $L$ is the likelihood constructed by \texttt{Lilith} using up-to-date signal strength measurements, $\mathcal{P}$ represents the set of parameters of the tested point and $\widehat{\text{2HDM}}$ the best-fit point of the model. 
The \texttt{Lilith} database 15.04 is used for this analysis. It contains all the latest Higgs signal strengths measurements from ATLAS~\cite{ATLAS-CONF-2015-007,Aad:2014eha,Aad:2014eva,Aad:2015vsa,Aad:2014xzb,Aad:2015gra,ATLAS-CONF-2015-006,Aad:2014iia,Aad:2014fia,Aad:2014xva} and CMS~\cite{Khachatryan:2014jba,Khachatryan:2014ira,Chatrchyan:2013mxa,Chatrchyan:2013iaa,Chatrchyan:2014nva,Chatrchyan:2013zna,Khachatryan:2014qaa,Khachatryan:2015ila,Chatrchyan:2014tja} as of April 2015 and a combined D$\O$ and CDF result~\cite{Aaltonen:2013xpo}. 


The updates in this version with respect to the one published in Phys.~Rev. D~{\bf 92}, 075004 (2015) [arXiv:1507.00933v3] are as follows. 
First, we include the updated bound on the charged Higgs mass in Type~II, $\mhpm>480$~GeV at 95\%~CL~\cite{Misiak:2015xwa}, based on the observed rates for the weak radiative $B$-meson decay, $\overline{B}\to X_s\gamma$ (instead of the previous bound of $\mhpm>300$~GeV). 
Second, we correct a bug in the evaluation of the ATLAS and CMS limits from $H,A\to\tau\tau$ searches~\cite{Aad:2014vgg,Khachatryan:2014wca}: the branching ratio for $A\to\tau\tau$ was set to the one of a SM Higgs with the same mass instead of the one computed in the 2HDM. 
This affects the intermediate-to-large $\tanb$ region of the Type~II models where an enhancement of the down-type fermion coupling occurs. In particular it eliminates part of the ``opposite-sign'' $C_D$ solution in Type~II. 
Third, we include the CMS constraint~\cite{Khachatryan:2015baw} on neutral Higgs bosons with masses between 25 GeV and 80 GeV, produced in association with a pair of $b$ quarks and followed by the decay into $\tau\tau$.
Finally, the CMS result~\cite{CMS-PAS-HIG-15-001} on the search for a new heavy resonance decaying to a $Z$ boson and a light resonance, followed by $Z\to \ell^+\ell^-$ and the light resonance decaying to $b\bar b$ or $\tau\tau$ is included. Sensitive to light resonances with masses above $\sim 35$~GeV, this analysis puts a very severe constraint on $gg\to A\to Zh$ with $Z\to \ell^+\ell^-$ and $h\to b\bar b$ in the $m_H=125$~GeV scenario~\cite{part2}. Here, in the $m_h=125$~GeV scenario, it contributes to the exclusion of $A$ with masses between $\sim 60$~GeV and $\sim 150$~GeV as will be shown in the next Section.


\section{\boldmath Results}\label{results-lighth}

\subsection{Parameters}

\begin{figure}[t!]\centering
\includegraphics[width=0.5\textwidth]{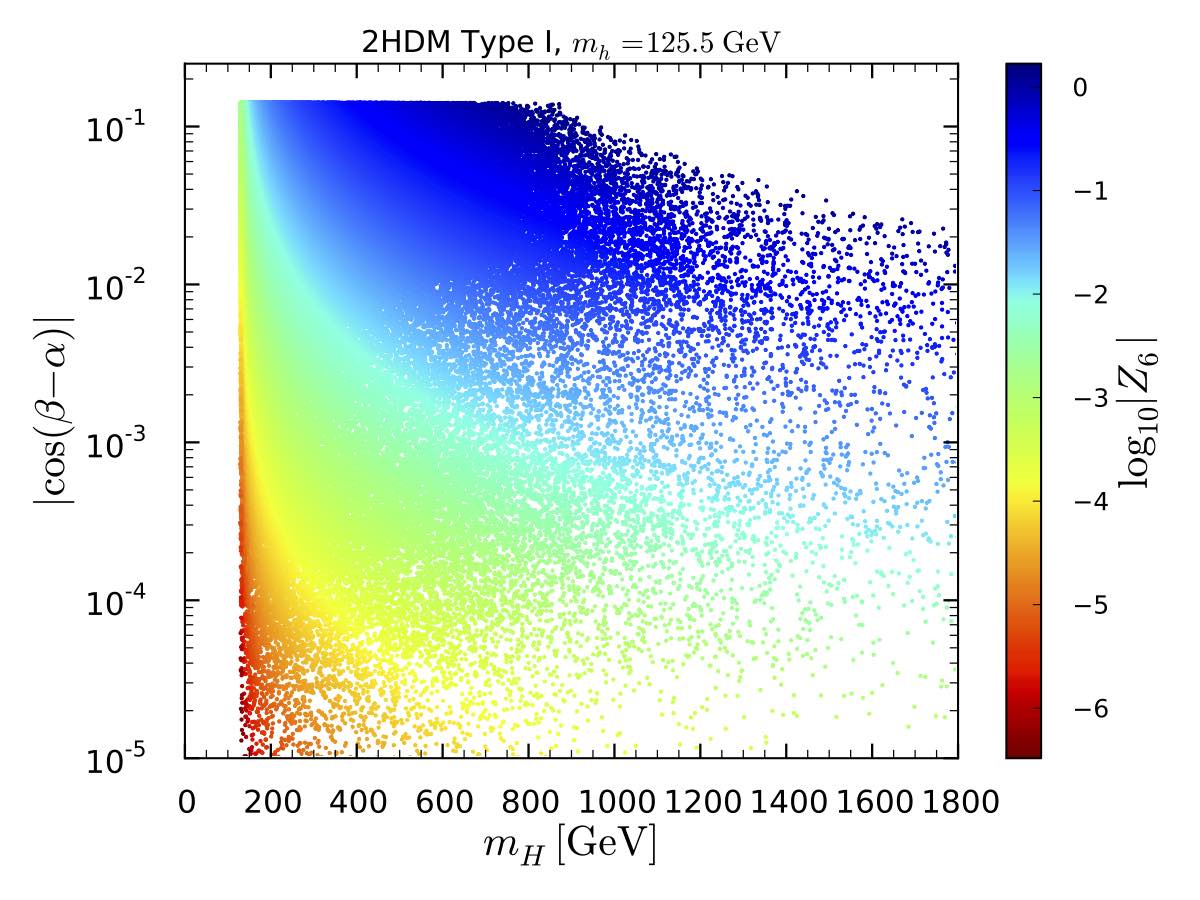}\includegraphics[width=0.5\textwidth]{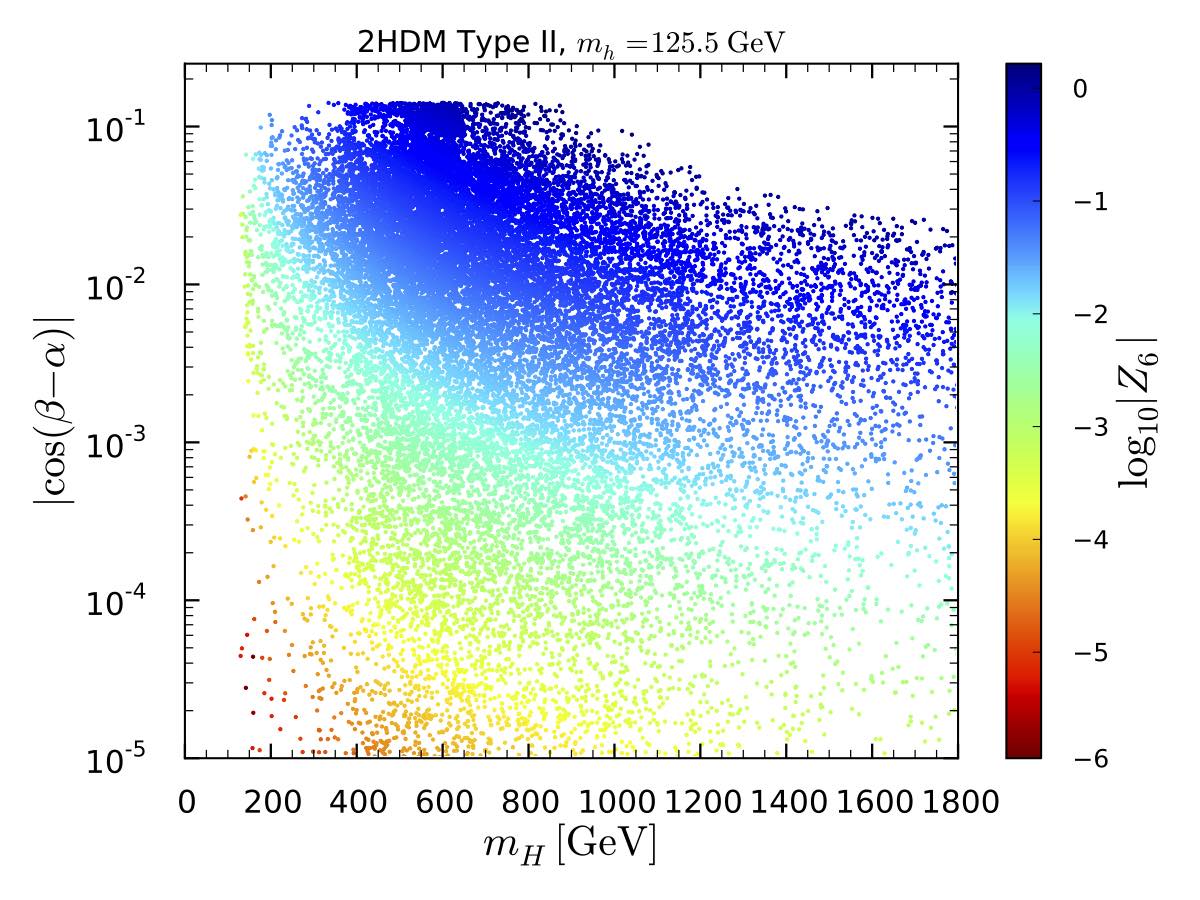}
  \caption{$|\cba|$ versus $m_H$ in Type~I (left) and Type~II (right) with $\log_{10}|Z_6|$ color code. Points are ordered from high to low $\log_{10}|Z_6|$ values.}
  \label{mH_cba_Z6_h125}
\end{figure}

Let us start by reviewing the relevant parameter space. 
Figure~\ref{mH_cba_Z6_h125} shows the crucial relation between $|Z_6|$, $|\cba|$ and $m_H$, illustrating the different ways alignment can occur with and without decoupling.\footnote{In this and subsequent figures, we give 3d information on a 2d plot by means of a color code in the third dimension. To this end, we must chose a definite plotting order. Ordering the points from high to low values in the third dimension, as done for $\log_{10}|Z_6|$ in Fig.~\ref{mH_cba_Z6_h125}, means that the highest values are plotted first and lower and lower values are plotted on top of them. As a consequence, regions with low values may (partly) cover regions with high values. The opposite is of course true for the ordering from low to high values. To avoid a proliferation of plots, in each figure we show only one ordering, trying to choose the one that gives most information. The figures with inverted plotting order are available upon request.}
As expected, $|Z_6|$ exhibits a clear dependence on the $H$--$h$ mass difference, see \eq{z6v}, and steeply drops towards zero in the limit $|\cba|\to 0$, \ie\ when the $h$ becomes purely SM-like. 
When $m_H$ is of the order of 1~TeV, one needs to be extremely close to $\sba=1$ to have small $|Z_6|$---for instance $|Z_6|\approx 10^{-3}$ requires $|\cba|\approx 6\times 10^{-5}$ for $m_H= 1$ TeV. In contrast, for a lighter $H$ the departure of $\sba$ from 1 can be more important---for instance the same $|Z_6|\approx 10^{-3}$ value requires $|\cba|\approx 2\times 10^{-3}$ for $m_H= 200$ GeV.
It is in principle always possible to obtain arbitrarily small values of $|Z_6|$ if one pushes $\sbma$ arbitrarily close to 1. For the purpose of the numerical analysis, we limit ourselves to $|\cba|\geq 10^{-5}$; we have checked that this captures well all features relevant for the $|\cba|\to 0$ limit.
Interestingly, as $m_H$ becomes larger, we observe that the decoupling limit sets a stronger upper limit on $|\cba|$ than the one set in the numerical scan ($|\cba|\lesssim 0.14$). Observing a heavy $m_H\gtrsim 850$ GeV at the LHC would provide a better-than-$1\%$ indirect determination of the $h$-coupling to electroweak gauge bosons in the framework of these scenarios.

\begin{figure}[t!]\centering
\includegraphics[width=0.5\textwidth]{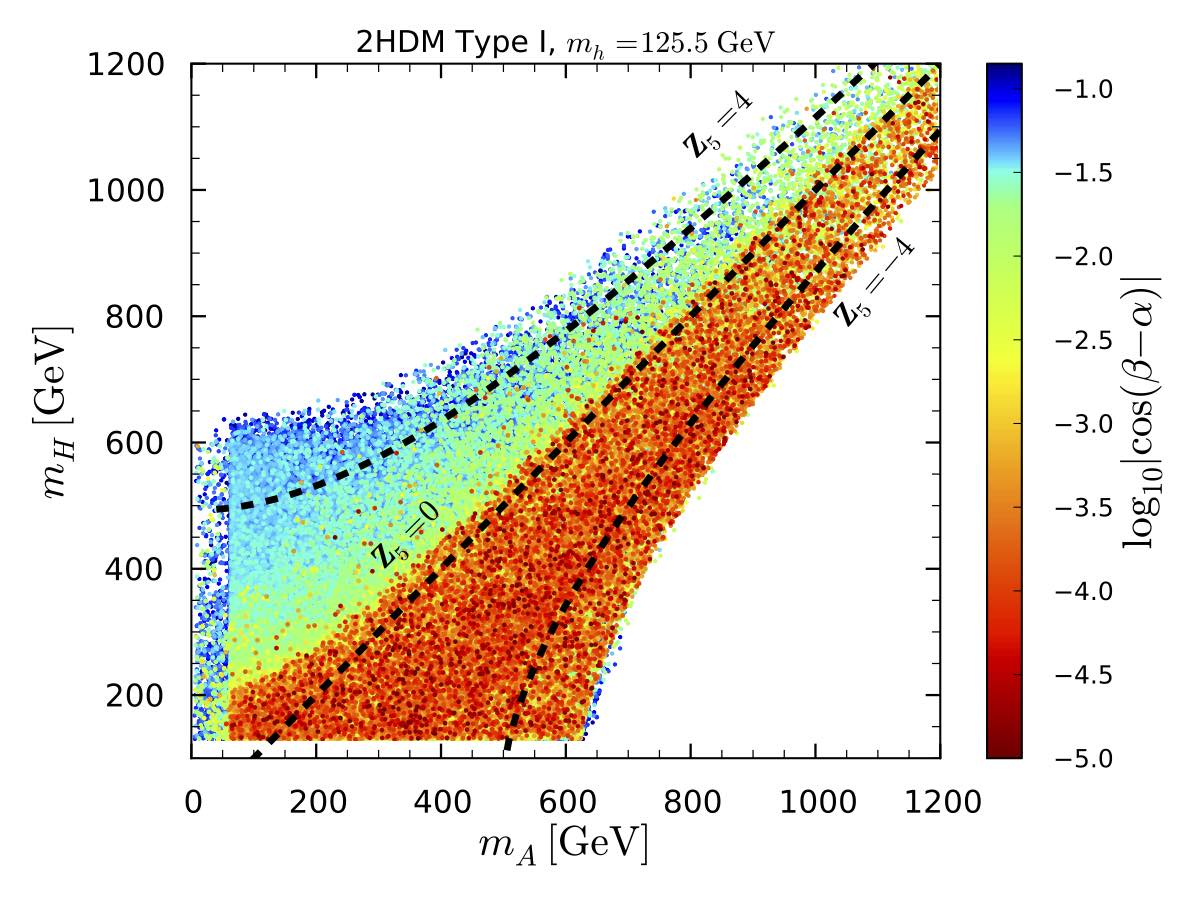}\includegraphics[width=0.5\textwidth]{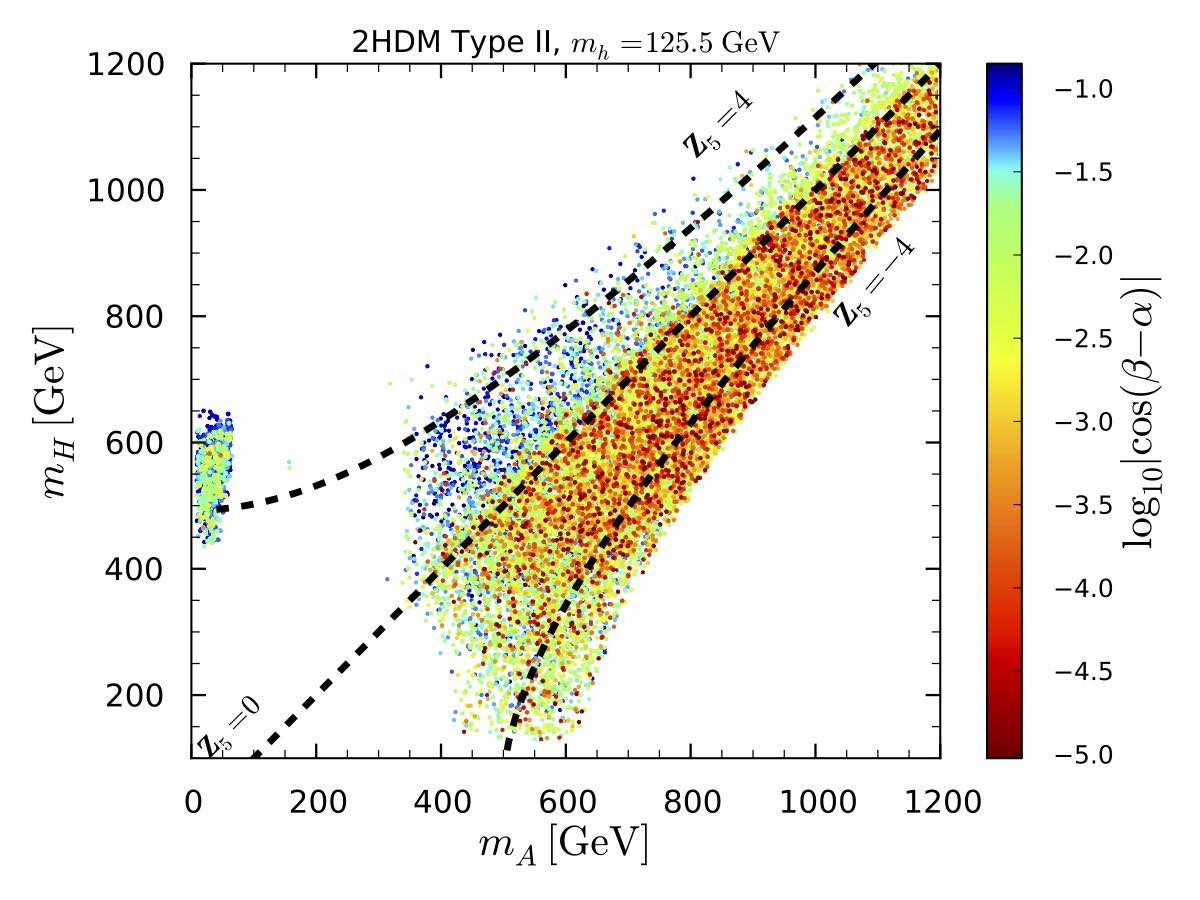}\\
\caption{$m_H$ versus $m_A$ in Type~I (left) and Type~II (right) with the color code indicating the value of $\log_{10}|\cba|$. Points are ordered  from high to low $\log_{10}|\cba|$. The dashed lines are isolines of $Z_5$=4 (upper line), 0 (middle line) and $-4$ (lower line) for $|\cbma|=0.015$ (varying $|\cbma|$ from $0$ to $0.14$ has no visible effect on them).}
  \label{mA_mH_Z6}
\end{figure}

The range of $m_A$ is also interesting. In principle $m_A$ can be above or below $m_{h,H}$, and even $m_A<m_h/2$ is possible and consistent with the data \cite{Bernon:2014nxa}. However, once $m_H$ is fixed, the allowed range of $m_A$ is limited (and vice versa) as illustrated in Fig.~\ref{mA_mH_Z6}. We see that in both Type~I and Type~II, if the scalar $H$ is heavy and decoupled, the same is true for the pseudoscalar $A$. Conversely, if  $H$ is light, say below 600~GeV, then also $A$ must be below about 800~GeV. 
Furthermore, it appears that for $|\cba|\lsim 10^{-3}$ (or, equivalently, small $|Z_6|$) $m_H<m_A$ is favored. This can be understood from \eq{z5v} [or \eq{hh2mass}]: since the $m_h^2\cbma^2$ (or $Z_6\cbma/\sbma$) term therein is always quite small, the mass ordering between $m_H$ and $m_A$ is largely determined by the sign of $Z_5$. The value of $Z_5$, in turn, is driven by $\lambda_5$ [cf.~\eq{Zeye}], which according to our numerical analysis tends to be negative for small $\cbma$. 
The absence of points over a large region of low $m_{H,A}$ in Type~II is in part due to the $H,A\to\tau\tau$ limits~\cite{Aad:2014vgg,Khachatryan:2014wca}, which eliminate a large swath of parameter space with $C^D_h\simeq -1$ and $m_A\approx 150$--$350$~GeV, and to the CMS $H\to ZA$ search~\cite{CMS-PAS-HIG-15-001} which eliminates points down to $m_A\simeq 60$~GeV (with a mild dependence on $m_H$). We note that the surviving points with $m_A\lesssim 60$~GeV have $\tan\beta<2$. In addition, the charged Higgs mass limit $\mhpm>480$~GeV in Type~II~\cite{Misiak:2015xwa} results in the elimination of the remaining quadrant with $m_{H,A}\lesssim 400$~GeV (actually up to $m_H\approx 430$~GeV for very light $m_A$).

\begin{figure}[t!]\centering
\includegraphics[width=0.5\textwidth]{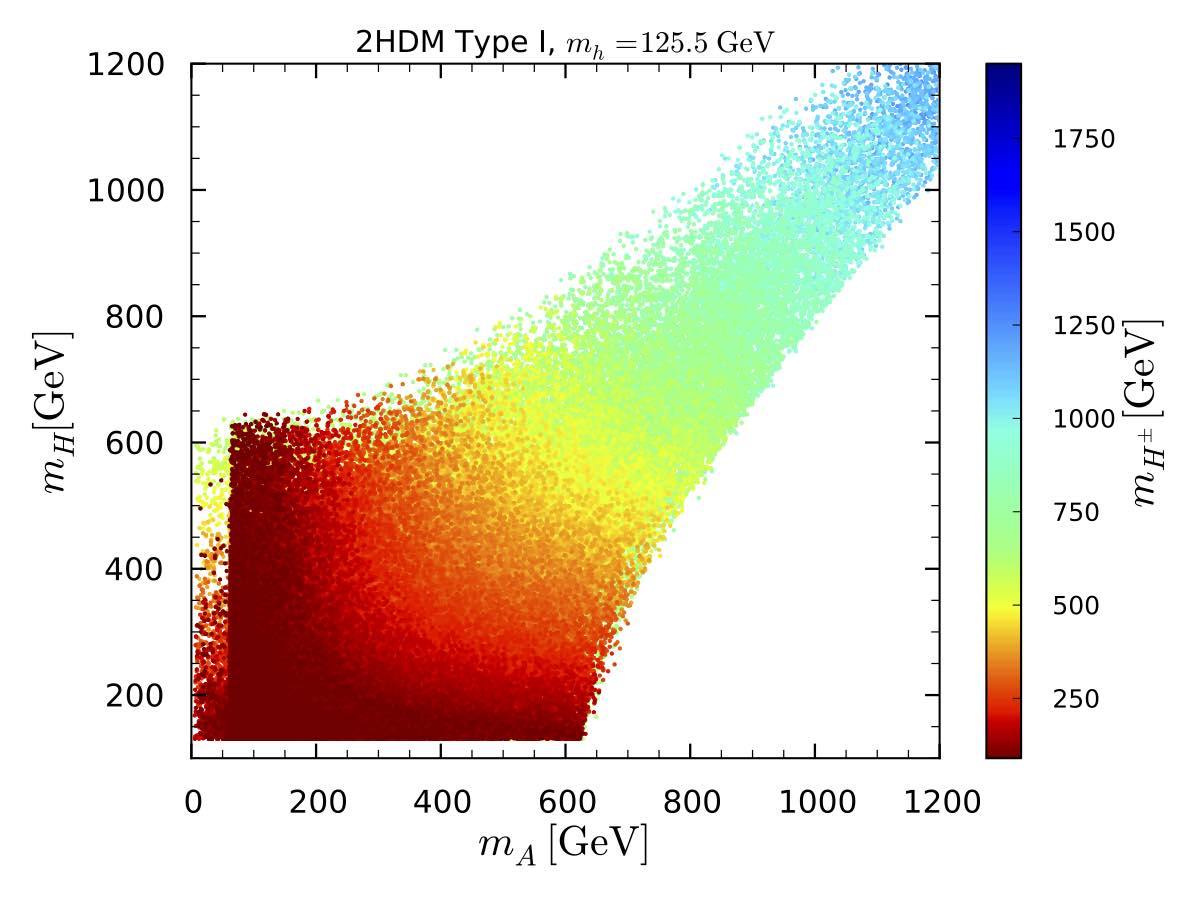}\includegraphics[width=0.5\textwidth]{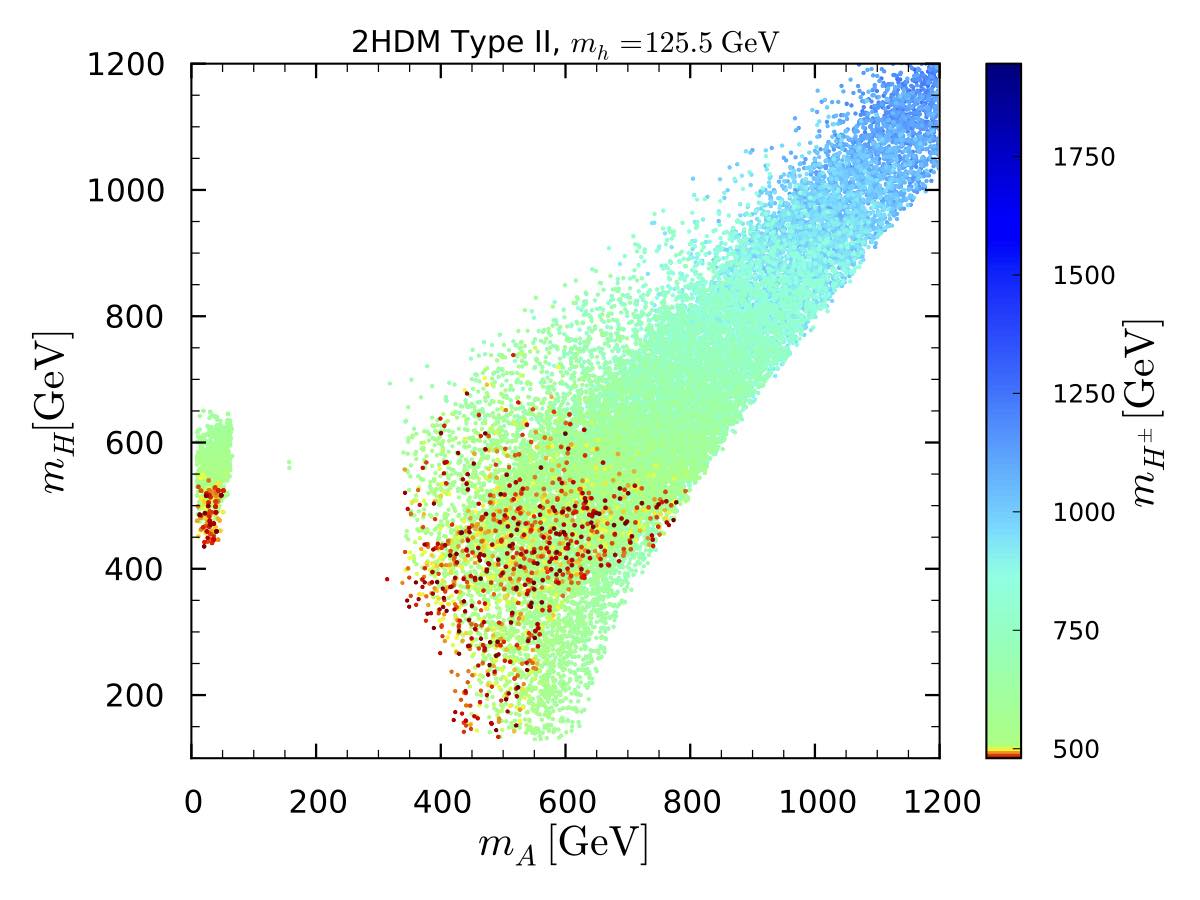}
\caption{$m_H$ versus $m_A$ in Type~I (left) and Type~II (right) with the color code indicating the value of $m_{H^\pm}$.  Points are ordered from high to low $m_{H^\pm}$. }
  \label{mA_mH_mC}
\end{figure}

The interrelation between $m_A$, $m_H$ and $m_{H^\pm}$ is illustrated in Fig.~\ref{mA_mH_mC}. 
The two panels show $m_H$ versus $m_A$ with  color-coding according to $m_{H^\pm}$, with the ordering going from high (blue) to low (red) $m_{H^\pm}$ values. While the correlation of $m_{H^\pm}$  with $m_H$ and $m_A$ is somewhat different in Type~I and Type~II, in both models a light charged Higgs below 500--600~GeV requires that the $H$ and $A$ also be not too heavy, with masses
below about 800~GeV. When inverting the plotting order of $m_{H^\pm}$ (not shown), we find that for any given $\mhpm$ there is a lower limit on $m_H$ and $m_A$: for $m_{H^\pm}\sim 1$~TeV, also $m_{H,A}$ are of that order. In turn, when $m_H$ and $m_A$ are in the non-decoupling regime,  $m_{H^\pm}$ cannot be much heavier. The absence of points in the light mass region $m_{H,A}\lesssim 400$~GeV in Type~II (but not for Type~I), already noted in the previous paragraph is due to the fact that in the Type~II model $B$-physics requires $\mhpm\gsim 480\gev$ and at low $m_A$ the precision electroweak $T$ parameter constraint would be violated if $m_H$ differs very much from $\mhpm$.
As also mentioned above, an additional band with $m_A\approx 150$--$350$~GeV is cut out by the $H,A\to\tau\tau$ limits. We will see later that this corresponds to a large extent to the ``opposite-sign'' $C_D$ solution with large $\tan\beta$ in Type~II.

\subsection{Couplings}

\begin{figure}[t!]\centering
\includegraphics[width=0.5\textwidth]{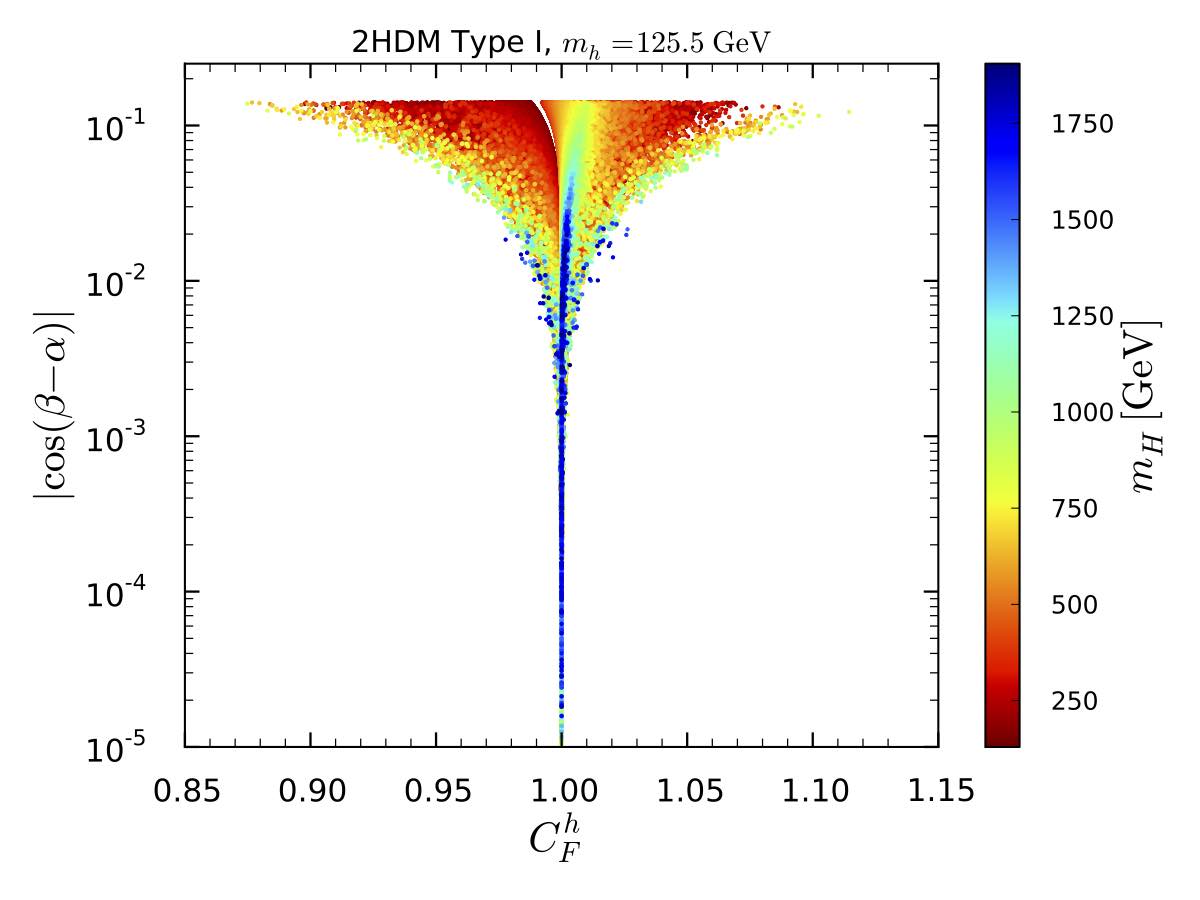}\includegraphics[width=0.5\textwidth]{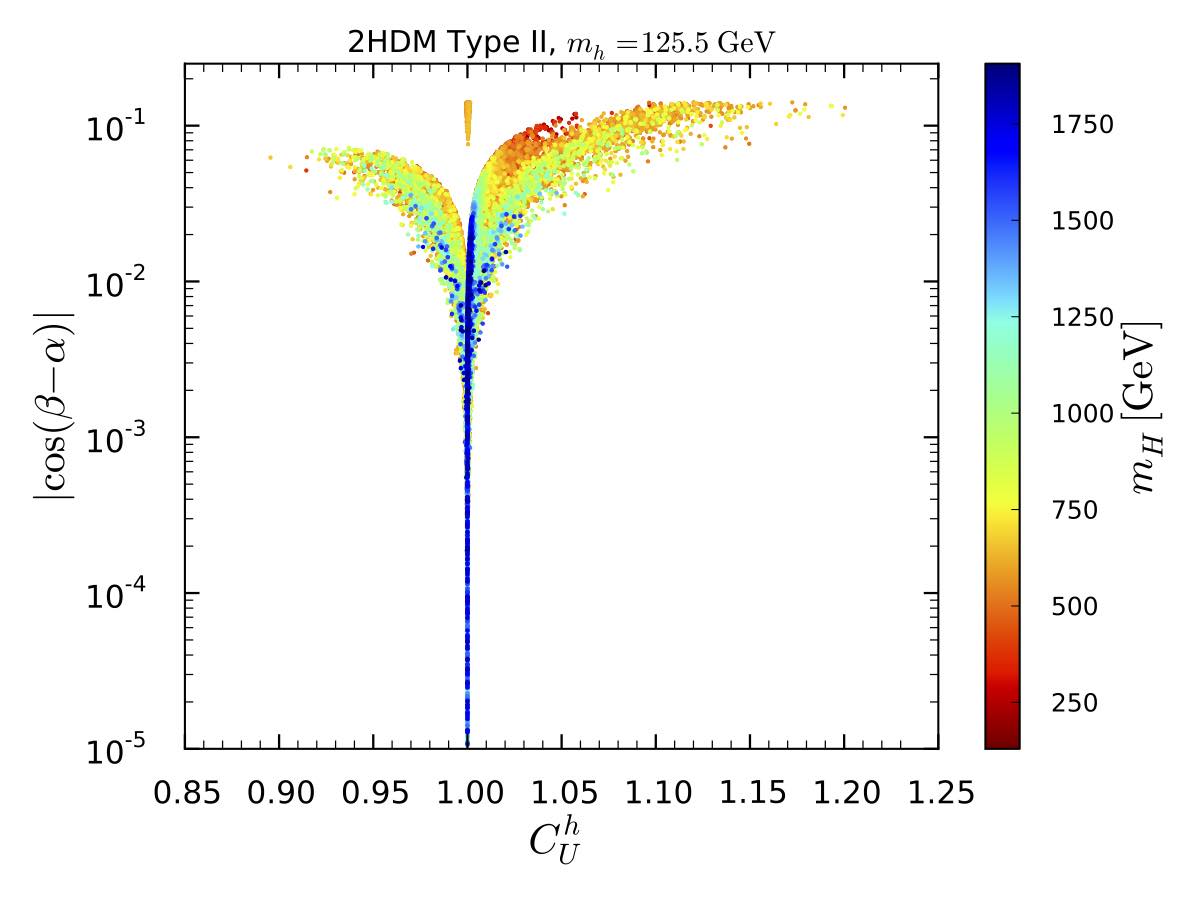}
  \caption{$|\cba|$ versus $C_F^h$ in Type~I (left) and $|\cba|$ versus $C_U^h$ in Type~II (right) with $m_H$ color code. Points are ordered from low to high $m_H$. The points with $C_U^h\approx 1$ and $|\cbma|>0.03$ are the points for which $C_D^h\approx -1$, i.e. the opposite-sign Yukawa coupling points, see Fig.~\ref{CD_h125}.} 
  \label{CU_h125}
\end{figure}

The next question to address is what variations in the couplings of the $125.5$~GeV state are still possible in the limit of approximate alignment where $C_V^h\approx 1$.  In particular,
recall that in the scan we impose $\sba>0.99$ with $m_h=125.5$~GeV, without requiring however that the other couplings of the $h$ be very SM-like. 
To answer this question, we first show in Fig.~\ref{CU_h125} the dependence of the reduced couplings to (up-type) fermions, see Table~\ref{tab:2hdm-couplings}, $C_F^h\equiv C_U^h=C_D^h$ in Type~I ($C_U^h$ in Type~II) on $|\cba|$. 
The mass of the heavier scalar $H$ is shown as a color code. We see that when $m_H$ is light, for only 1\% deviation from unity in $C_V^h$, $C_U^h$ can deviate as much as about 10\% (20\%) from unity in Type~I (Type~II). Inverting the plotting order of $m_H$ (not shown), it is interesting to note that these deviations are largest for $m_H\approx 700$--800~GeV while slightly more constrained for lighter $m_H$. On the other hand, in the decoupling limit the deviations in $C_U^h$ are more constrained, with a maximum of $5\%$ for $m_H\gtrsim 1.2$ TeV in both Type~I and Type~II.
It is also interesting to observe how quickly alignment leads to SM-like couplings: for $|\cba|\lsim 10^{-2}$ the deviations in 
$C_U^h$ are limited to just a few percent no matter the value of $m_H$. 

\begin{figure}[t!]\centering
\includegraphics[width=0.49\textwidth]{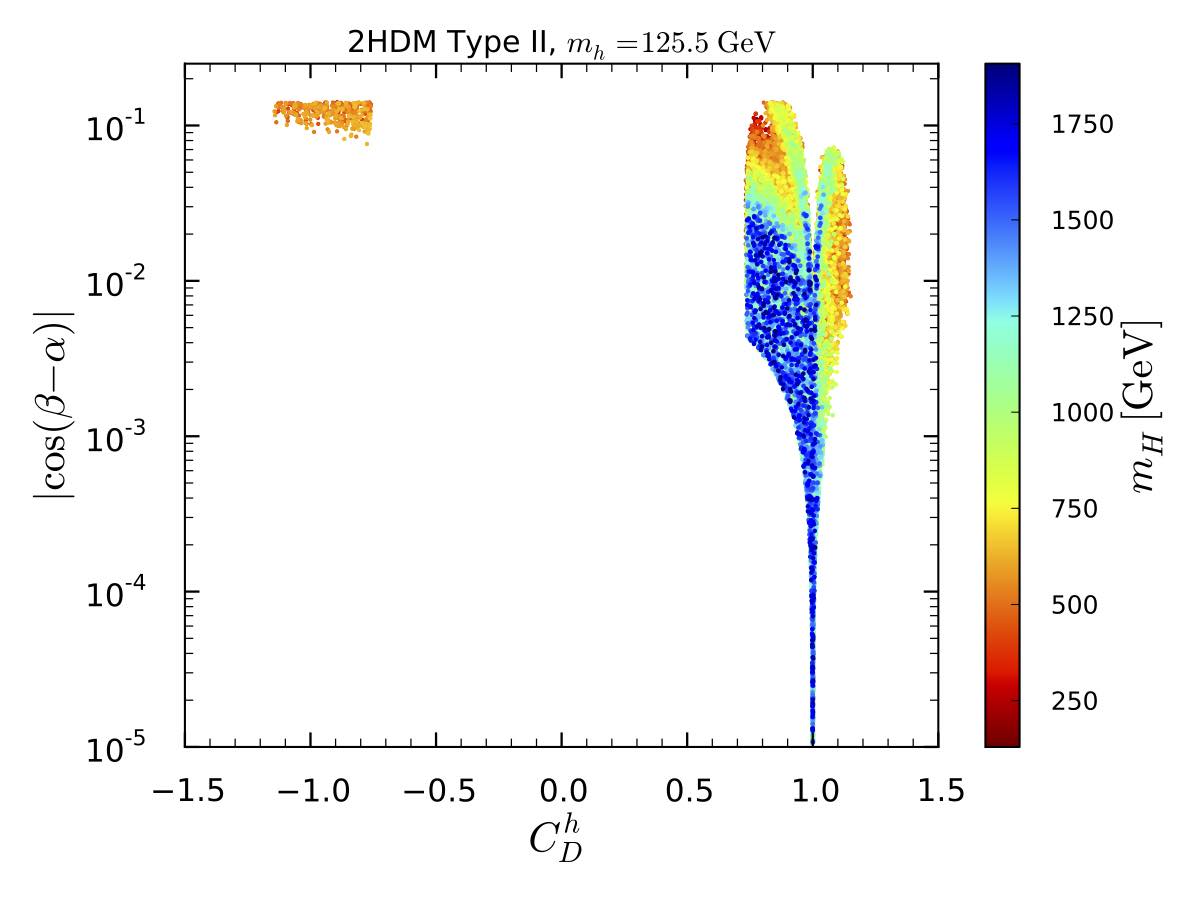}\includegraphics[width=0.49\textwidth]{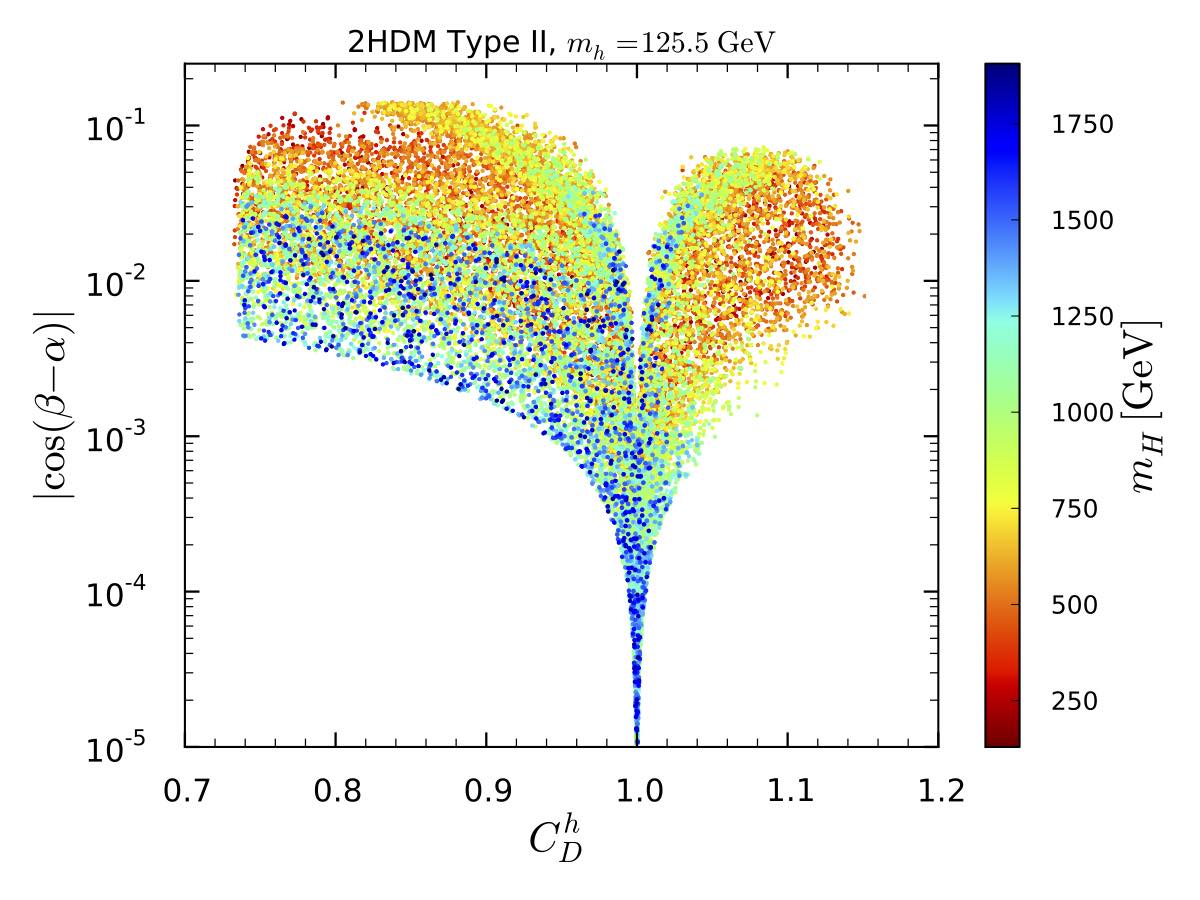}\\
  \caption{$|\cba|$ versus $C^h_D$ in Type~II with $m_H$ color code for the full $C_D^h$ range (left) and zooming on the $C^h_D>0$ region (right). Points are ordered from low to high $m_H$.}
  \label{CD_h125}
\end{figure}

The situation is quite different for the coupling to down-type fermions, $C_D^h$, in Type~II, see Fig.~\ref{CD_h125}. 
First of all, the possible deviations are larger than for  $C_U^h$, with $C_D^h$ ranging from about 0.70 to 1.15 even for $|\cba|\sim10^{-2}$. Indeed, this is an example of the delayed alignment limit discussed below \eq{yukid4}; one needs $|\cba|$ as low as about $3\times 10^{-4}$ to have $C_D^h$ within 2\% of unity. This drives the whole phenomenology of the scenario: as we will see, sizable deviations of $C_D^h$ from 1 lead to possible large deviations in the signal strengths even for quite small $|\cba|$. 
Inverting the plotting order of $m_H$ (not shown), we note, however, that for any given $|\cbma|$ of a few times $10^{-3}$ or smaller, $C_D^h$ is limited to be closer to 1 when $m_H$ is small than in the decoupling case with large $m_H$.

Moreover, $C_D^h=1$ is not possible unless $|\cba|$ is very small (again a few times $10^{-3}$ or smaller) as a consequence of the lower bound $\tb\geq 0.5$ imposed in the analysis. Large positive deviations of $C_D^h$, up to $\sim 1.12$, would indicate $m_H\lesssim 750$~GeV. On the contrary, $C_D^h$ values which are substantially smaller than 1 can be achieved in both the decoupling and non-decoupling regimes except for a small island of points located around $C_D^h\approx 0.8$ and $|\cba|\approx 0.1$ that is achieved only for $m_H\lesssim 400$~GeV.
Thus, for instance, a discovery of a light $H$ state in association with a measured value of $C_D^h\sim 0.8$ would give an indirect way to probe sub-percent deviation of $C_V^h$ in this Type~II scenario.

Finally, for light $m_H$ the sign of $C_D^h$ relative to $C_V^h$ and $C_U^h$ can be opposite to the corresponding SM value. This is realized for not so small values of $|\cba|\geq 0.07$, \textit{i.e.}~at the boundary of what we consider as the alignment limit, for $330 \gev \leq m_H\leq 660 \gev$, $350 \gev \leq m_A\leq 660 \gev$ and $0.22 \leq |Z_6| \leq 0.90$. For the points in this region, the up-type coupling is very close to 1, corresponding to the few isolated points observed in the right panel of Fig.~\ref{CU_h125}.
As discussed in \cite{Ferreira:2014naa}, the eventual LHC Run~2 precision will allow one to either confirm or eliminate the opposite-sign coupling possibility using precise signal rate measurements of the $h$ in a few channels. Should the 
opposite-sign coupling be confirmed, one would expect to also see $A$ signals (plus perhaps $H$ signals) in the above mass range, thereby providing a confirmation of this scenario. (The cross sections for $A$ and $H$ signals will be discussed in Section~\ref{cross-sections}.) It should also be noted here that this region is much reduced by the correction of the $H,A\to\tau\tau$ constraints as compared to the previous version [arXiv:1507.00933v3].

\begin{figure}[t!]\centering
\includegraphics[width=0.5\textwidth]{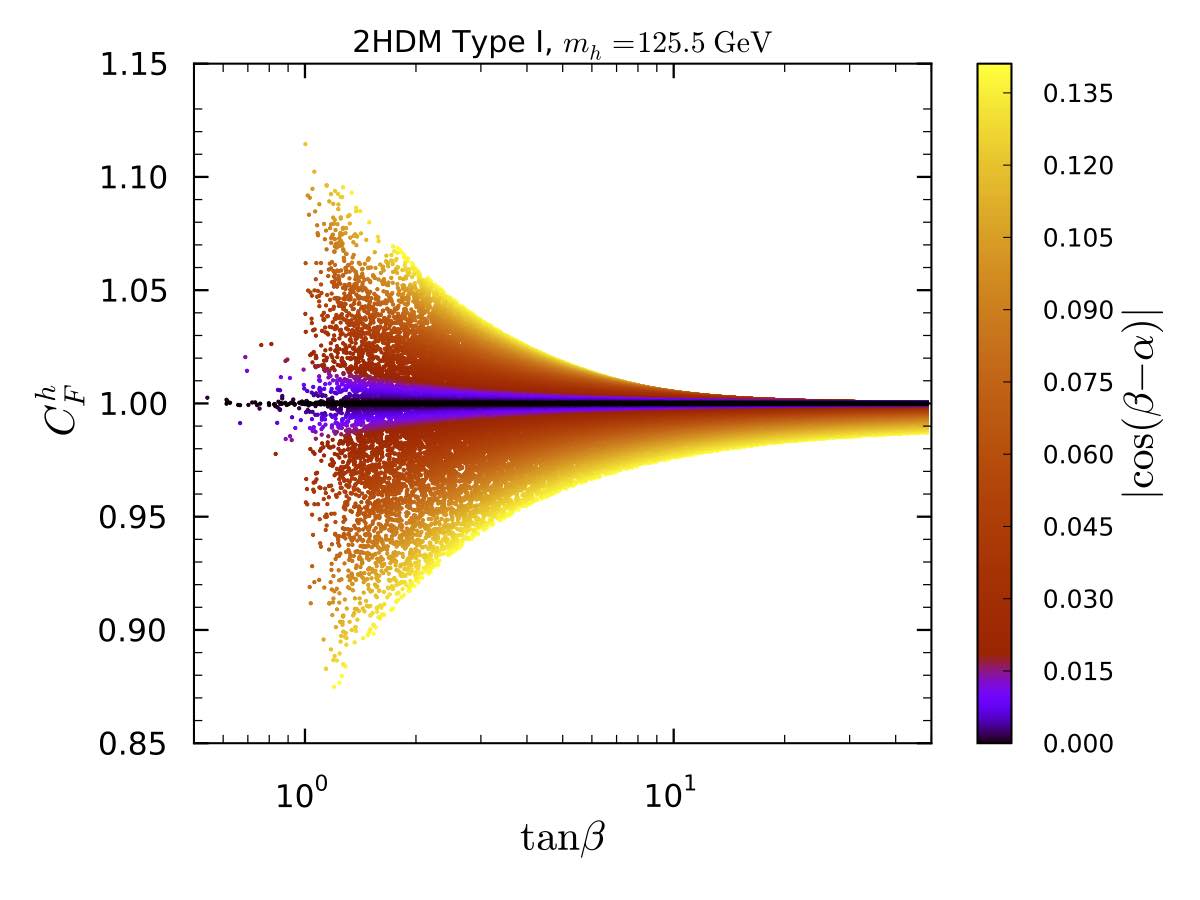}
\includegraphics[width=0.5\textwidth]{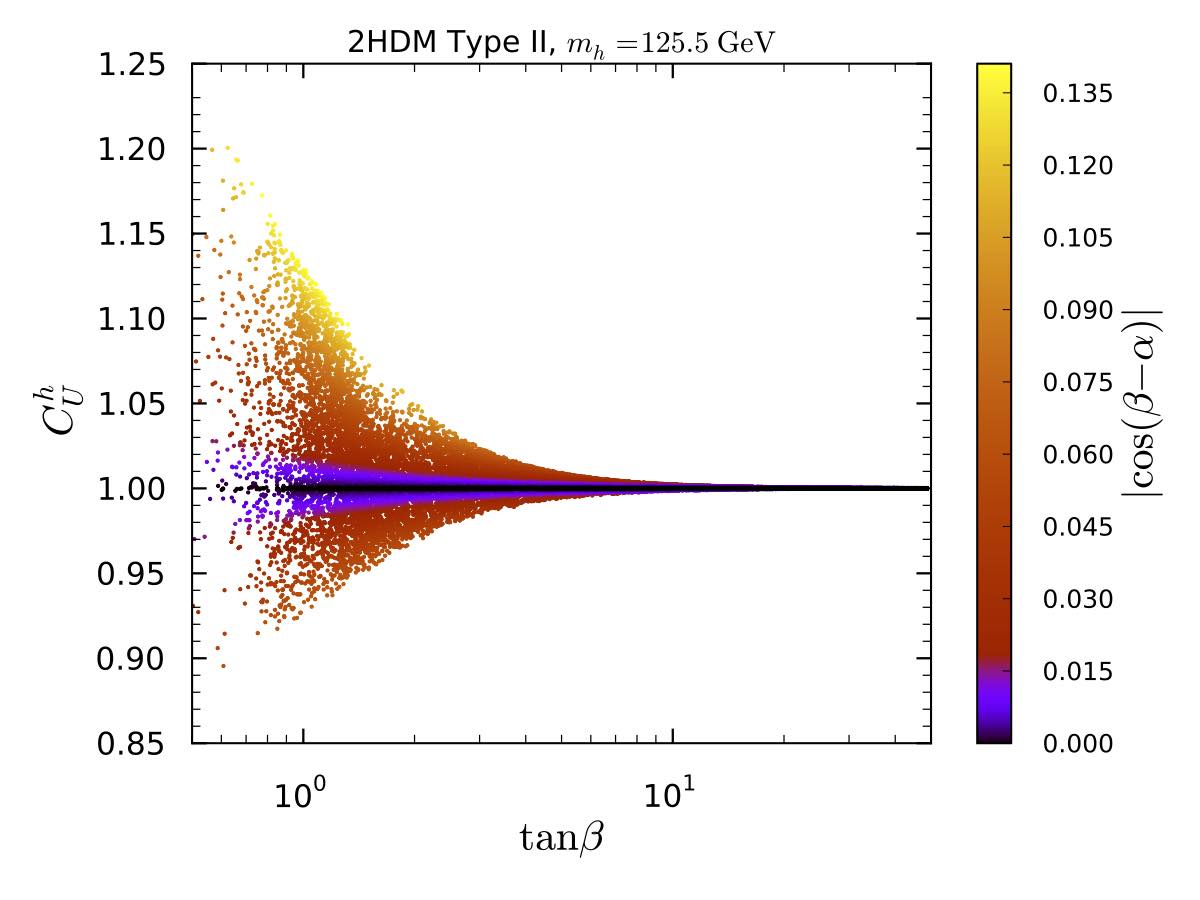}\includegraphics[width=0.5\textwidth]{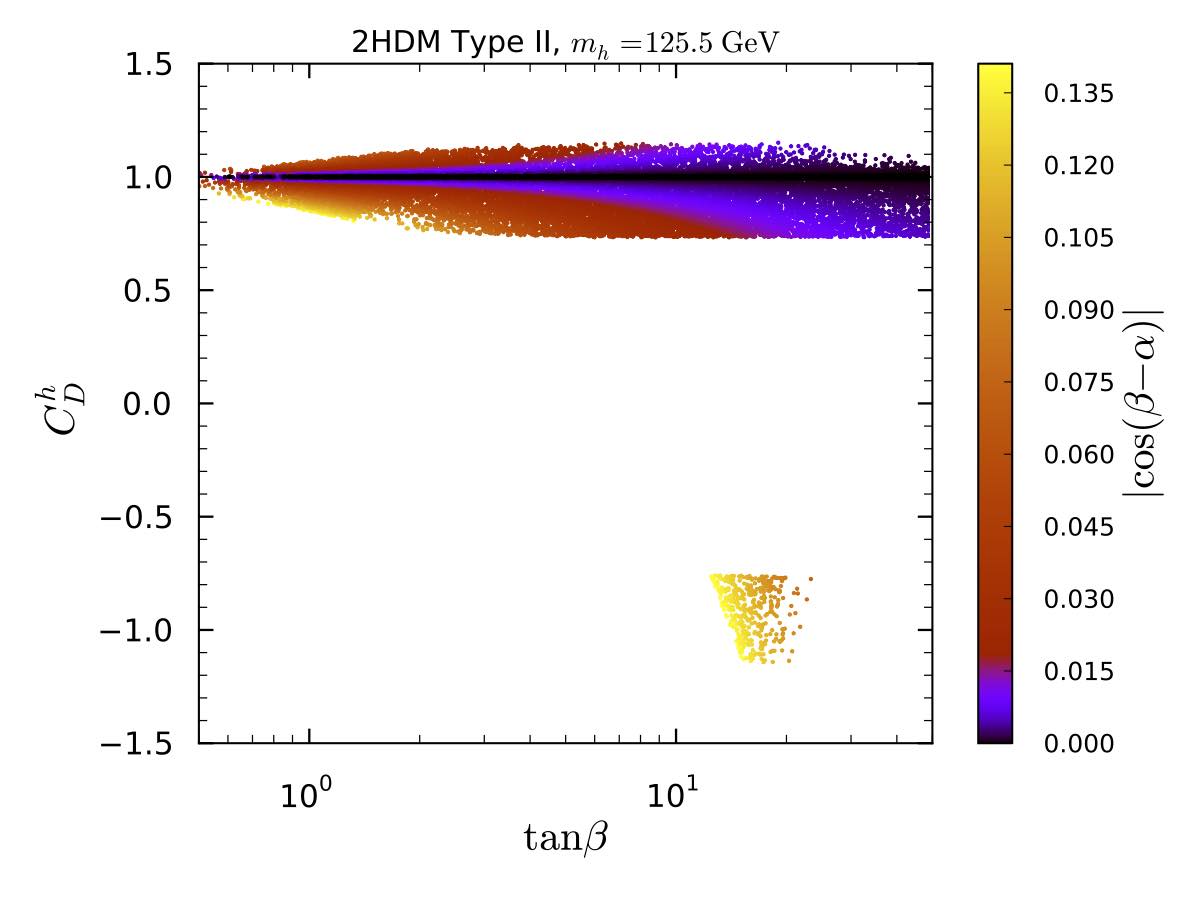}
  \caption{Fermionic couplings versus $\tan\beta$ in Type~I (upper panel) and Type~II (lower panels) with $|\cba|$ color code. Points are ordered from high to low $|\cba|$.}
  \label{tb_CF_h125}
\end{figure}

The $\tan\beta$ dependence of the fermion couplings of $h$ is shown in Fig.~\ref{tb_CF_h125}. We see that large $\tan\beta$ leads to $C_F^h$ very close to 1 in Type~I and $C_U^h$ very close to 1 in Type~II. However in Type~II,
at large $\tan\beta$, small $\cba$ is not enough to drive $C_D^h\to1$: the approach to SM-like coupling is delayed, as discussed in Section~2 in the text below Table~\ref{tab:2hdm-couplings}. Note also that the opposite-sign $C_D^h$ solution in Type~II requires $\tan\beta\gtrsim 10$ and $C_V^h\sim 0.9994$ (which is experimentally indistinguishable from exact alignment).

\begin{figure}[t!]\centering
\includegraphics[width=0.5\textwidth]{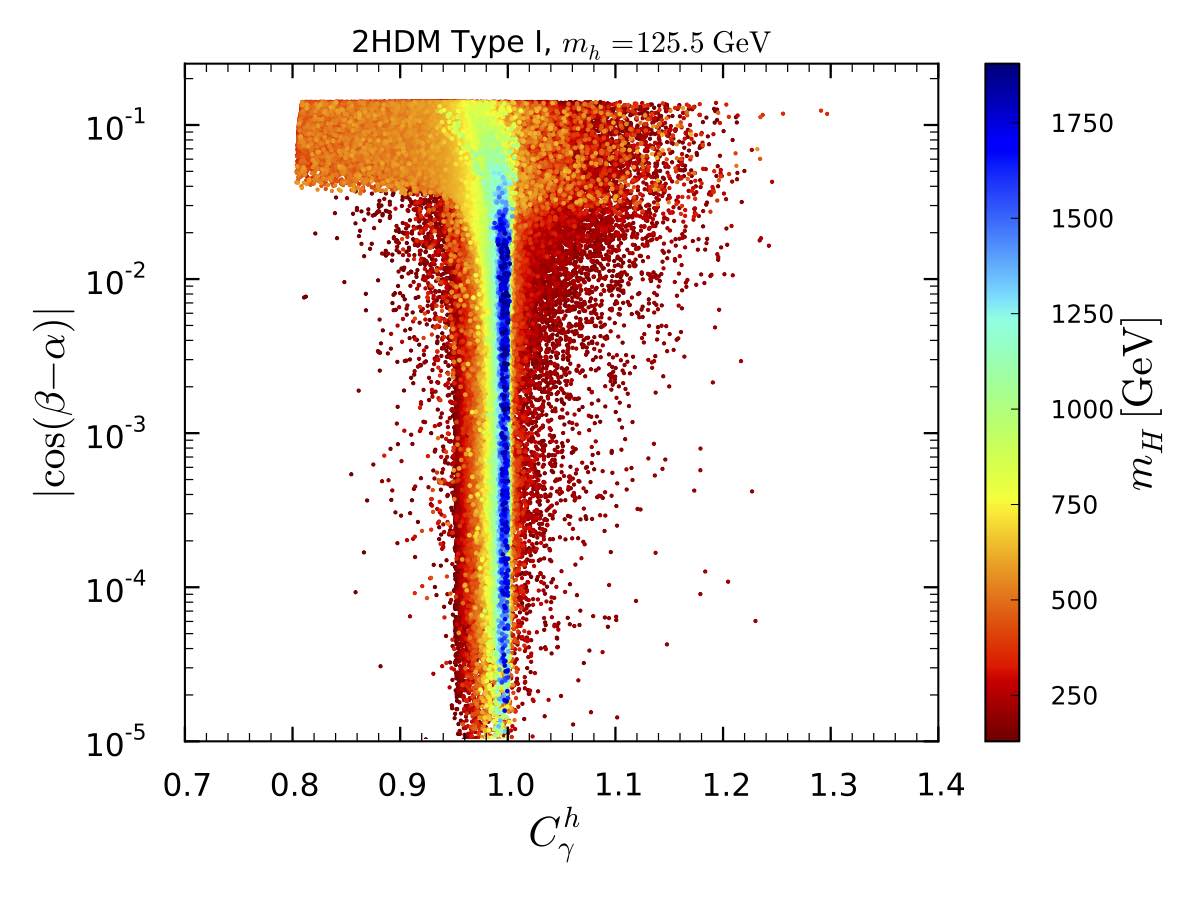}\includegraphics[width=0.5\textwidth]{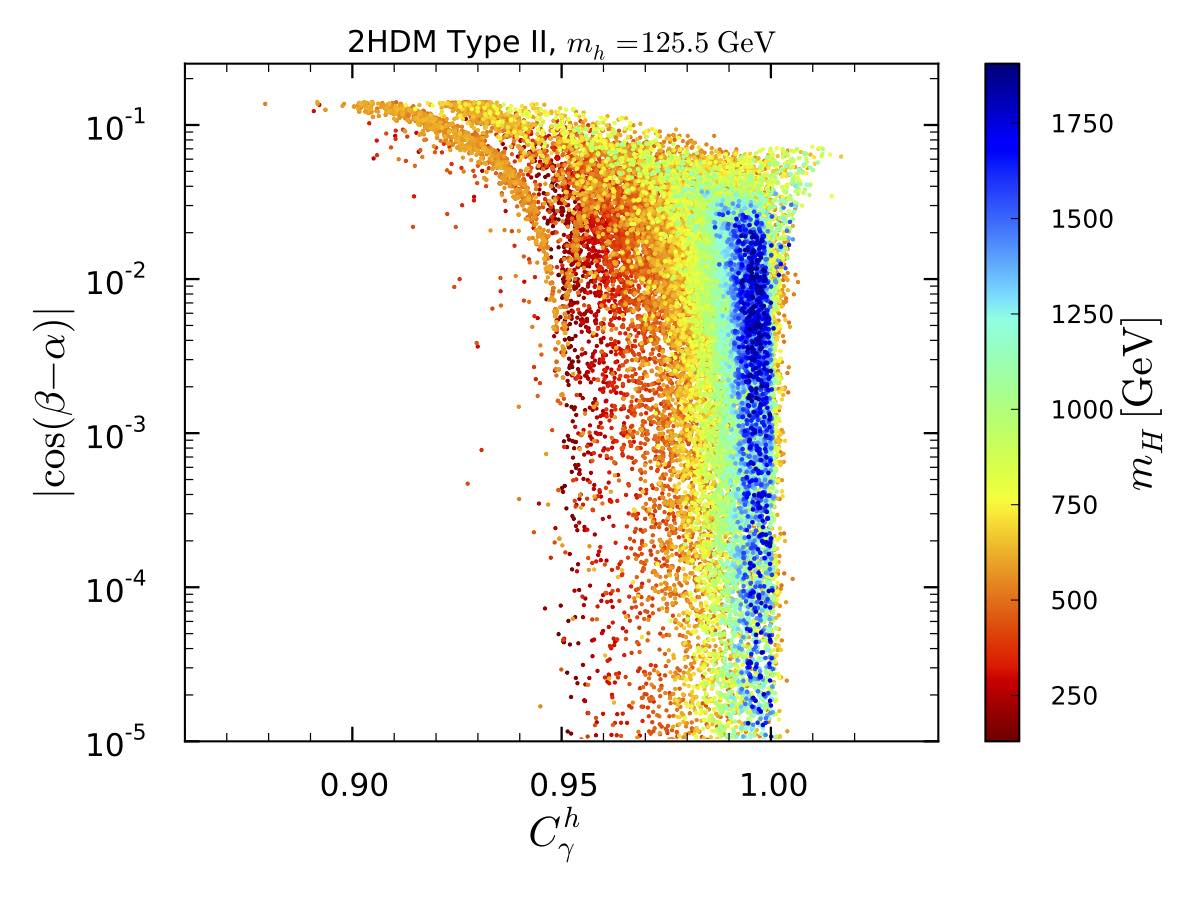}\\
  \caption{$|\cba|$ versus $C_\gamma^h$ in Type~I (left) and Type~II (right) with $m_H$ color code. Points are ordered from low to high $m_H$.
} \label{Cgluongamma_h125}
\end{figure}

The loop-induced coupling to photons, $C_\gamma^h$, is presented in Fig.~\ref{Cgluongamma_h125}. Even at very small $\cba$, $C_\gamma^h$ can deviate substantially from 1. This is due to the charged-Higgs contribution to the $h\gamma\gamma$ coupling. 
This contribution can be large with either sign, positive or negative, in Type~I, while in Type II  \textit{large} contributions are always negative and suppress $C^h_\gamma$~\cite{Ferreira:2014naa}. Note in particular the Type II points with $C_\gamma^h\sim 0.95$ associated with the opposite-sign $C_D^h$ cases for which the charged Higgs loop contribution does not decouple and always leads to a suppression. 
Regarding the loop-induced coupling to gluons, in the Type~I model, $C_g^h$, is equal to $C_F^h$ (up to NLO),  the dependence of which on $|\cba|$ was presented in Fig.~\ref{CU_h125}. In the case of Type~II, $C_g^h$ and $C_U^h$ are very similar despite the difference between up and down-type couplings, this being due to the fact that the $b$-loop contribution to $C_g^h$ is rather small. The one exception in the case of Type~II arises for the opposite-sign scenario for which the $b$-loop contribution changes sign and interferes {\it constructively} with the $t$-loop contribution. In this case, $C_g^h$ is always enhanced, $C_g^h\sim 1.06$~\cite{Ferreira:2014naa}.

\begin{figure}[t!]\centering
\includegraphics[width=0.5\textwidth]{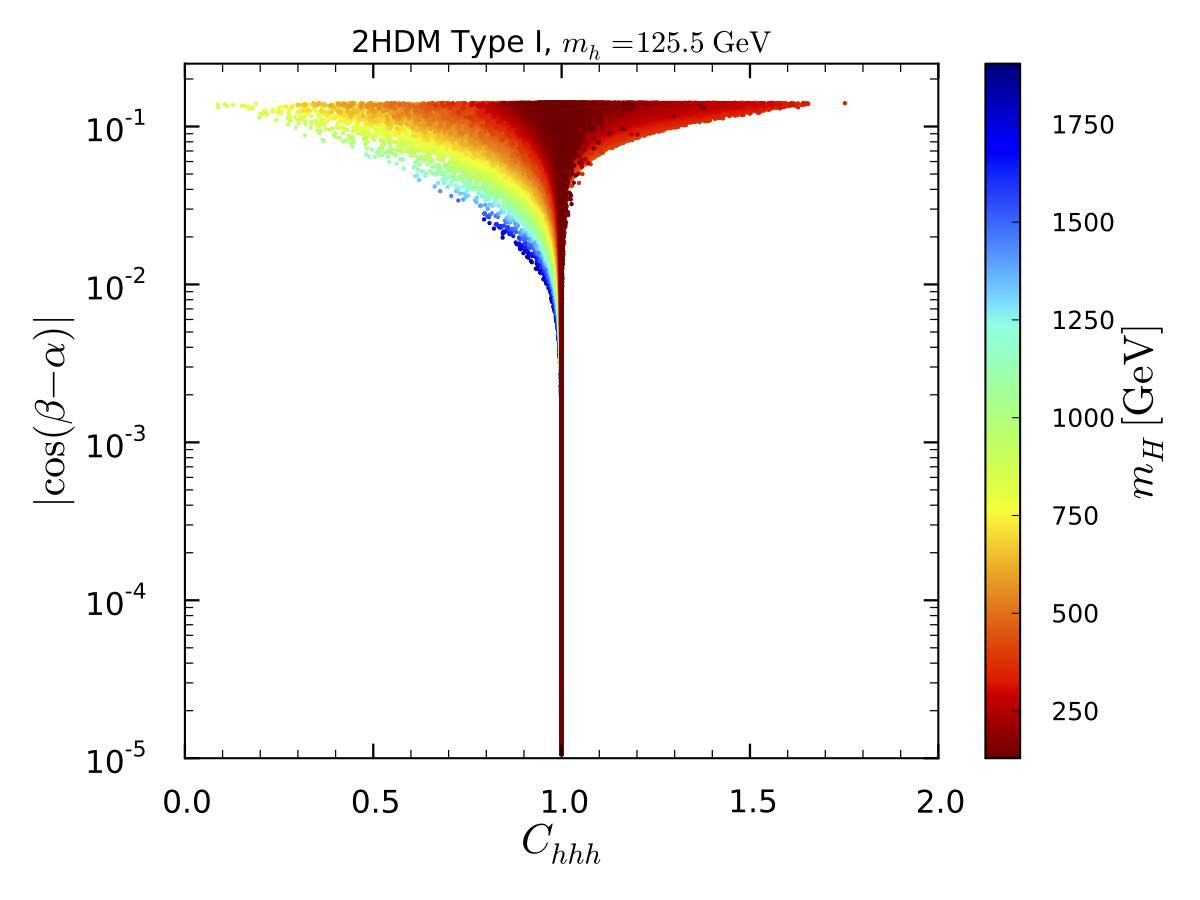}\includegraphics[width=0.5\textwidth]{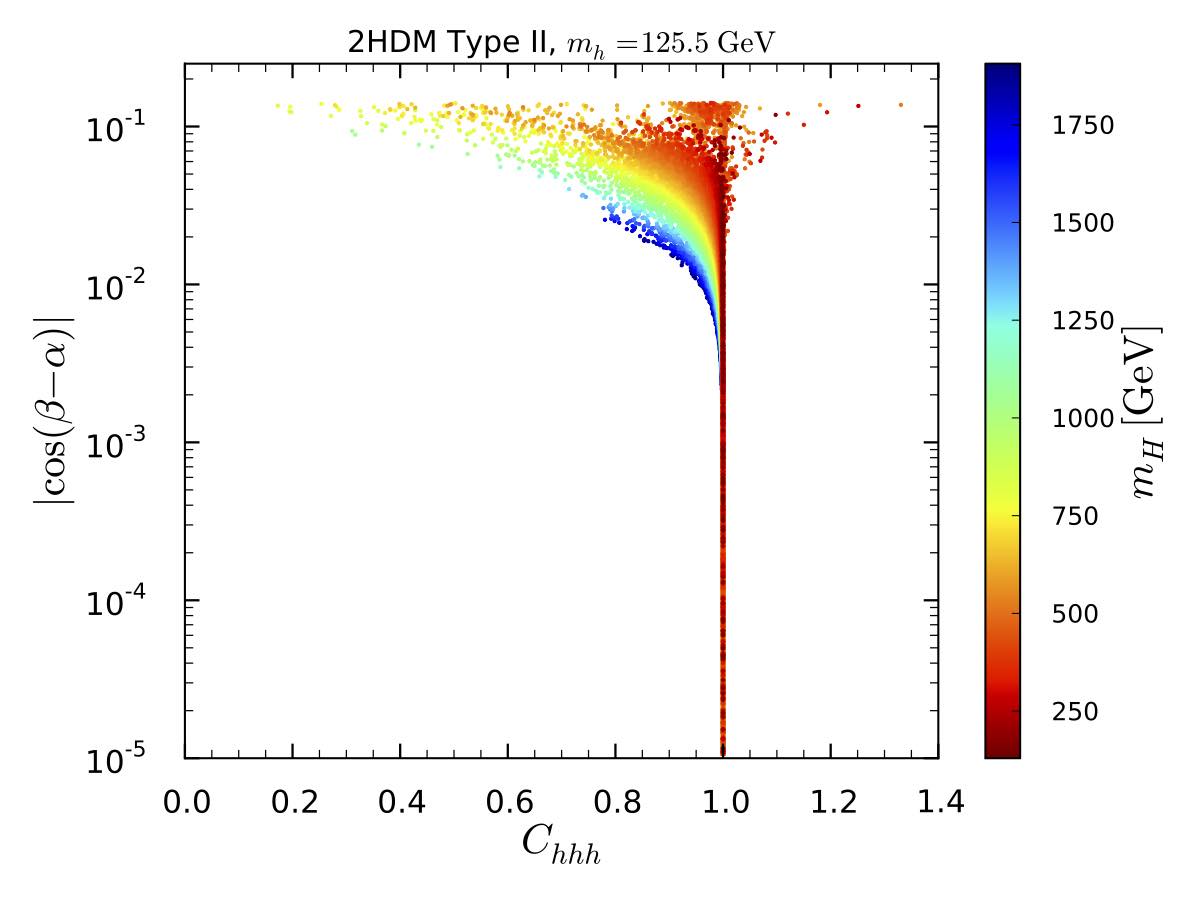}
\caption{$|\cba|$ versus the reduced triple Higgs coupling $C_{hhh}$ in Type~I (left) and Type~II (right) with $m_H$ color code. Points are ordered from high to low $m_H$ values.}
  \label{mH_Chhh_h125}
\end{figure}

While the exceedingly small deviations in $C_V^h$ that we consider here will most likely not be directly accessible at the LHC, precision measurements of the other couplings together with a measurement of, or a limit on, $m_{H,A}$ can be used for consistency checks and for eventually pinning down the model. 
Of special interest in this context is also the triple Higgs coupling. The dependence of $C_{hhh}\equiv g_{hhh}/g_{hhh}^{\rm SM}$ on $\cba$ and $m_H$ is shown in Fig.~\ref{mH_Chhh_h125}. 
It is quite striking that large values of $C_{hhh}>1$ (up to $C_{hhh}\approx 1.7$ in Type~I and up to $C_{hhh}\approx 1.35$ in Type~II) can be achieved in the non-decoupling regime, roughly $m_H\lesssim 600\gev$, for $|\cba|$ values of the order of $0.1$, whereas for heavier $m_H$,  $C_{hhh}$ is always suppressed as compared to its SM prediction.  These features were explained in the discussion below Eq.~\eqref{hhhalign}.\footnote{This cannot be seen directly in Fig.~\ref{mH_Chhh_h125}, but we verified that points with $m_H>630$~GeV never have $C_{hhh}>1$.}
Note also that for $m_H\sim 1$~TeV, $C_{hhh}$ approaches the SM limit of 1 as $|\cbma|$ decreases more slowly than is the case for  lighter $m_H$; 
substantial deviations $C_{hhh}<1$ are possible as long as $|\cba|$ is roughly greater than a few times $10^{-2}$.
This comes from the $(2Z_6/Z_1)\cba$ term in Eq.~\eqref{hhhalign}: since, in the convention where $\sbma\geq 0$, $Z_6\cba$ is always negative, cf.~Eq.~\eqref{sgnz6}, and since $Z_6$ can be sizable when $m_H\sim 1$~TeV, see Fig.~\ref{mH_cba_Z6_h125}, this can lead to a suppression as extreme as $C_{hhh}\approx 0.1$. (For $m_H\gg 1$~TeV the deviations are smaller in part because the possible range of $\cba$ is limited as seen in Fig.~\ref{mH_cba_Z6_h125}.)
For very light $m_H$, on the other hand, $Z_6$ is much smaller and hence the deviations with $C_{hhh}<1$ are more limited.  For $m_H\lesssim 250\gev$ we find $C_{hhh}\approx 0.80$--$1.40$ in Type~I and $C_{hhh}\approx 0.95$--$1.13$ in Type~II.
This is at the limit of what can be measured, as the expected precision is about 50\% at the high-luminosity options of the LHC and the ILC with 500~GeV,  
and about 10--20\% at a 1--3~TeV $e^+e^-$ linear collider with polarized beams \cite{Dawson:2013bba}. 

\begin{figure}[t!]\centering
\includegraphics[width=0.5\textwidth]{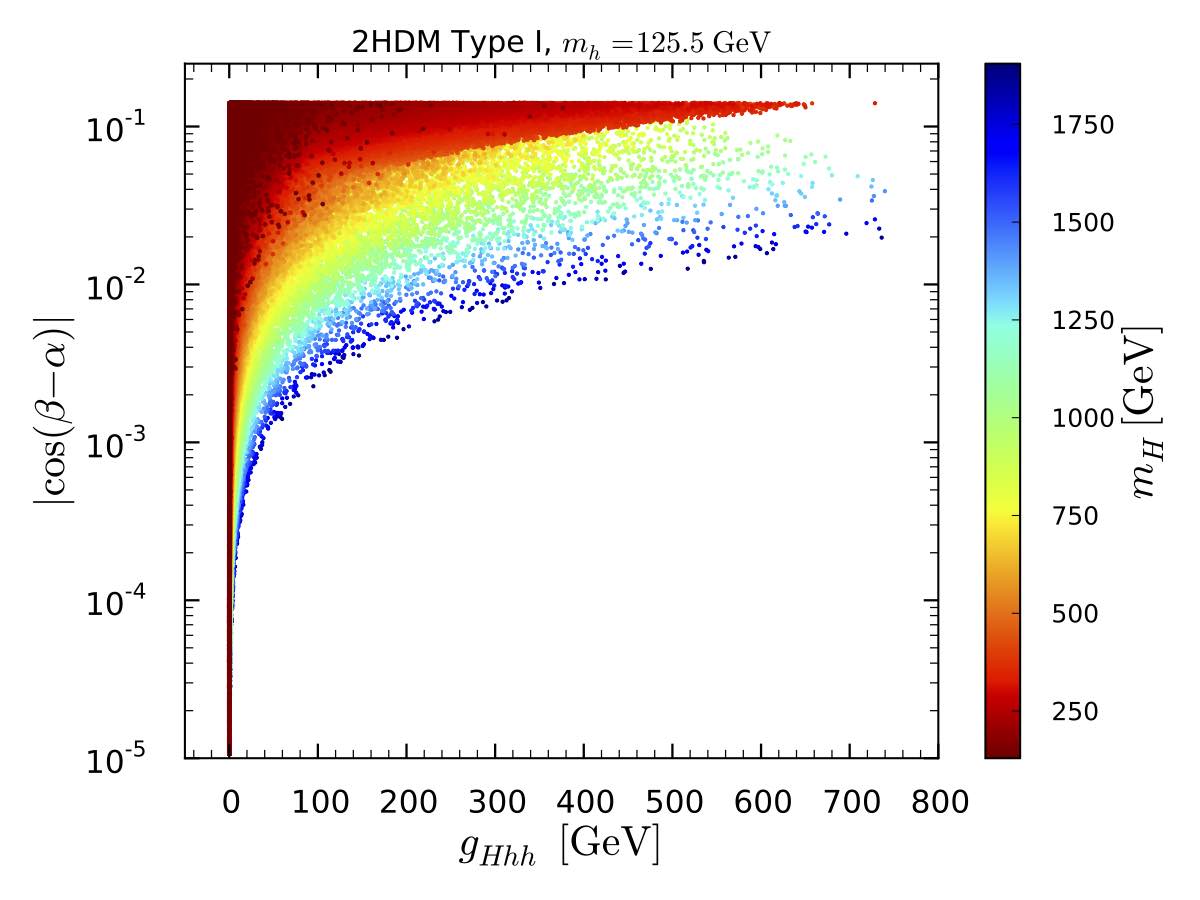}\includegraphics[width=0.5\textwidth]{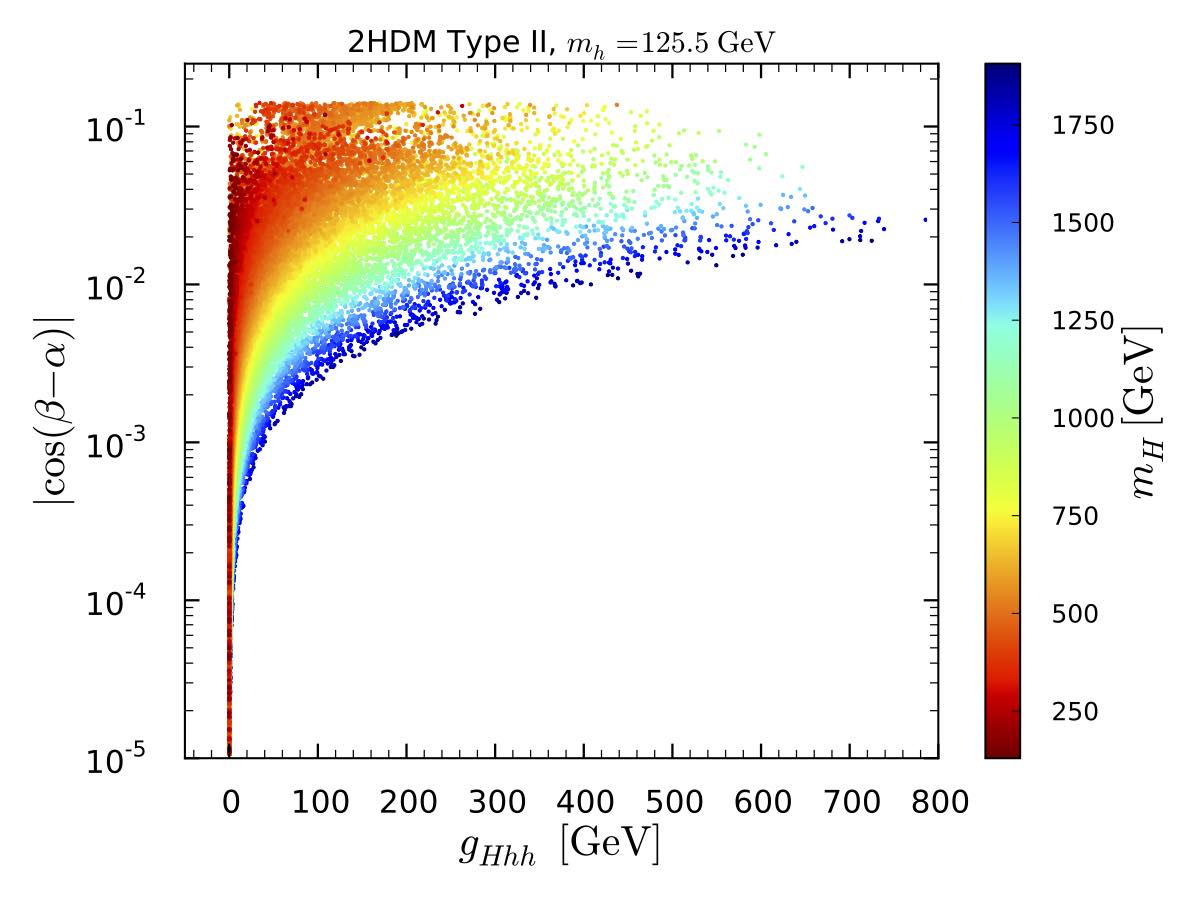}
\caption{$|\cba|$ versus the triple Higgs coupling $g_{Hhh}$ in Type~I (left) and Type~II (right) with $m_H$ color code. Points are ordered from high to low $m_H$ values.}
  \label{mH_CHhh_h125}
\end{figure}

The relation between the triple Higgs coupling $g_{Hhh}$, $|\cba|$ and $m_H$ is presented in Fig.~\ref{mH_CHhh_h125}. In Type~I, large values of $g_{Hhh}$ can be achieved in the non-decoupling regime for $|\cba|$ of the order $10^{-1}$. This is also true in Type~II, though the range of $g_{Hhh}$ is somewhat smaller.  
We observe moreover that for given $|\cba|\lesssim 10^{-1}$, 
the achievable $Hhh$ coupling grows with $m_H$. Nonetheless, as will be shown in Section~\ref{cross-sections},  
the $H\to hh$ decay is mostly relevant below the $t\bar{t}$ threshold. 
Moreover, in the exact alignment limit, the $Hhh$ coupling vanishes.

\subsection{Signal strengths}\label{signalstrengths}

The variations in the couplings to fermions discussed above have direct consequences for the signal strengths of the SM-like Higgs boson. 
Since the results depend a lot on the fermion coupling structure, we examine this separately for Type~I and Type~II. 

\begin{figure}[t!]\centering
\includegraphics[width=0.5\textwidth]{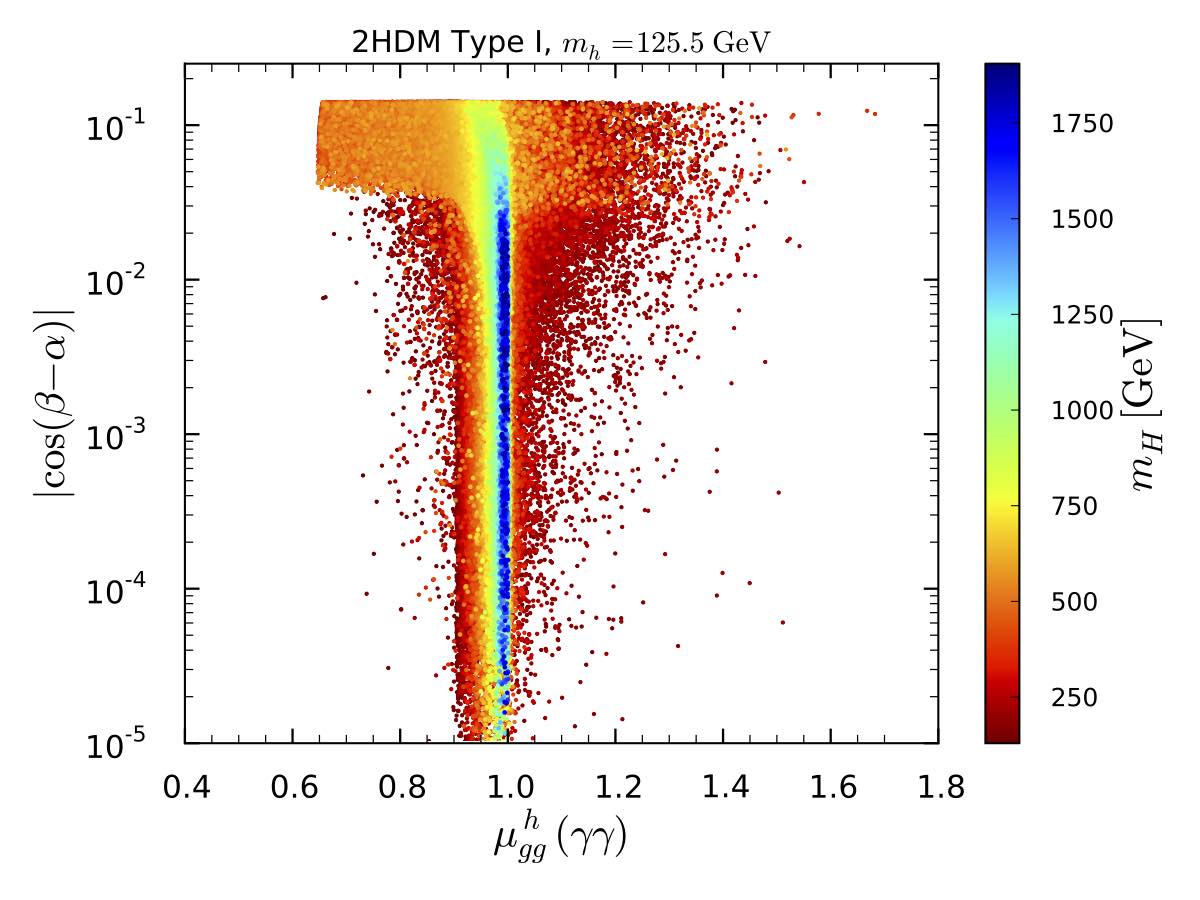}\includegraphics[width=0.5\textwidth]{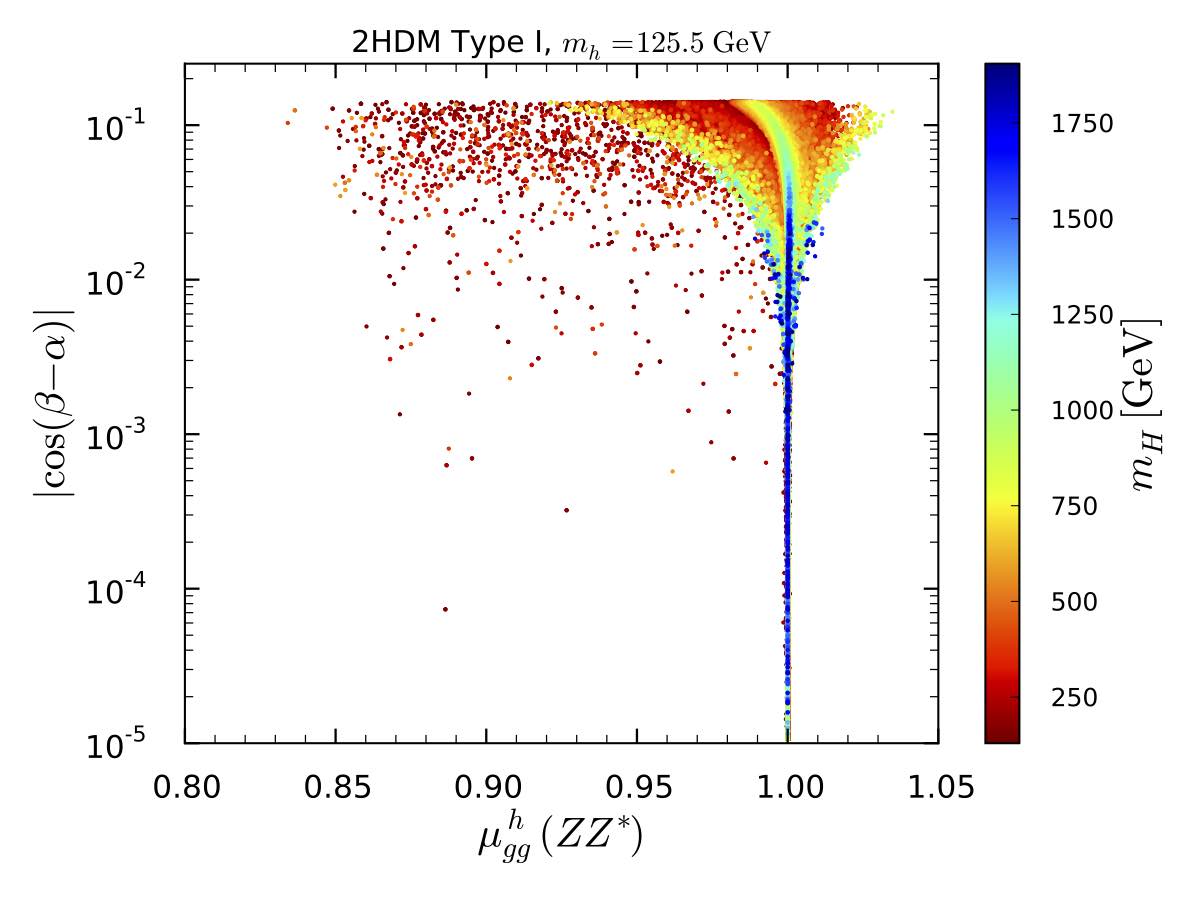}  \caption{Signal strengths in Type~I for the 125.5 GeV state, for $gg\to h\to\gamma\gamma$ (left) and $gg\to h\to ZZ^*$ (right) with $m_H$ color code. Points are ordered from low to high $m_H$ values.  Points with $\mu_{gg}^h(ZZ^*)<0.92$ are ones for which $h\to AA$ decays are present, so that the total $h$ width is increased, which suppresses this particular channel's rate.}
  \label{mu_mh_h125_type1}
\end{figure}

Let us start with Type~I. Figure~\ref{mu_mh_h125_type1} shows the signal strengths for 
gluon-gluon fusion and decay into $\gamma\gamma$ ($\mu_{gg}^h(\gamma\gamma)$, left panel), and 
decay into $ZZ^*$ ($\mu_{gg}^h(ZZ^*)$, right panel).  
Recalling that $C_F^h$ varies between 0.87 and 1.11 in Type~I and comparing with Fig.~\ref{Cgluongamma_h125}, 
it is clear that the variation in $\mu_{gg}^h(\gamma\gamma)$ comes to a large extent from the charged Higgs contribution to the $\gamma\gamma$ loop. Even for $|\cba|\to 0$, large deviations from 1 can occur due to a sizable charged Higgs contribution or the presence of a light pseudoscalar $m_A<m_h/2$ that increases the SM-like Higgs total width.
On the other hand, in the decoupling limit, the charged Higgs loop is small and $C^h_{\gamma}$ is largely determined by the relative size of the top and bottom loops compared to the $W$ loop (which enters with opposite sign).  On the contrary, $C_{g}^h$ is solely determined by the size of the $t$ and $b$ loop contributions. One finds numerically that the $h\gam\gam$ coupling is more suppressed than the $hgg$ coupling is enhanced, so that $\mu^h_{gg} (\gamma\gamma) \lesssim1$ in the decoupling regime.

\begin{figure}[t!]\centering
\includegraphics[width=0.5\textwidth]{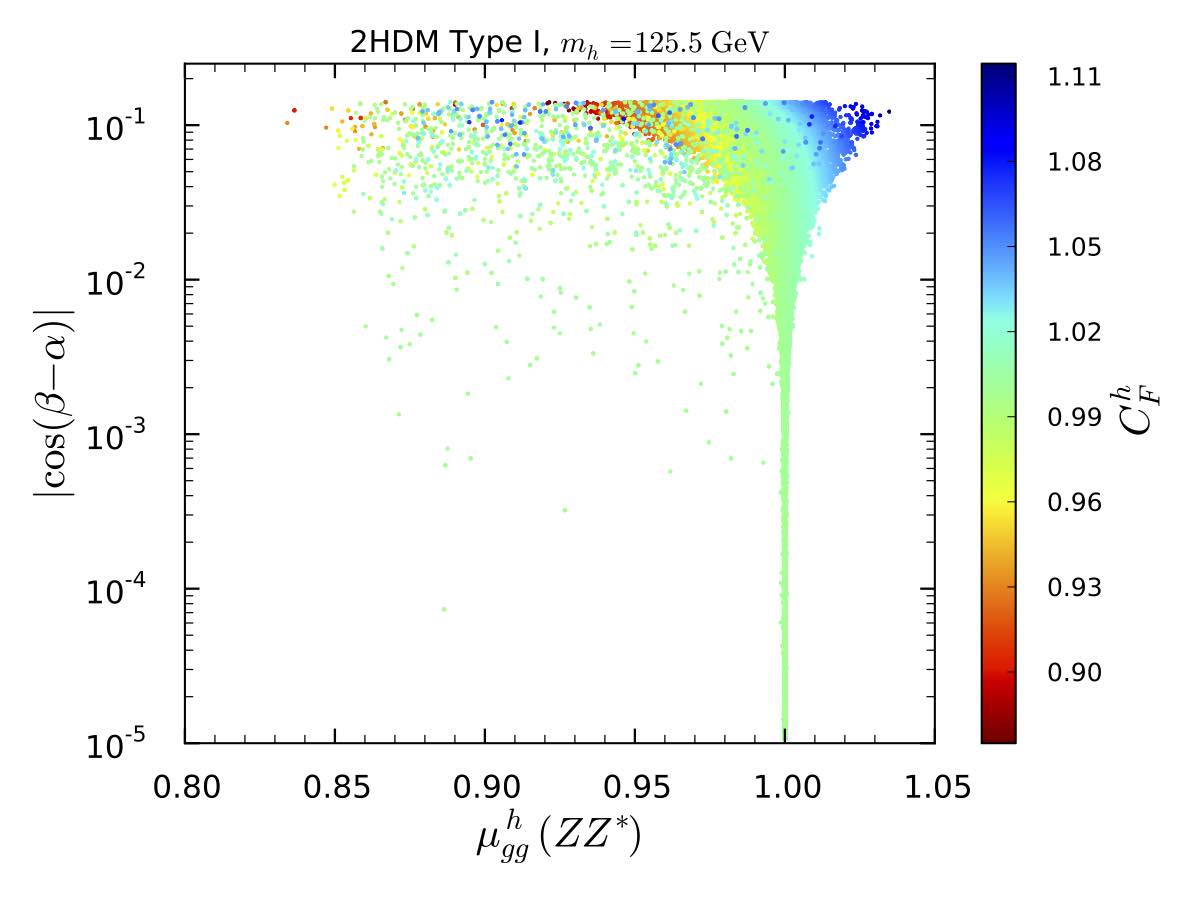}\includegraphics[width=0.5\textwidth]{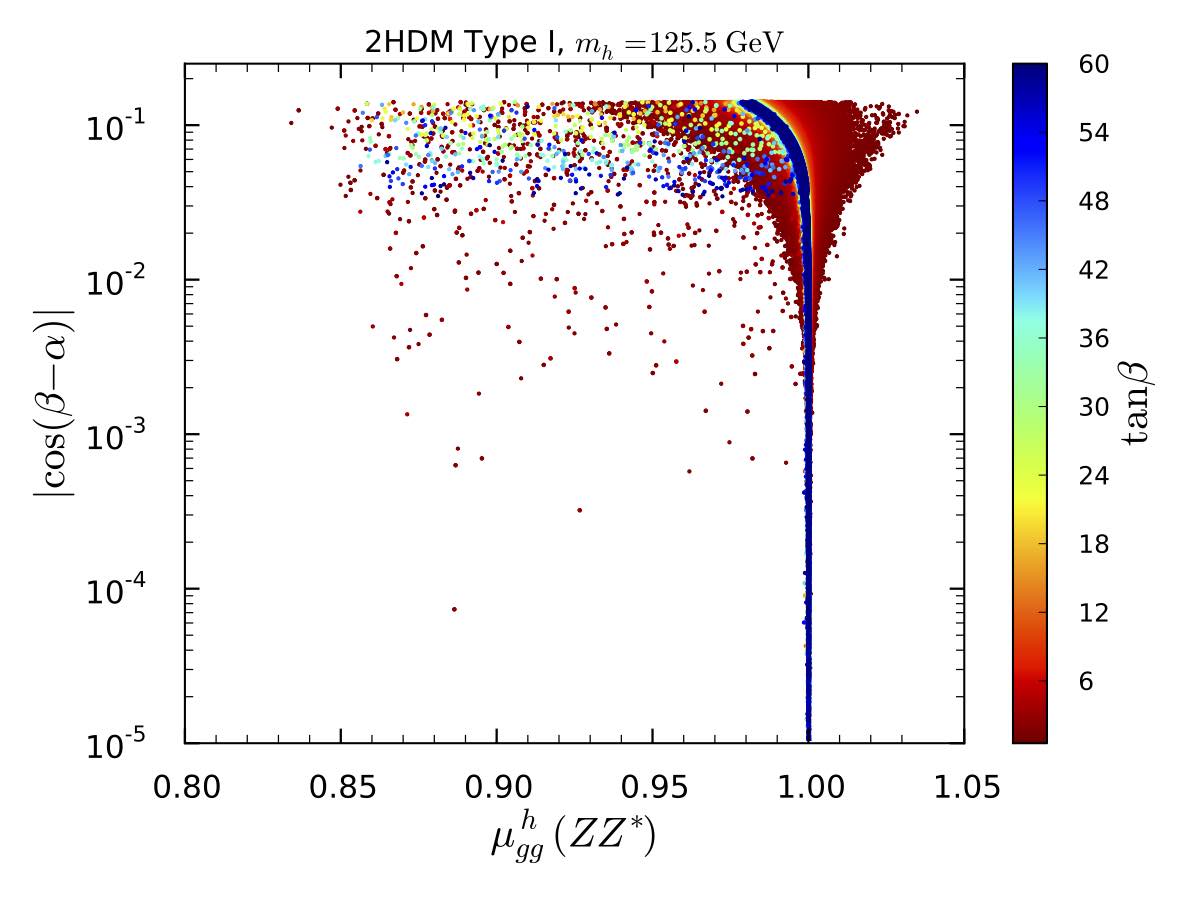}
  \caption{Signal strength for $gg\to h\to ZZ^*$ in Type~I for the 125.5 GeV state with $C_F^h$ (left) and $\tan\beta$ (right) color code. Points are ordered from low to high $C_F^h$ and $\tan\beta$ values.}
  \label{mu_CFtb_h125_type1}
\end{figure}

In contrast, $\mu_{gg}^h(ZZ^*)$ shows less variation, $\mu_{gg}^h(ZZ^*)=[0.92,1.04]$ if the $h\to AA$ decay channel is closed, with small excursions around 1 allowed in the decoupling limit. It also exhibits a less distinct dependence on $m_H$ compared to $\mu_{gg}^h(\gam\gam)$. The reason is that $\mu_{gg}^h(ZZ^*)$ is driven by $C_F^h$ and $\tan\beta$, as illustrated in Fig.~\ref{mu_CFtb_h125_type1}. The dependence on $C_F^h$ is clear as larger (smaller) $C_F^h$ leads to larger (smaller) cross section for $gg\to h$. The dependence on $\tan\beta$ results from an interplay between the top (which drives the $gg\to h$ cross section) and bottom (which drives the total $h$ width) Yukawa couplings both given by $C_F^h=\sba+\cba/\tb$. The scattered points with suppressed $\mu_{gg}^h(ZZ^*)$ are those where the $h\to AA$ decay mode is open and increases the total width. An analogous picture emerges for the VBF-induced $h\tau\tau$ signal strengths, since $\mu_{\rm VBF}^h(\tau\tau)=\mu_{gg}^h(ZZ^*)$ in Type~I.

\begin{figure}[t!]\centering
\includegraphics[width=0.5\textwidth]{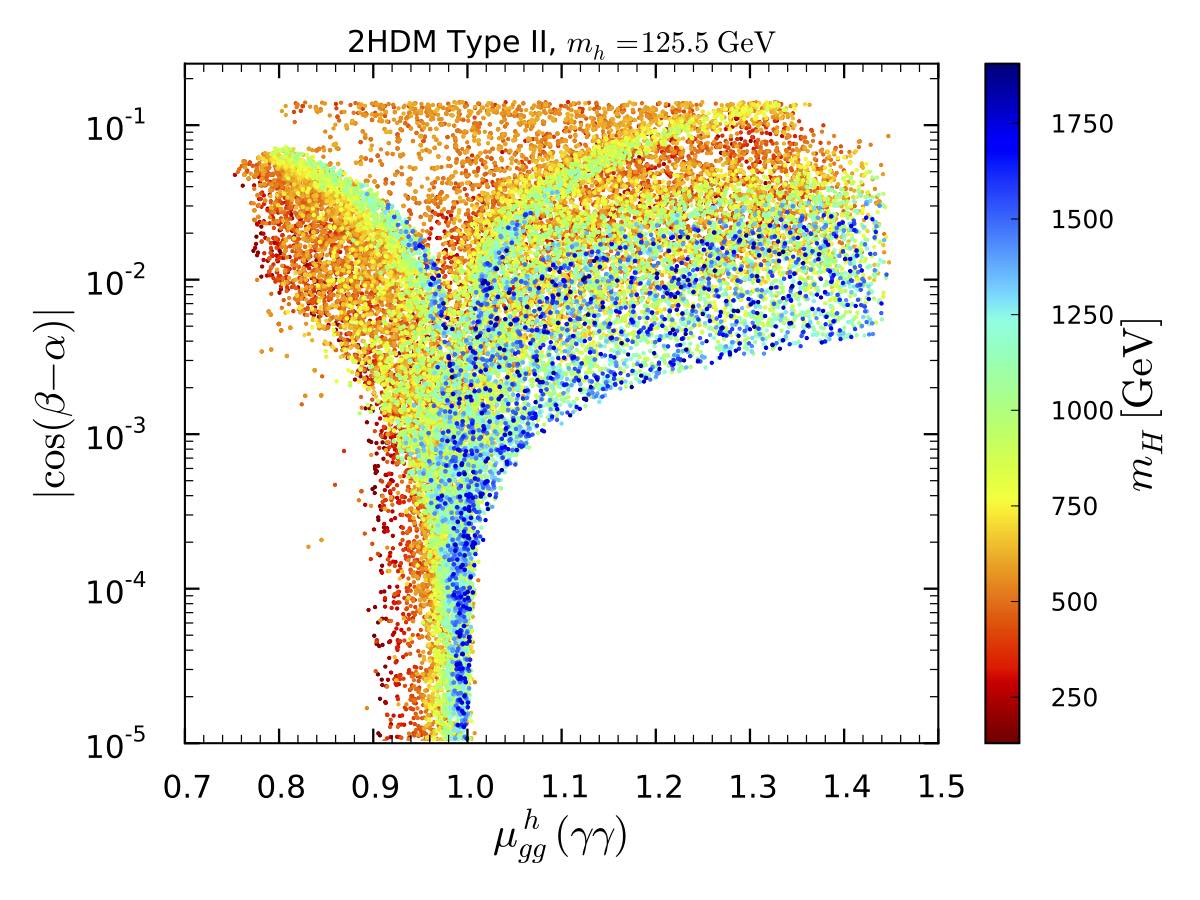}\includegraphics[width=0.5\textwidth]{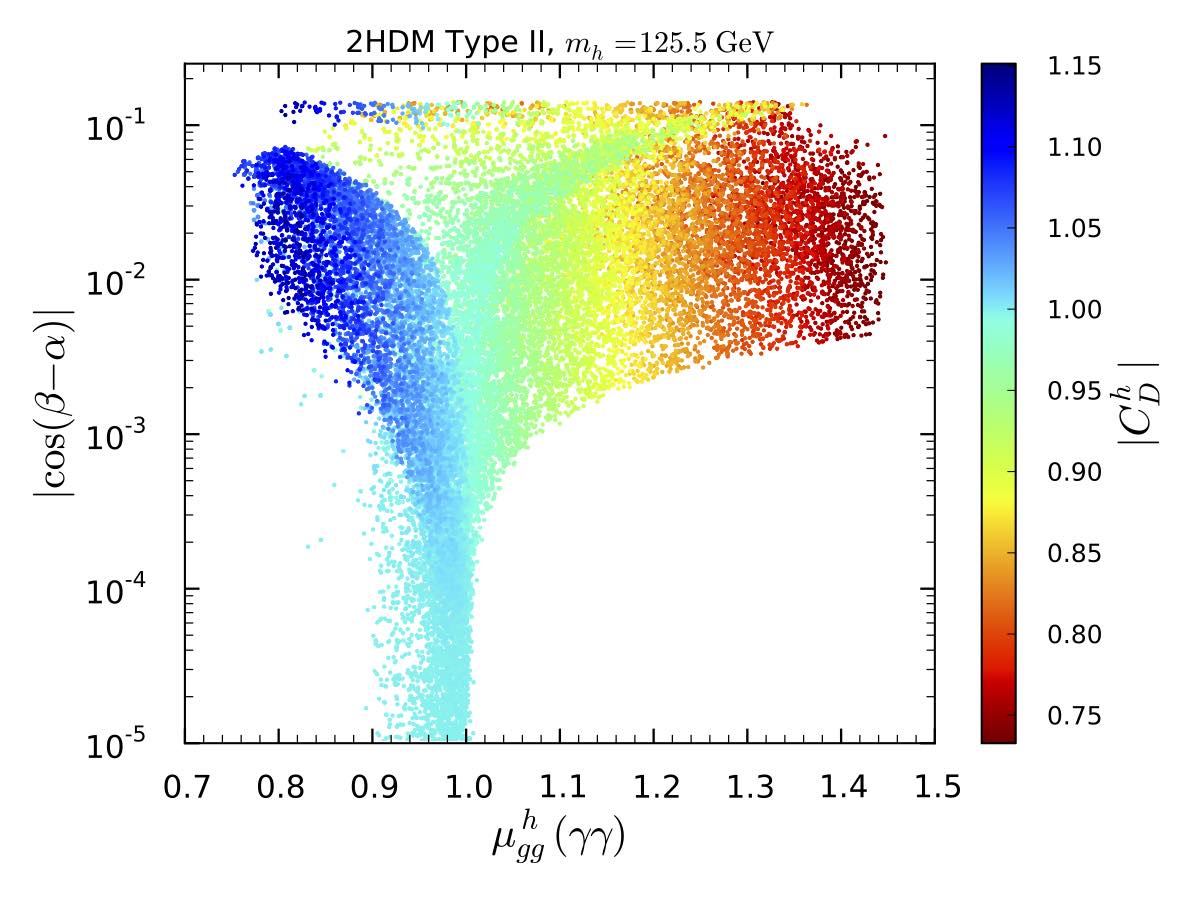}\\
\includegraphics[width=0.5\textwidth]{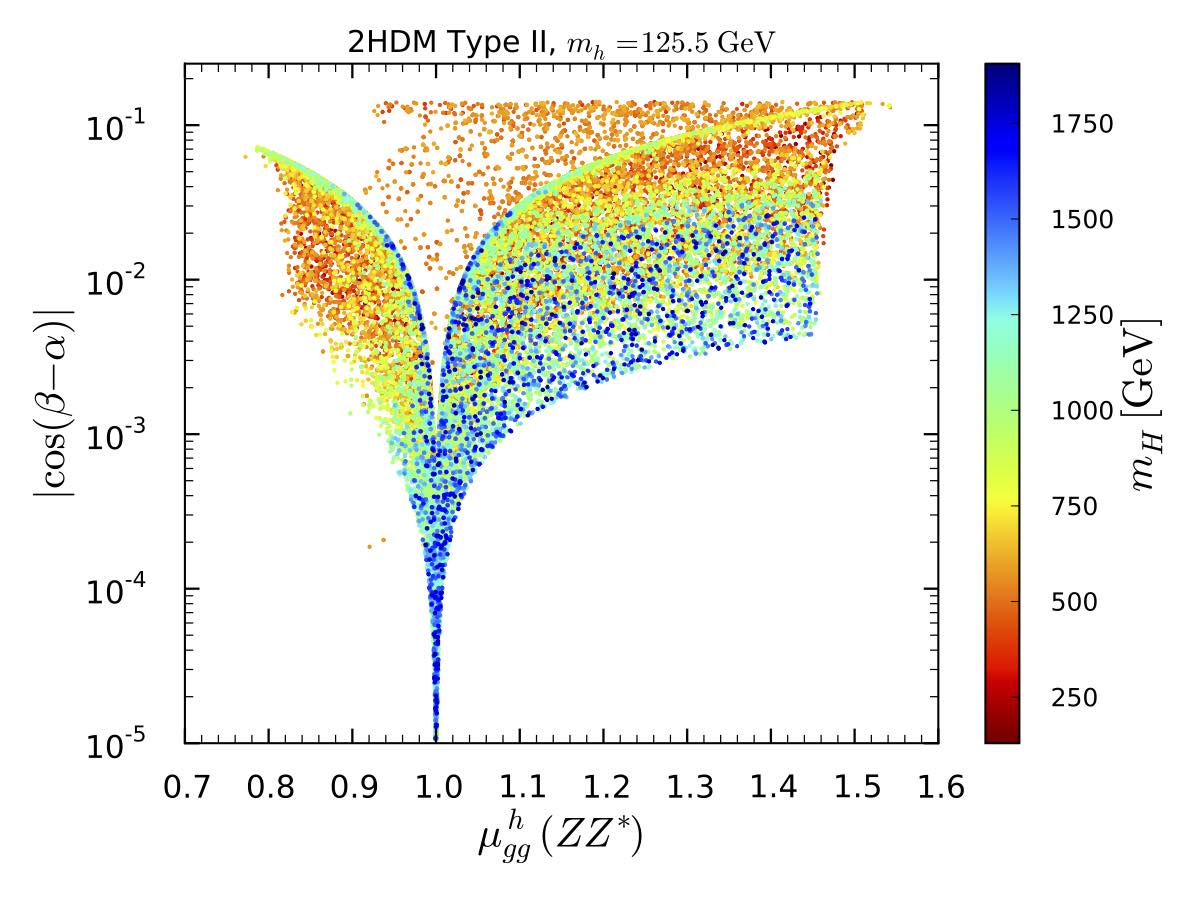}\includegraphics[width=0.5\textwidth]{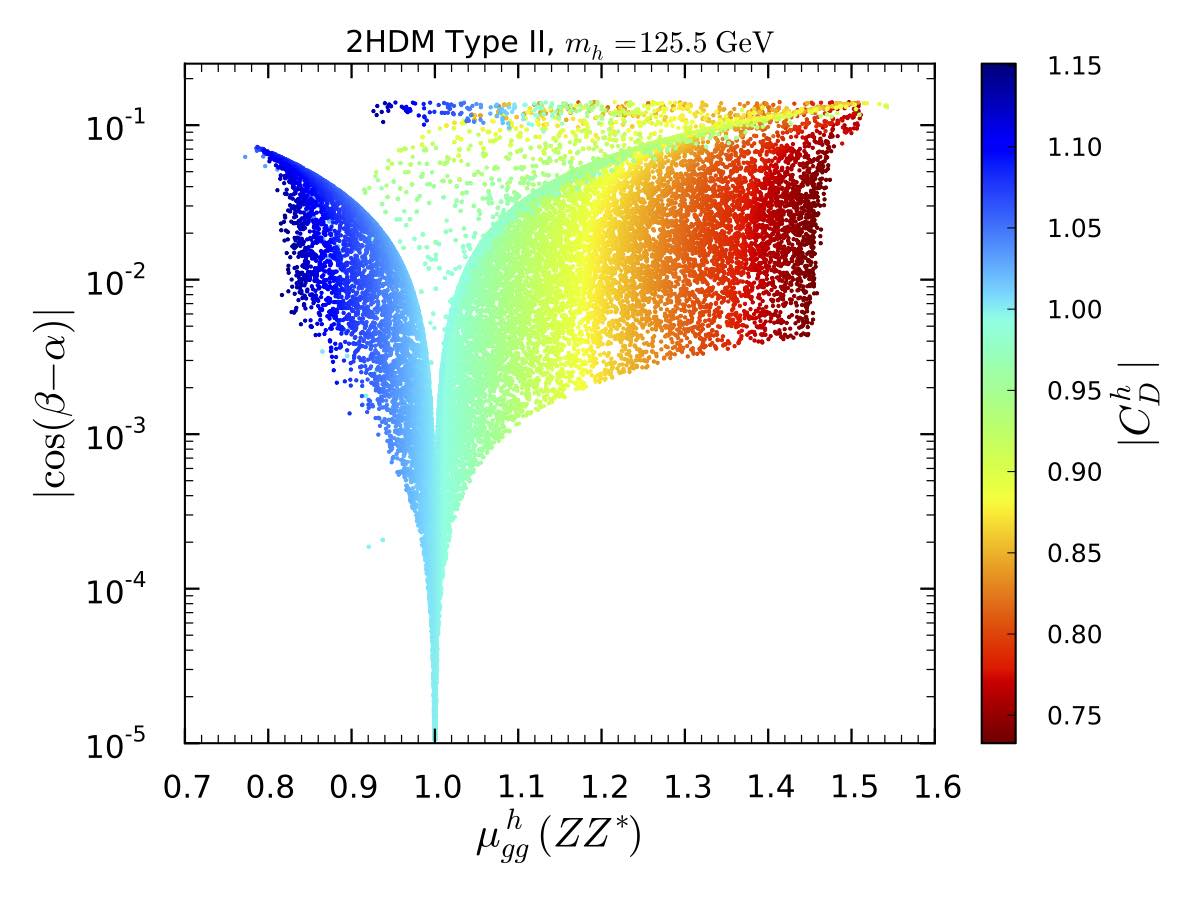}\\  \includegraphics[width=0.5\textwidth]{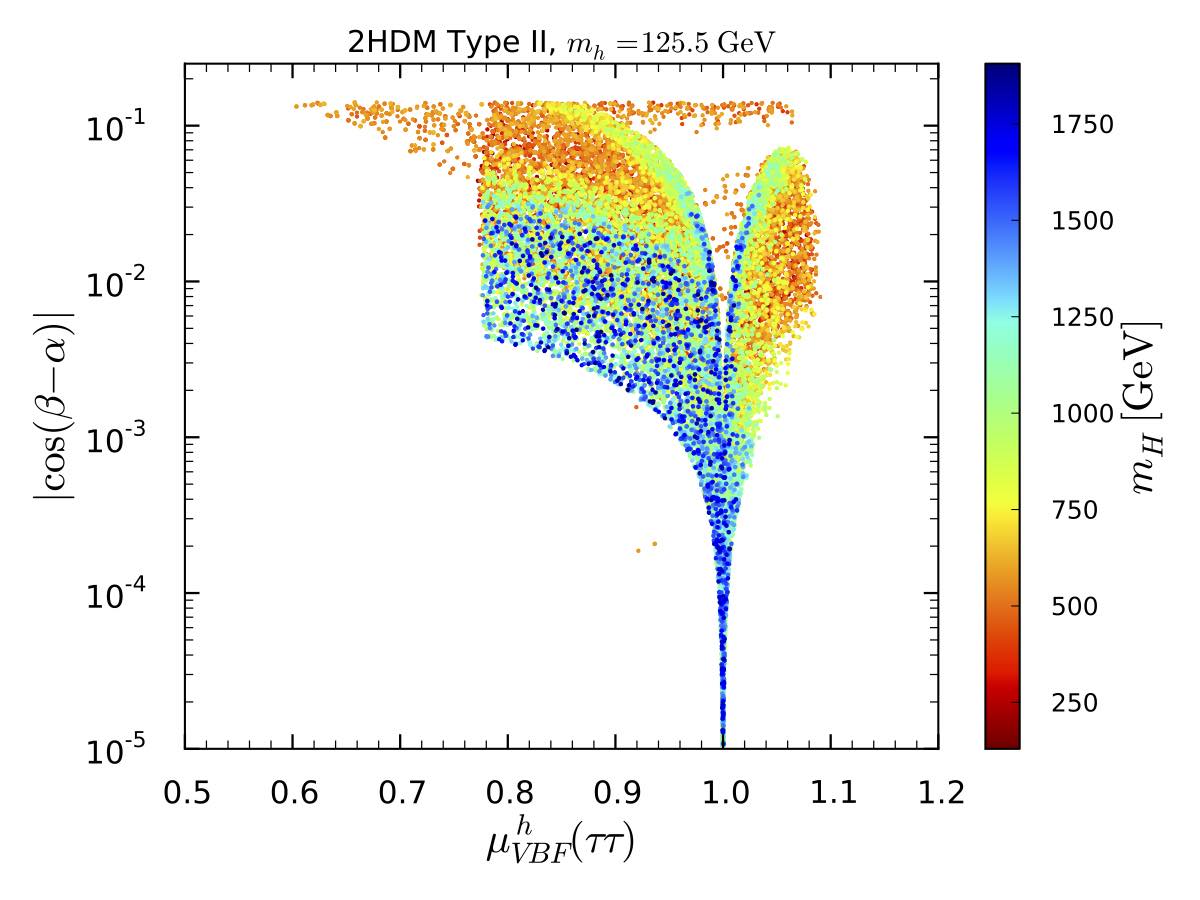}\includegraphics[width=0.5\textwidth]{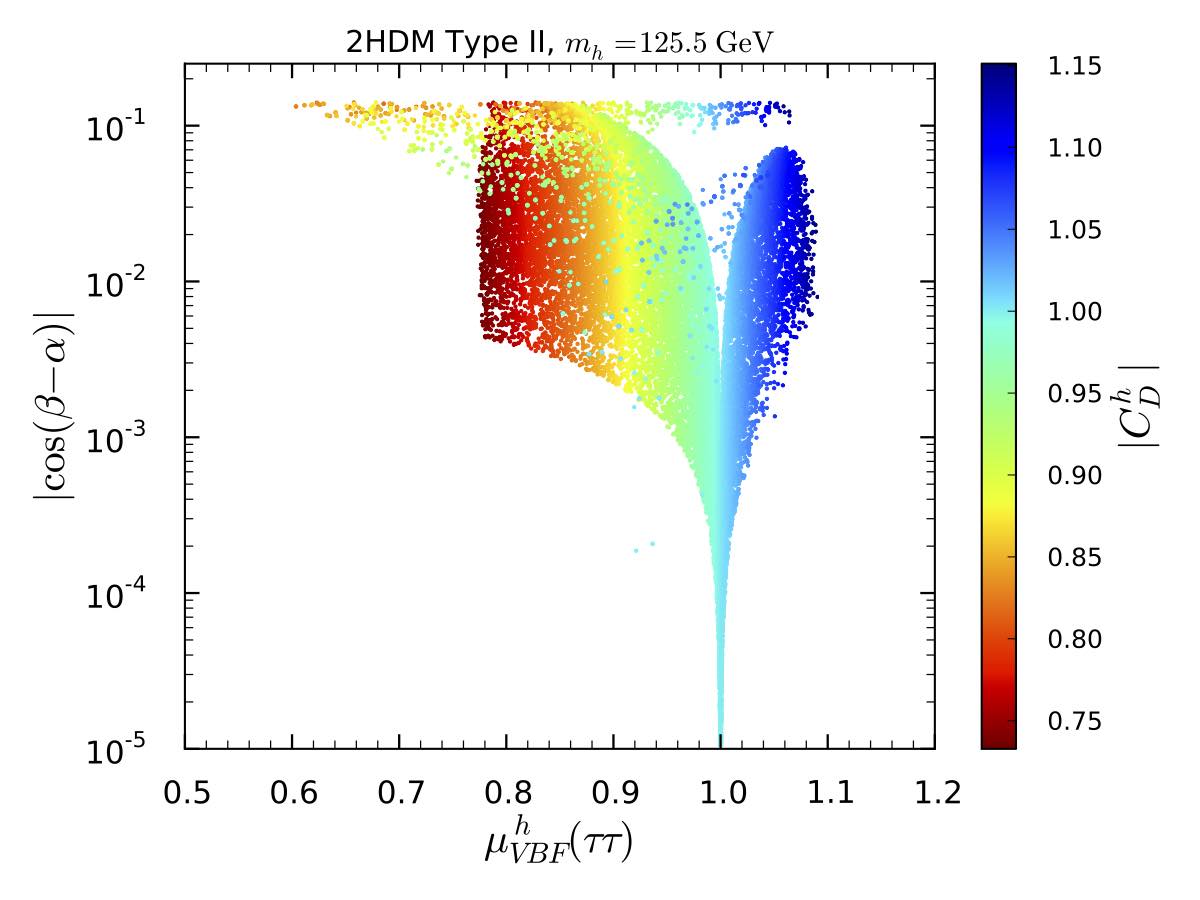}\\  \caption{Signal strengths in Type~II for the 125.5 GeV state with $m_H$ (left) and $|C_D^h|$ color code. Points are ordered from low to high $m_H$ and $|C_D^h|$ values.}
  \label{mu_mh_h125_type2}
    \vskip 0.4in
\end{figure}

In Type~II, we find that the situation is quite different. Here, the signal strengths are driven by both the top quark coupling, which impacts $C_g^h$, and by the bottom Yukawa coupling $C_D^h$, which also enters $C_g^h$ and, often of greatest importance, determines the $h\to b\bar b$ decay width. In Fig.~\ref{mu_mh_h125_type2} we therefore show the signal strengths $\mu_{gg}^h(\gamma\gamma)$, $\mu_{gg}^h(ZZ^*)$ and $\mu_{\rm VBF}^h(\tau\tau)$ in Type~II 
comparing the dependence on $m_H$ (left panels) to the dependence on $|C_D^h|$ (right panels). Note that the $m_H$ dependence of the signal strengths reflects the $m_H$ dependence of $C_D^h$ in Fig.~\ref{CD_h125}. 
As a consequence, $\mu_{gg}^h(\gamma\gamma)$ and $\mu_{gg}^h(ZZ^*)$ can be enhanced in the decoupling regime, with values going as high as 1.4--1.5 (mainly due to suppression of the total $h$ width), to be compared to the current model-independent 95\% CL limits of $\mu_{gg}^h(\gamma\gamma)\in [0.76,1.69]$ and $\mu_{gg}^h(ZZ^*)\in [0.71,1.80]$. Suppression is also possible, reaching a level of $\mu_{gg}^h(\gamma\gamma)=$0.74--0.76  for low $m_H$ if $|\cbma| > 0.01$ but limited to 0.9 for large $m_H\gtrsim 1250$ GeV.  For all $m_H$, the amount of possible suppression decreases systematically with decreasing $|\cbma|$.
For $\mu_{\rm VBF}^h(\tau\tau)$ the behaviour is exactly opposite. For completeness we note that the horizontal bar at $|\cbma|\sim 10^{-1}$ is the $C_D^h<0$ region, and the scattered points are those where the $h\to AA$ decay is open.\footnote{This region has been sizably affected by the updated constraints as described at the end of Section~\ref{setup}.} 
Finally note that as $|\cbma|$ decreases, the signal strengths in Type~II converge to 1 much more slowly than in Type~I. This is a consequence of the delayed alignment of $C_D^h$ to 1 in Type~II when $\tanb$ is large.
An additional effect arises in $\mu^h_{gg}(\gamma\gamma)$ due to the charged Higgs loop contribution to the $\hl\to\gamma\gamma$ amplitude.  In 
particular, there exists an intermediate range of charged Higgs masses\footnote{In this intermediate mass region, the charged Higgs mass is given by \eq{chhiggs}, where $Y_2\sim\mathcal{O}(v^2)$ and $Z_3\gsim 1$ such that the upper bound of $Z_3$ is constrained by its unitarity bound.}
for which $g_{hH^+ H^-}\sim -2 m_{H^\pm}^2/v$ [cf.~\eq{hlhphm}], which yields a constant non-decoupling contribution that suppresses the $\hl\to\gamma\gamma$ amplitude~\cite{Ferreira:2014naa} (see also \cite{Bhattacharyya:2013rya,Bhattacharyya:2014oka}). 
Indeed, even for values of $|\cba|$ as low as $10^{-4}$, this signal strength does not converge to 1 until $m_H$ (and thus $m_{H^\pm}$) is above about 1~TeV.

\begin{figure}[t!]\centering
\includegraphics[width=0.5\textwidth]{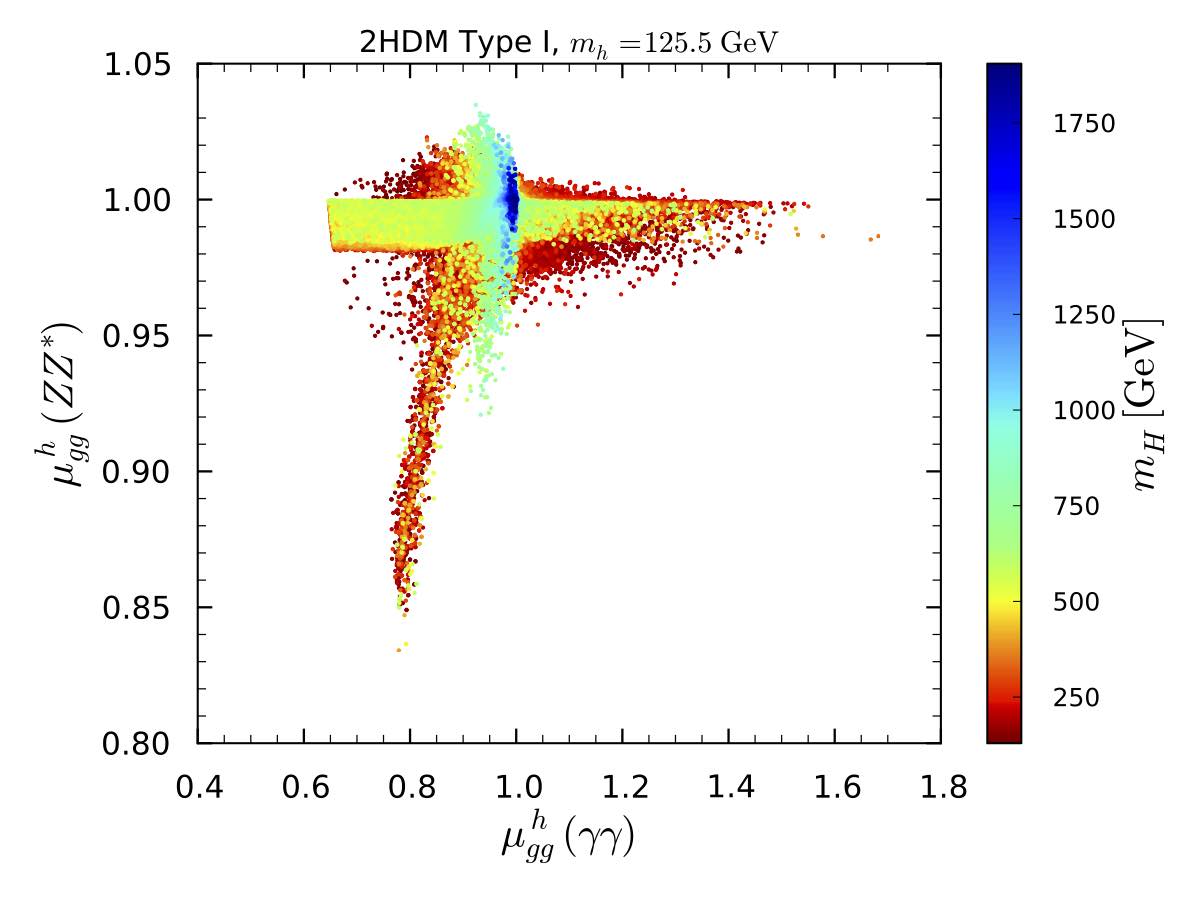}\includegraphics[width=0.5\textwidth]{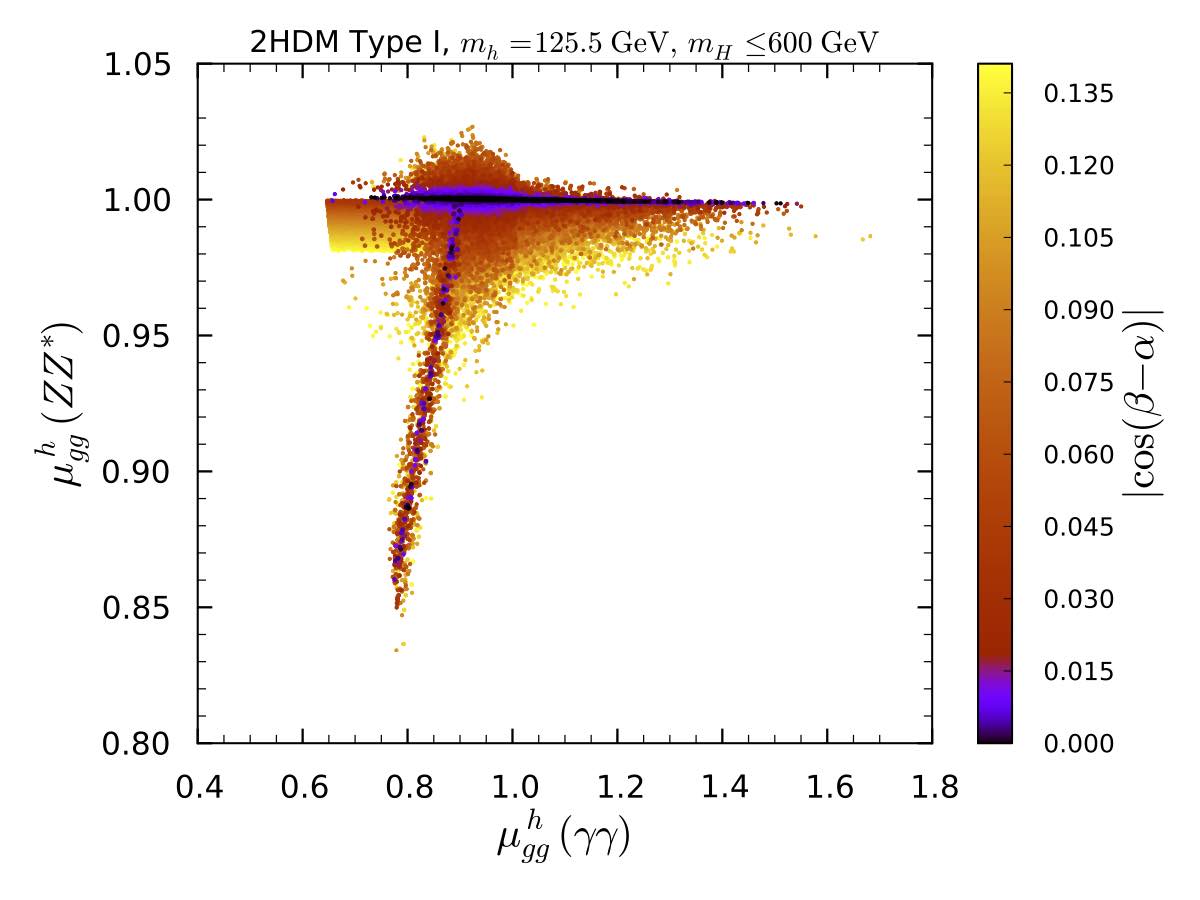}\\
\includegraphics[width=0.5\textwidth]{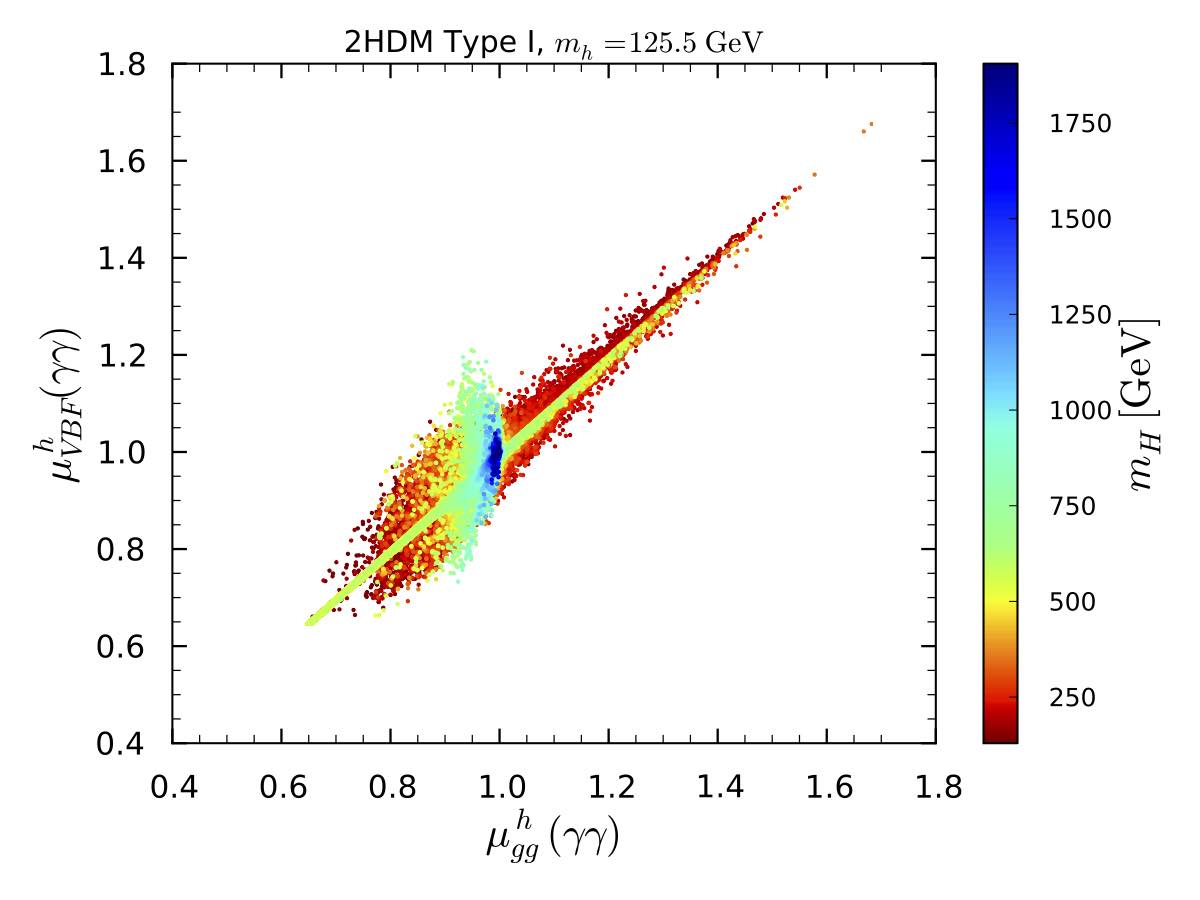}\includegraphics[width=0.5\textwidth]{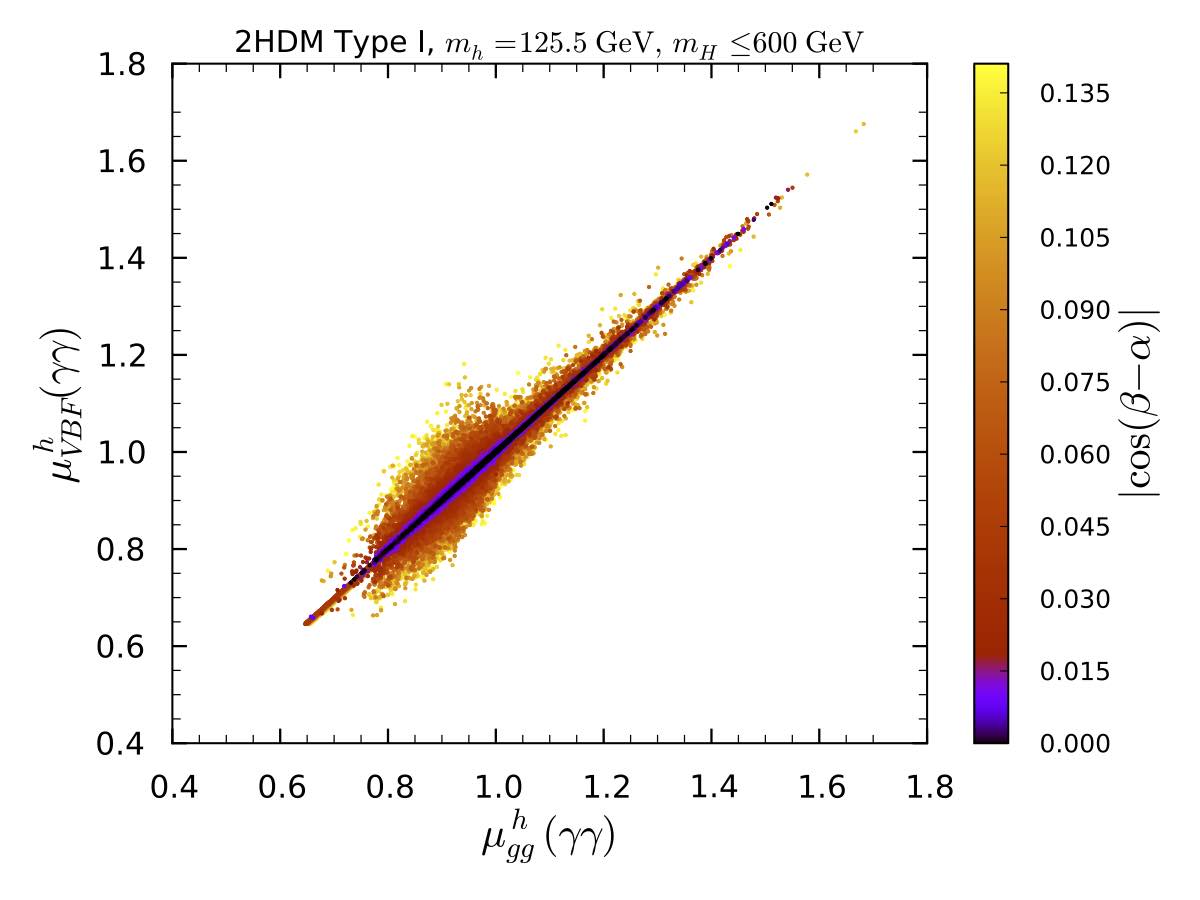}
\caption{Correlations of signal strengths in Type~I, on the left illustrating the dependence on $m_H$, on the right 
illustrating the dependence in $|\cba|$. Points are ordered from low to high $m_H$ values (left) and high to low $|\cba|$ values (right).}
  \label{mu_mu_h125_type1}
\end{figure}

\begin{figure}[t!]\centering
\includegraphics[width=0.5\textwidth]{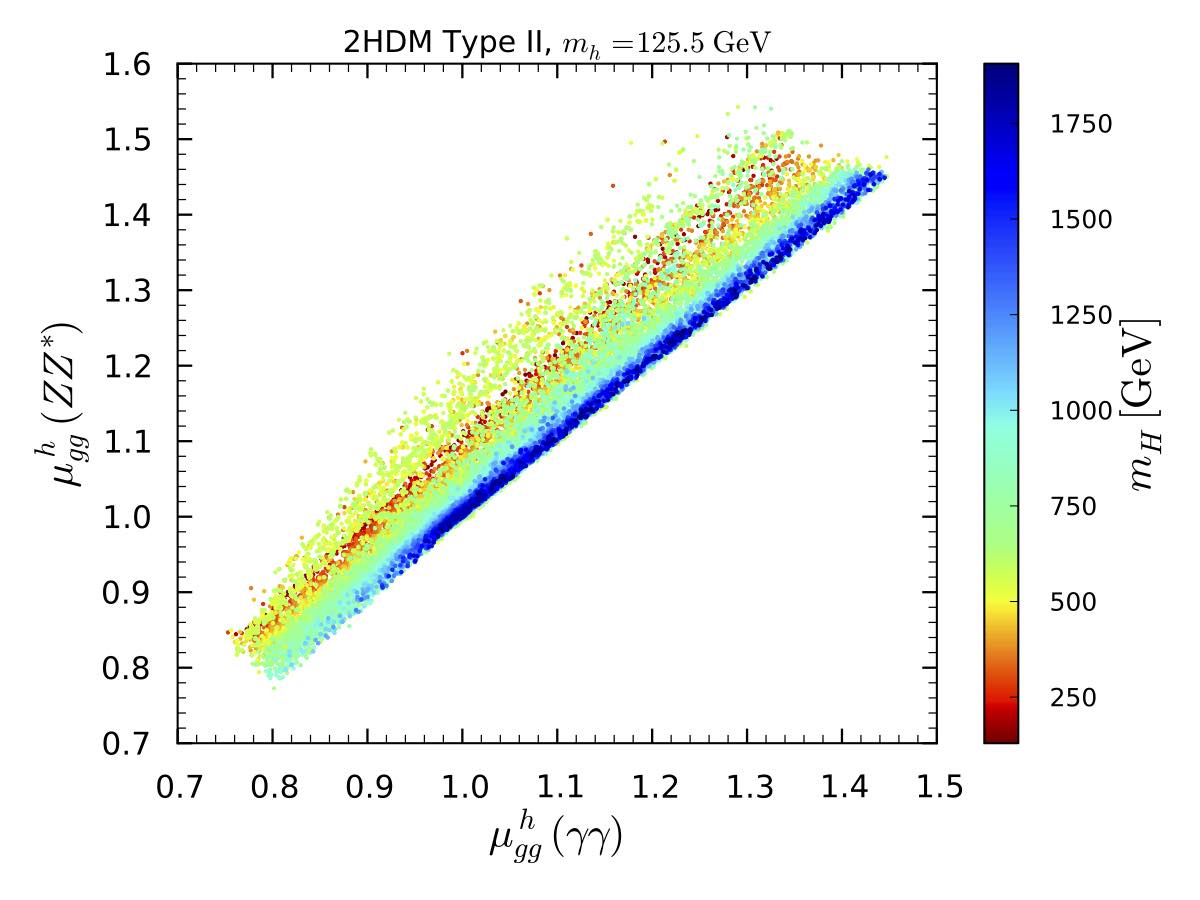}\includegraphics[width=0.5\textwidth]{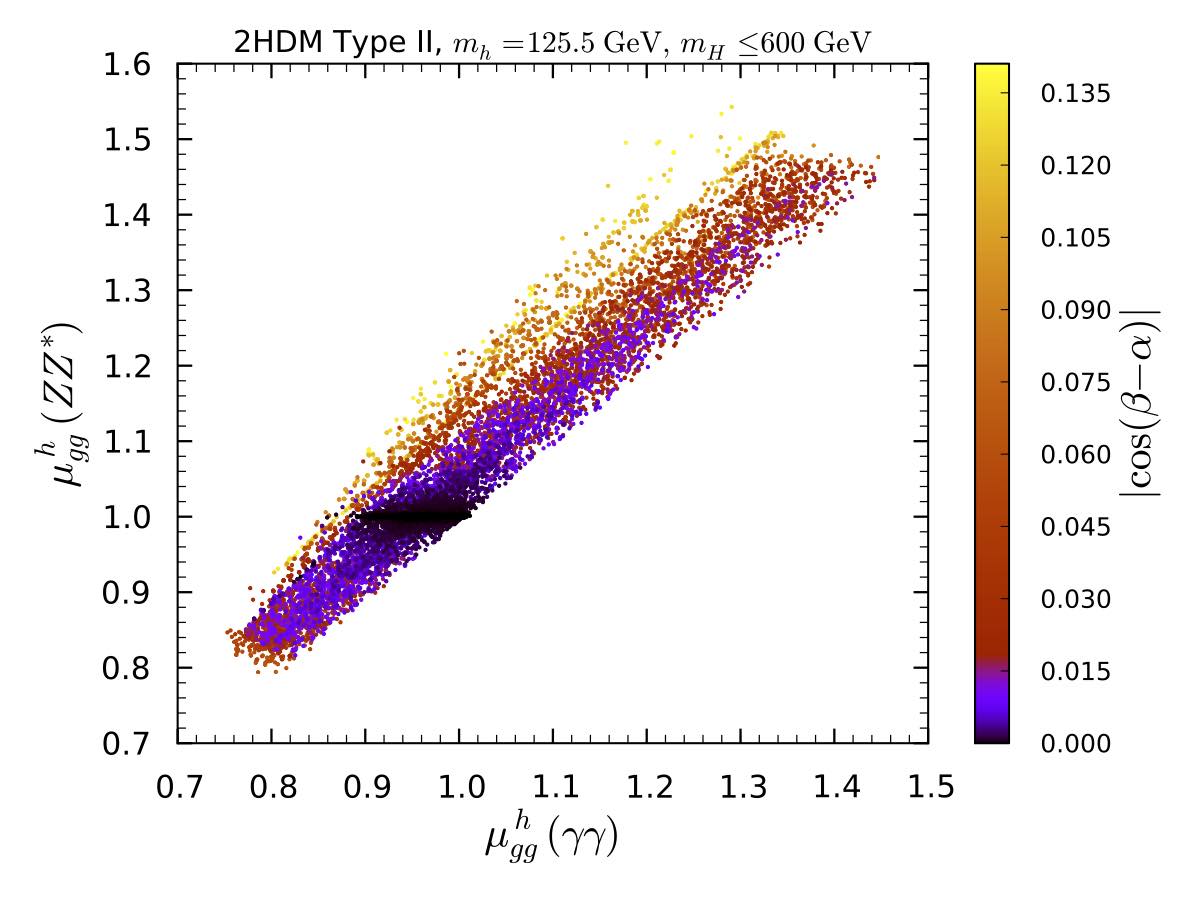}\\
\includegraphics[width=0.5\textwidth]{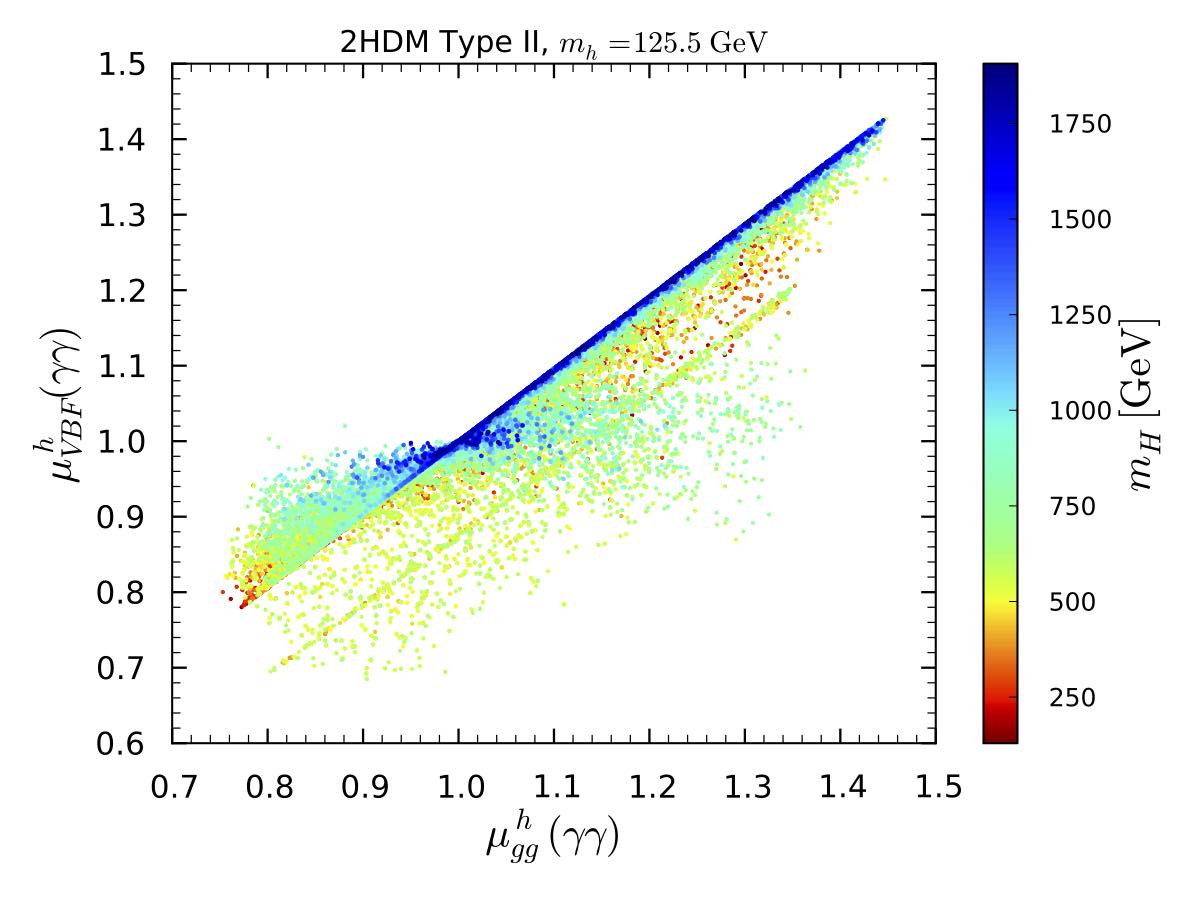}\includegraphics[width=0.5\textwidth]{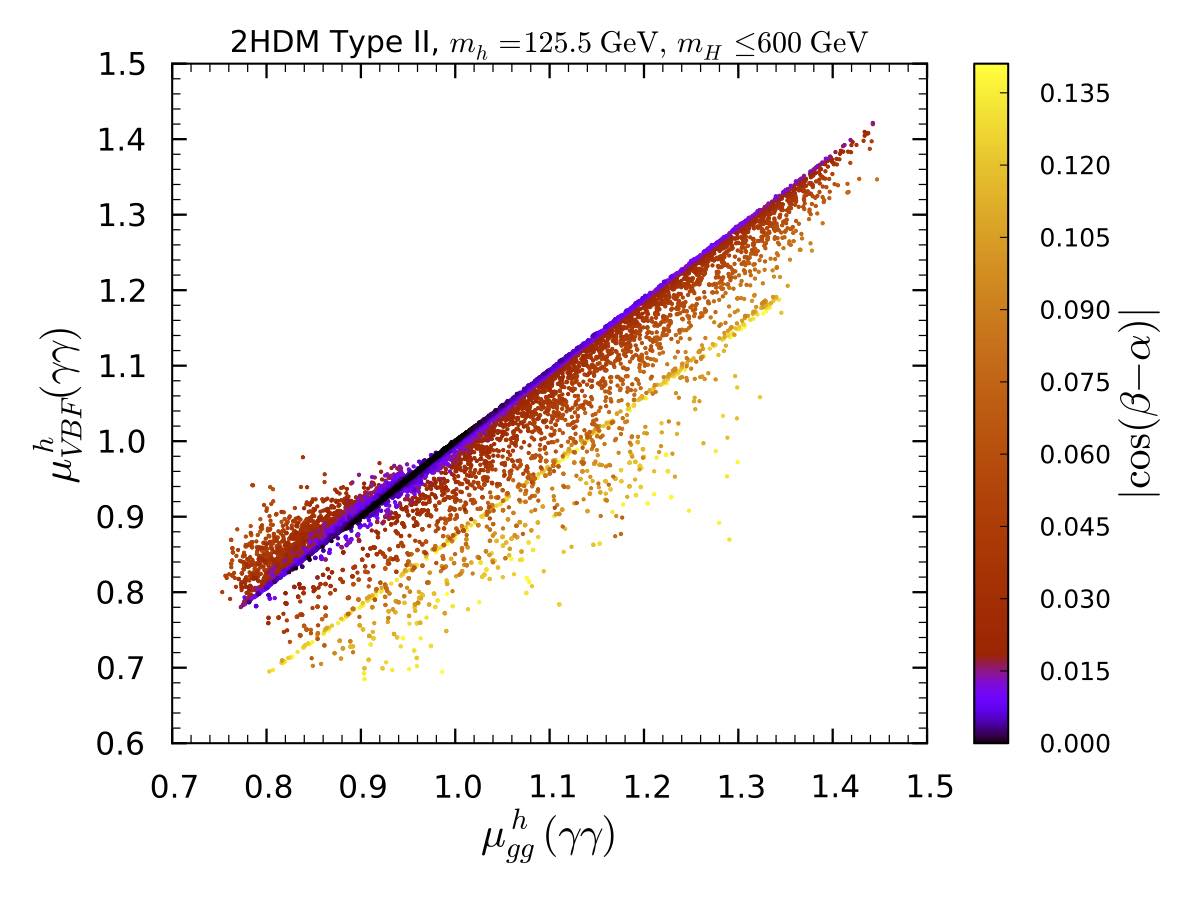}\\
\includegraphics[width=0.5\textwidth]{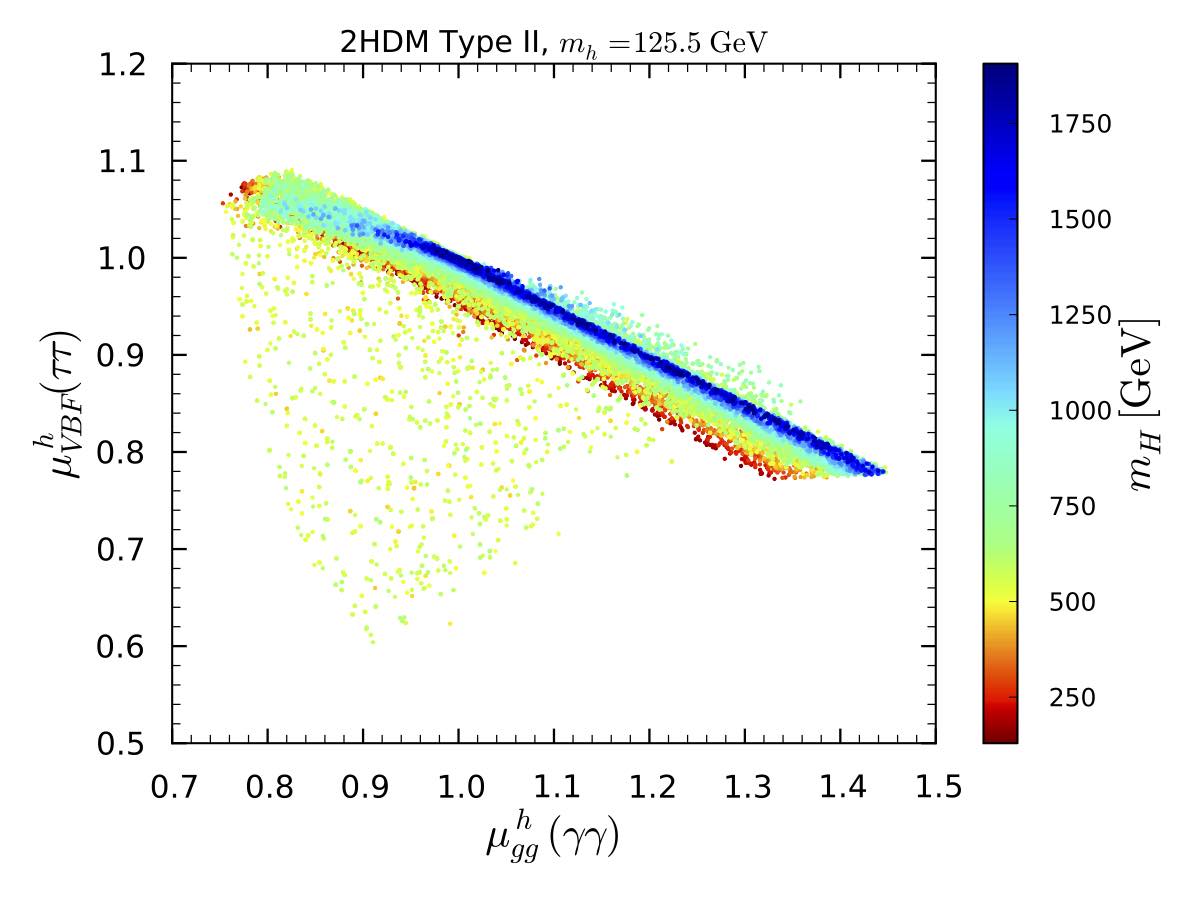}\includegraphics[width=0.5\textwidth]{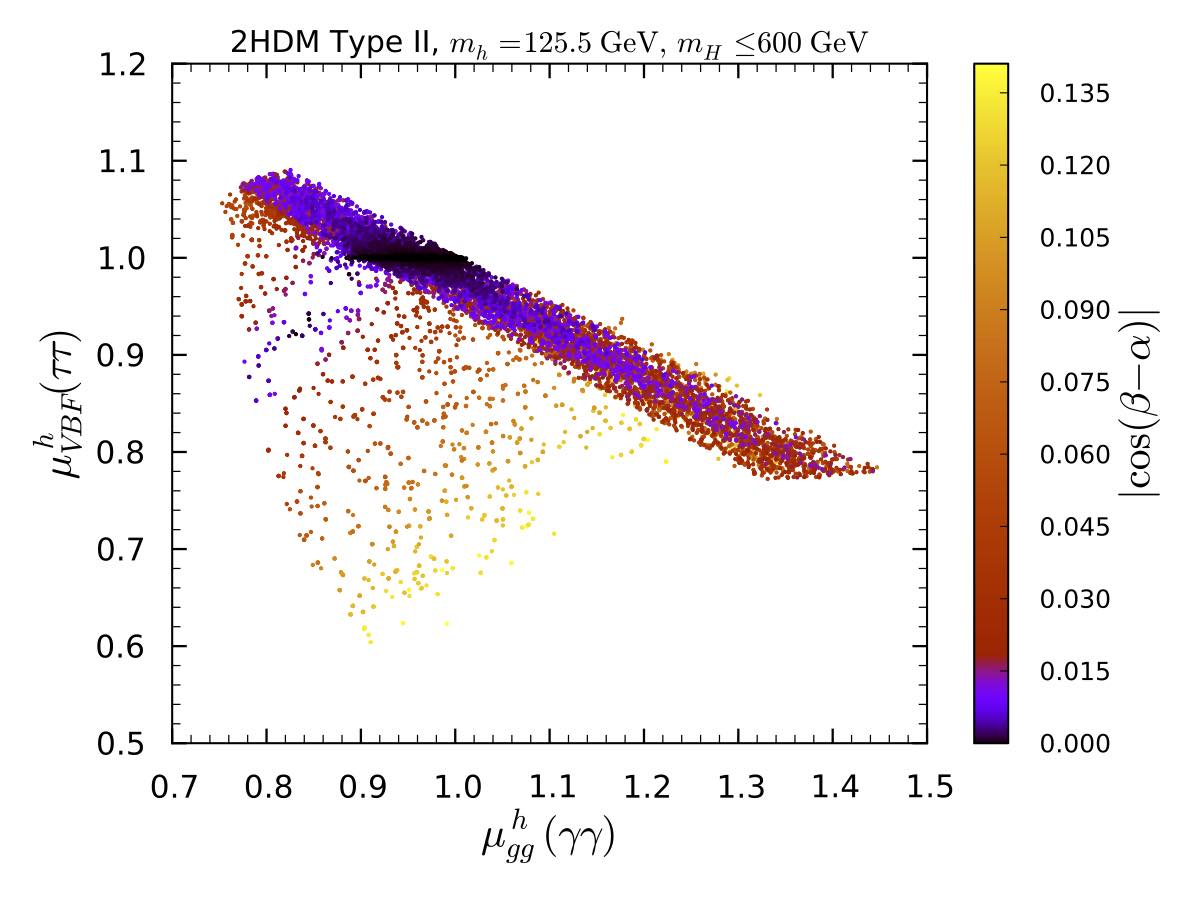}  
\caption{Correlations of signal strengths in Type~II, on the left illustrating the dependence on $m_H$, on the right 
illustrating the dependence in $|\cba|$. Points are ordered from low to high $m_H$ values (left) and high to low $|\cba|$ values (right).}
  \label{mu_mu_h125_type2}
  \vskip 0.4in
\end{figure}

Putting everything together we find quite distinct correlations of signal strengths in both Type~I and Type~II that depend on whether the additional Higgs states are decoupled or not. This is illustrated in Fig.~\ref{mu_mu_h125_type1} for Type~I and in Fig.~\ref{mu_mu_h125_type2} for Type~II. In both figures, the panels on the left show the dependence on $m_H$ while the panels on the right show the dependence on $|\cba|$ for the non-decoupling regime with $m_H\le 600\gev$. 
We note that there are definite combinations of signal strengths that cannot be reached in the decoupling regime. 
A measurement of such values would be a very strong motivation to look for additional light Higgs states. In turn, when the masses of additional light Higgs states are measured, signal strength correlations as shown in Figs.~\ref{mu_mu_h125_type1} and \ref{mu_mu_h125_type2} can help pin down the model. 
Furthermore, for $m_H\le 600\gev$ even in the apparent alignment limit $|\cba|\to 0$ there can be deviations in the signal strengths from unity that cannot be mimicked by decoupling.

Examples for Type~I are the suppression of both $\mu_{gg}^h(\gamma\gamma)$ and $\mu_{gg}^h(ZZ^*)$, or the combination $\mu_{gg}^h(\gamma\gamma)>1$ with $\mu_{gg}^h(ZZ^*)\approx 1$. The former case is also present in Type~II for light $m_H$, while the latter does not occur at all in Type~II. More concretely, in the decoupling regime of Type~II,  $\mu_{gg}^h(\gamma\gamma) \approx \mu_{gg}^h(ZZ^*)$, whereas for light $m_H$ one can have $\mu_{gg}^h(\gamma\gamma) < \mu_{gg}^h(ZZ^*)$ even if $|\cbma|$ is very small (comparing Fig~\ref{mu_mu_h125_type2}, top row, left vs. right). Another example is the simultaneous suppression or enhancement of $\mu_{gg}^h(\gamma\gamma)$  and $\mu_{\rm VBF}^h(\gamma\gamma)$ in Type~I, that is not possible in the decoupling regime (cf. Fig~\ref{mu_mu_h125_type1}, bottom left). In Type~II, one can have a simultaneous enhancement, up to 1.45 of $\mu_{gg}^h(\gamma\gamma)$ and $\mu_{\rm VBF}^h(\gamma\gamma)$ in the decoupling regime, but simultaneous suppression is limited to $\sim 0.9$--$0.95$ (cf. Fig~\ref{mu_mu_h125_type2}, middle left);  simultaneous suppression to a level of $\sim 0.8$ is however possible in the alignment limit for $m_H\lsim 300\gev$, i.e. well away from the decoupling regime. Precise enough signal strength measurements could therefore provide strong hints that we are in the alignment without decoupling regime of a 2HDM even if no additional Higgs states are discovered at that time.

\subsection{Cross sections for $H$ and $A$ production}\label{cross-sections}

\begin{figure}[t!]\centering
\includegraphics[width=0.51\textwidth]{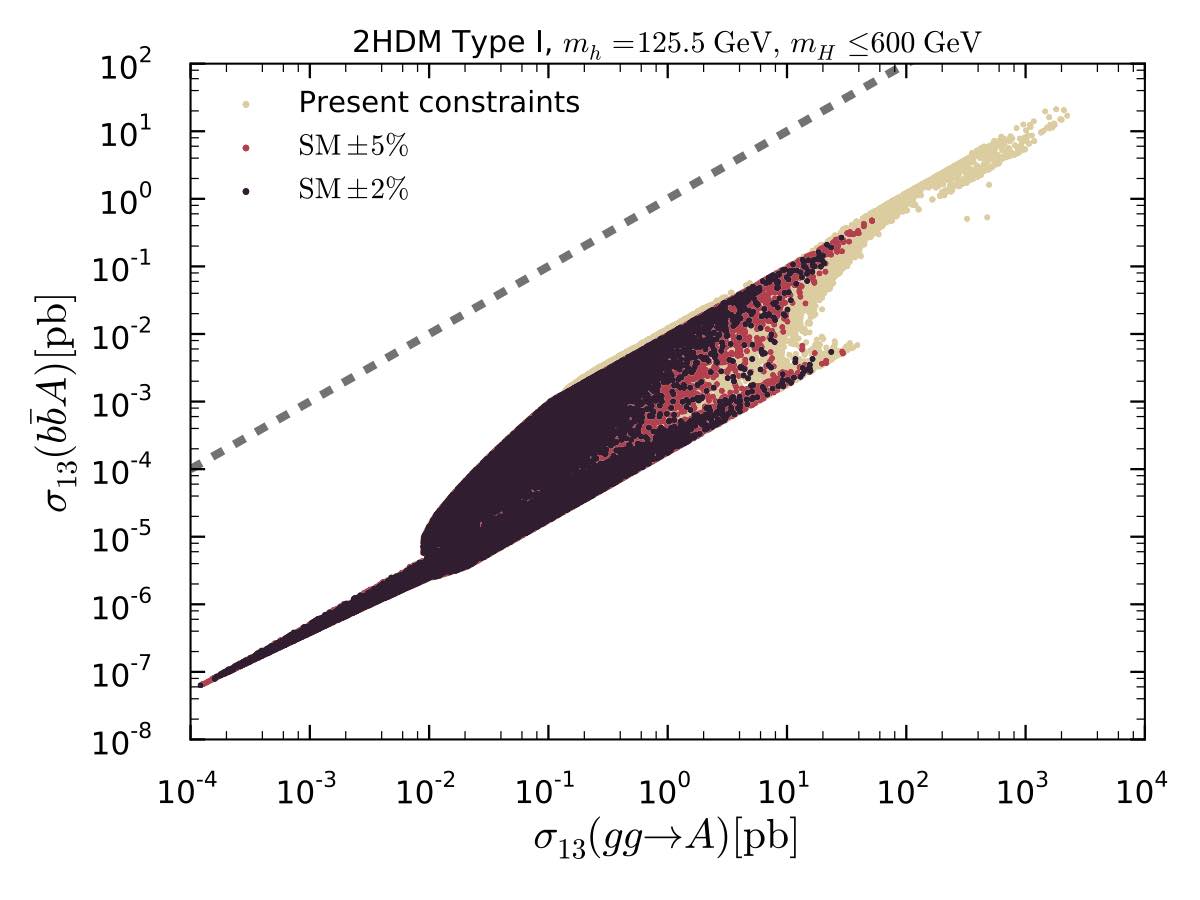}\includegraphics[width=0.51\textwidth]{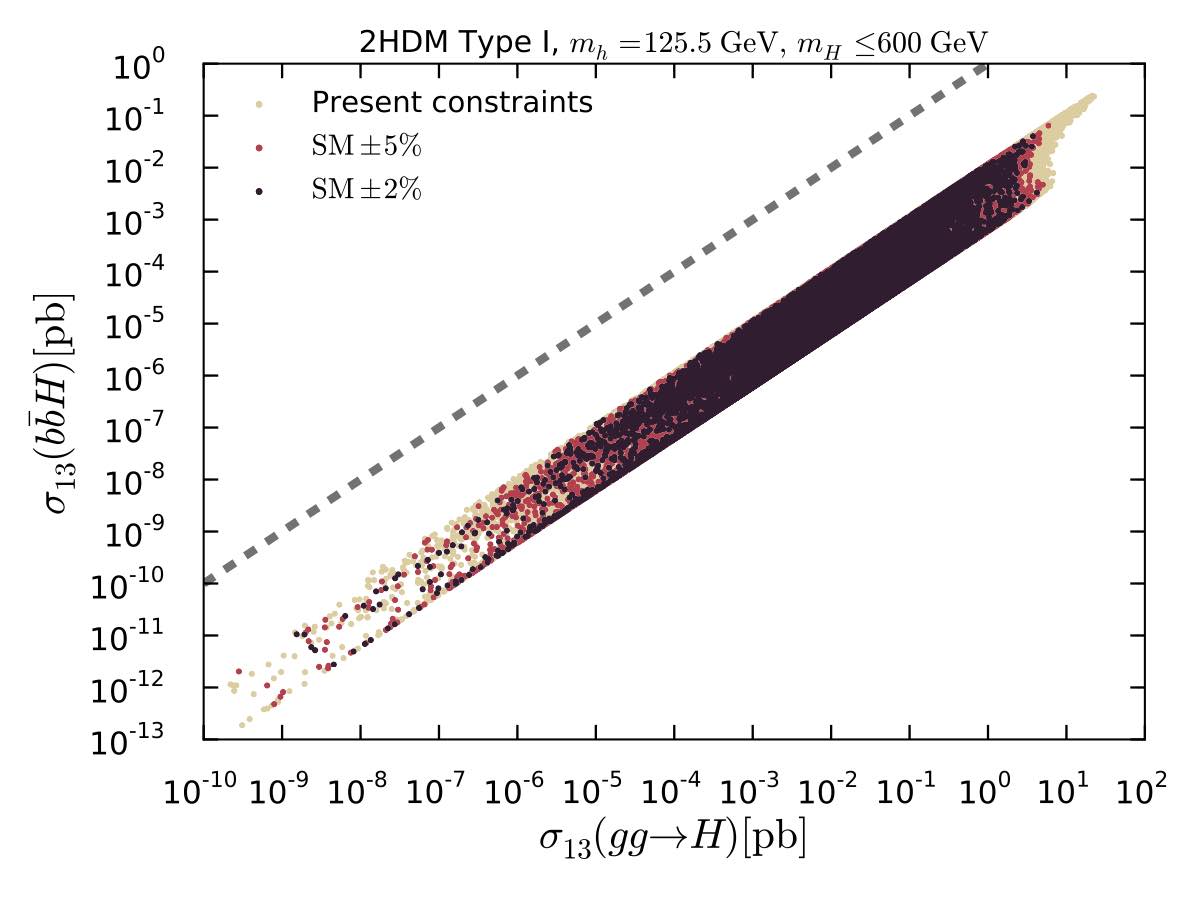}
  \caption{$\sigma(b\bar{b}X)$ versus $\sigma(gg\to X)$ for $X=A$ (left) and $X=H$ (right) in Type~I at the 13~TeV LHC for points satisfying all present constraints (in beige) as well as  points for which the signals strengths from Eq.~\eqref{mu5} are within $5\%$ and $2\%$ of the SM predictions (in red and dark red, respectively). The dashed lines indicate $\sigma(b\bar{b}X)=\sigma(gg\to X)$.}
  \label{xsec_correlation_I_13}
\end{figure}

\begin{figure}[t!]\centering
\includegraphics[width=0.51\textwidth]{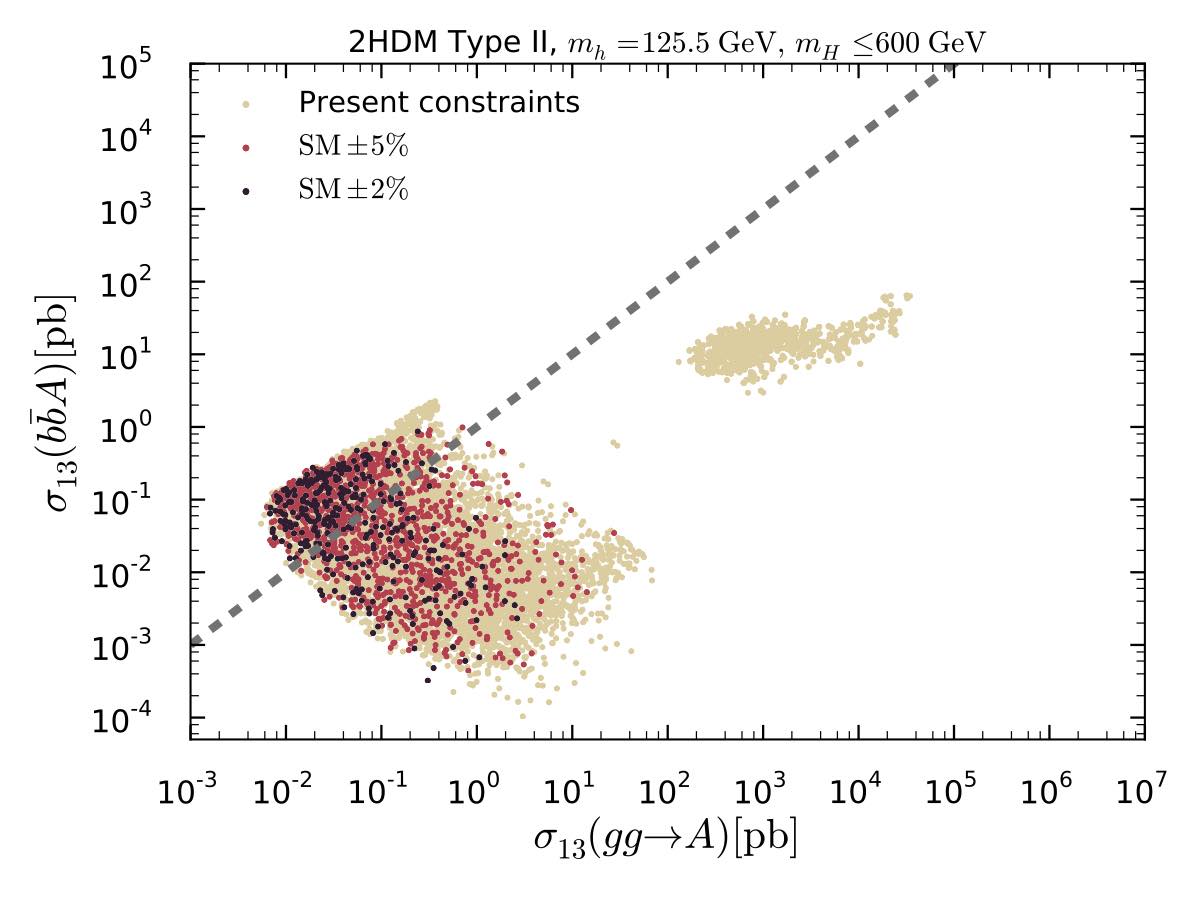}\includegraphics[width=0.51\textwidth]{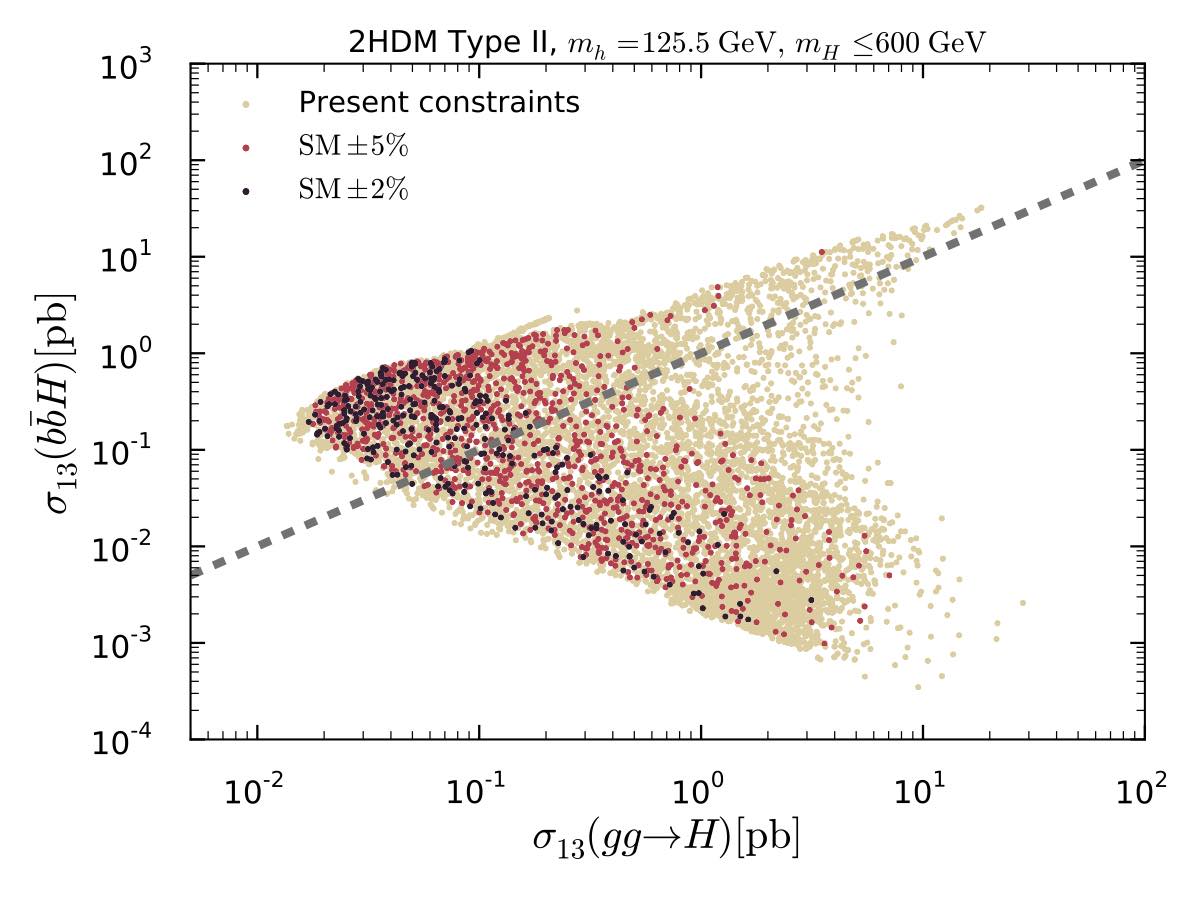}
\caption{As Fig.~\ref{xsec_correlation_I_13} but for Type~II.}
  \label{xsec_correlation_II_13}
\end{figure}

Let us now turn to the prospects of discovering the additional neutral states. 
The two largest production modes at the LHC are gluon fusion, $gg\to X$, and the associated production with a pair of $b$-quarks, $b\bar{b}X$, with $X=A,H$.
The correlations of the $gg\to X$ and $b\bar{b}X$ cross sections at the 13 TeV LHC  in the non-decoupling regime $m_H\leq 600$ GeV  are shown in Fig.~\ref{xsec_correlation_I_13} for the Type~I model and in Fig.~\ref{xsec_correlation_II_13} for the Type~II model. 
We show the points that pass all present constraints (in beige) and highlight those that have a very SM-like 125~GeV Higgs state by constraining all the following signal strengths to be within $5\%$ or $2\%$ of their SM values, respectively, denoted as SM$\pm 5\%$ (red) and SM$\pm 2\%$ (dark red):
\beq
\label{mu5}
   \mu_{gg}^h(\gamma\gamma),\ \mu_{gg}^h(ZZ^*),\ \mu_{gg}^h(\tau\tau),\ \mu_{VBF}^h(\gamma\gamma),\ \mu_{VBF}^h(ZZ^*),\ \mu_{VBF}^h(\tau\tau),\ \mu_{VH}^h(b\bar{b}),\ \mu_{t\bar{t}}^h(b\bar{b}) \,.
\eeq

We start the discussion with production of $A$ in Type~I, shown in the left panel of Fig.~\ref{xsec_correlation_I_13}. 
There is a strong correlation between the two production modes, gluon fusion and $b\bar{b}$ associated production, which stems from the fact that the relevant couplings are the same up to a sign: $C_U^A=-C_D^A=\cot\beta$. The larger spread in $\sigma(b\bar{b}A)$ observed for $\sigma(gg\to A)>10^{-2}$~pb comes from the fact that for $m_A\lesssim 2m_t\gev$ the $b\bar{b}A$ cross section grows faster with decreasing $m_A$ than that of  $gg\to A$. Therefore, along a line of fixed $\sigma(gg\to A)$ in the plot, a point with higher $\sigma(b\bar{b}A)$ has a smaller $m_A$. Note also that there is an interference of the top and bottom loop diagrams in $gg\to A$ which changes sign depending on $m_A$. 
Overall, however, $\sigma(gg\to A)$ is always at least two orders of magnitude larger than $\sigma(b\bar{b}A)$.  

The points with largest cross sections, $\sigma(b\bar{b}A)\approx10$ pb and $\sigma(gg\to A)\approx1000$ pb, correspond to the case $m_A<m_h/2$ which was studied in detail in~\cite{Bernon:2014nxa}. One feature of this region is that $\mu_{gg}^h(\gamma\gamma)$ and $\mu_{gg}^h(ZZ^*,WW^*)$ always differ from each other by about $10\%$. Constraining all $h$ signal strengths of Eq.~\eqref{mu5} within $5\%$ of unity therefore eliminates these points. Other points with high cross sections, but not in the very light pseudoscalar region, would also be eliminated by the SM$\pm 5\%$ or SM$\pm 2\%$ requirements. However, in this non-decoupling regime of $m_H\leq 600\gev$, points with sizeable cross sections up to 0.2~pb for $\sigma(b\bar{b}A)$ and up to about 40~pb for $\sigma(gg\to A)$ still remain even at the SM$\pm2\%$ level. 
At this same SM$\pm2\%$ level, the smallest $\sigma(gg\to A)$ is about 0.1~fb. 

Regarding production of the scalar $H$ in Type~I, shown in the right panel of Fig.~\ref{xsec_correlation_I_13}, the correlation is even stronger between $\sigma(b\bar{b}H)$ and $\sigma(gg\to H)$ since both are driven by the same fermionic coupling $C_F^H=\sin\alpha/\sin\beta$. Note that, as in the $A$ case, the gluon-fusion cross section is always larger than that for $b\bar{b}$ associated production. Sizable cross sections are still allowed under the SM$\pm 2\%$ constraint, which implies that in the non-decoupling regime there is a strong possibility of detecting a non-SM-like scalar state at the LHC. The structure of $C_F^H$ is however such that the coupling can equally well be very much suppressed, leading to extremely small cross sections. We will come back to this below. 

The corresponding results for Type~II are presented in Fig.~\ref{xsec_correlation_II_13}. 
In contrast to Type~I,  both $b\bar{b}$ associated production and gluon--gluon fusion modes for Type~II are in principle important since either can be dominant in different regions of the parameter space. There is only modest correlation between the two production modes due to the more complex structure of the Type~II fermionic couplings.
For A production, one clearly sees the $m_A < m_h/2$ region as the detached scattered points with very large cross sections. As for Type~I, these points disappear under the SM$\pm 5\%$ constraint. Still, even for SM$\pm 2\%$, cross sections for $b\bar{b}A$ close to 1~pb and for $gg\to A$ of a few pb can be achieved (although not simultaneously). 
For $H$ production a similar picture emerges, with the maximal cross sections however being a factor of a few smaller than for $A$ production.
The minimal cross sections in this $m_H<600\gev$ non-decoupling regime for the $A$ and $H$ are correlated in a way that is very favorable for discovery during Run~2 of the LHC.  For example,  if $\sigma(gg\to A)$ takes on its minimum SM$\pm2\%$ value of a few fb then $\sigma(b\bar b A)\gsim 20$ fb, whereas if $\sigma(b\bar bA)$ takes on its minimal value of few$\times 10^{-1}$ fb then $\sigma(gg\to A)\approx 300$ fb. 
These cross section levels imply that the $A$ should  be discoverable in at least one of the two production modes even in the extreme alignment limit. Analogous arguments hold for $H$ production.

\begin{figure}[t!]\centering
\includegraphics[width=0.51\textwidth]{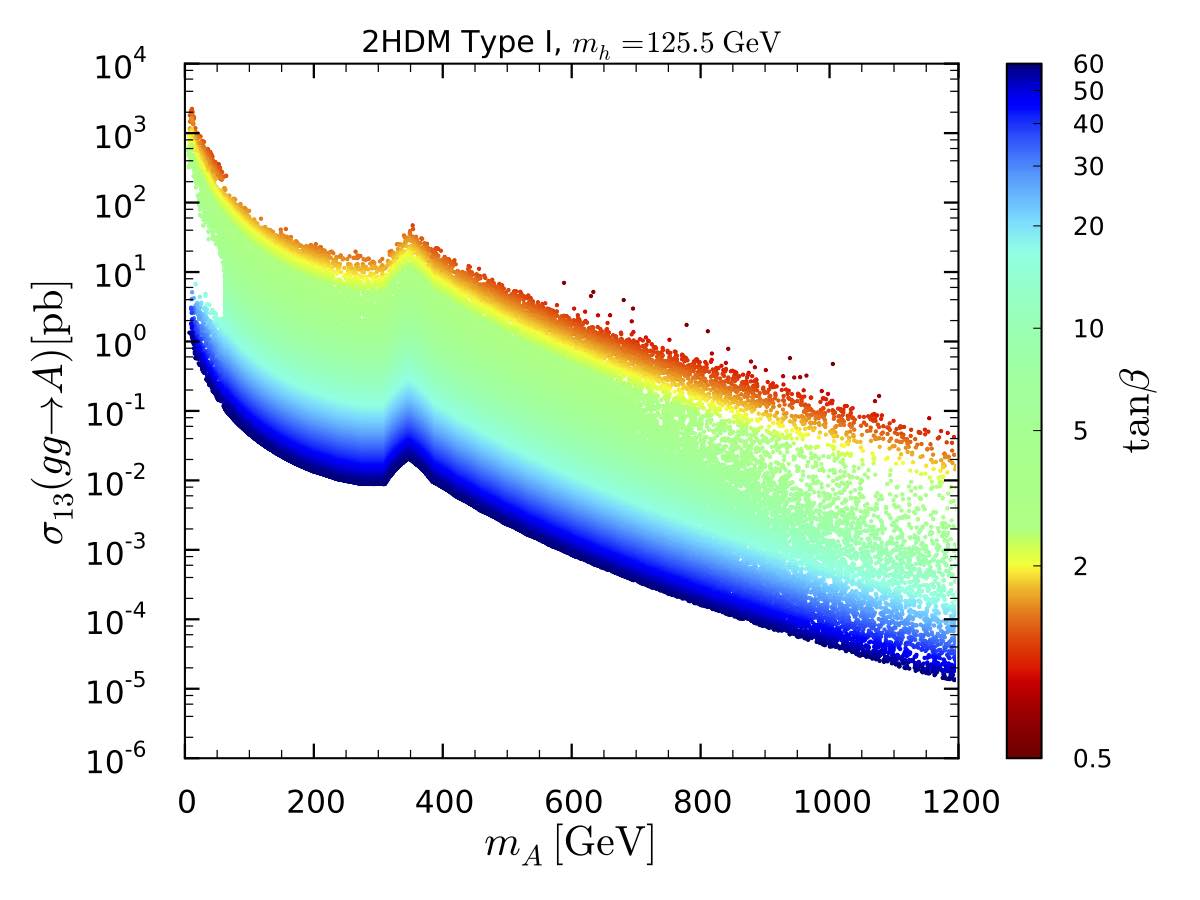}\includegraphics[width=0.51\textwidth]{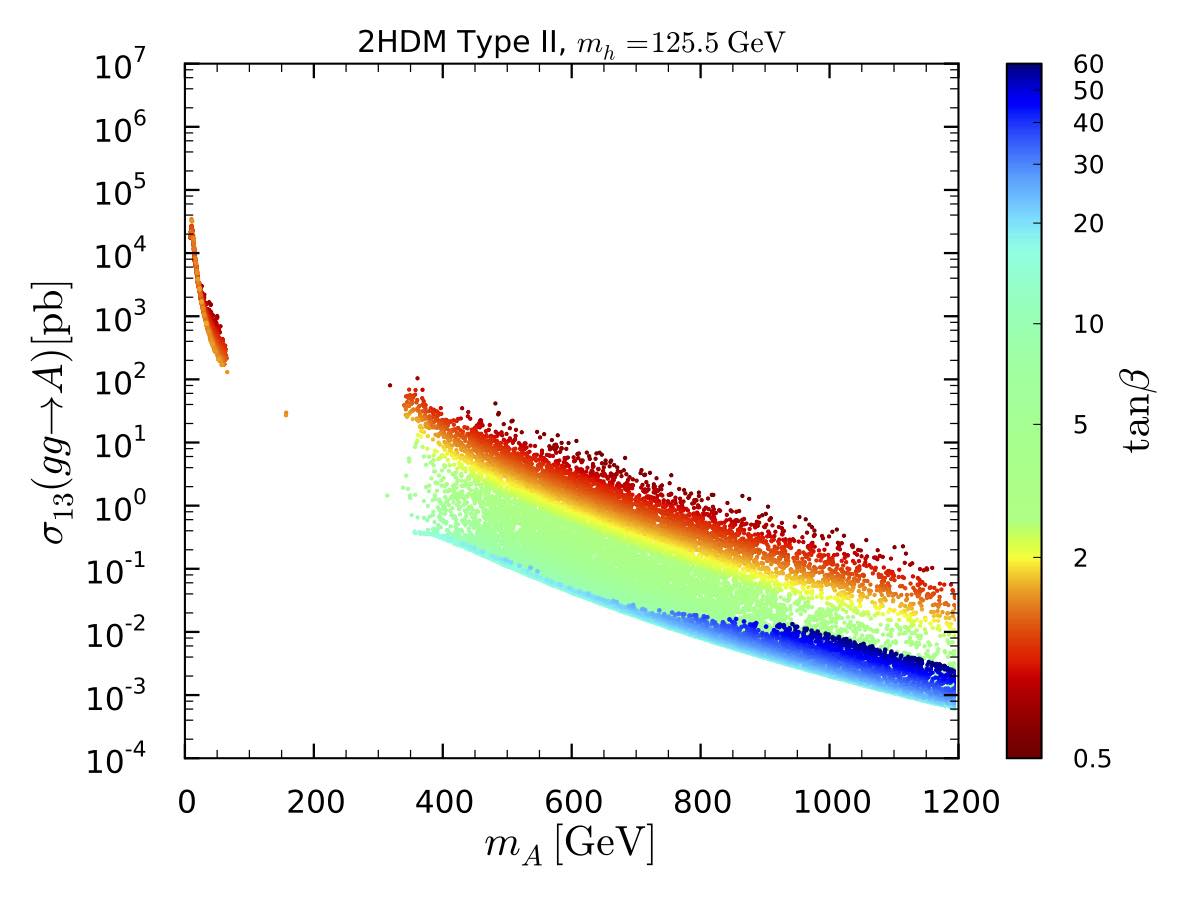}
\includegraphics[width=0.51\textwidth]{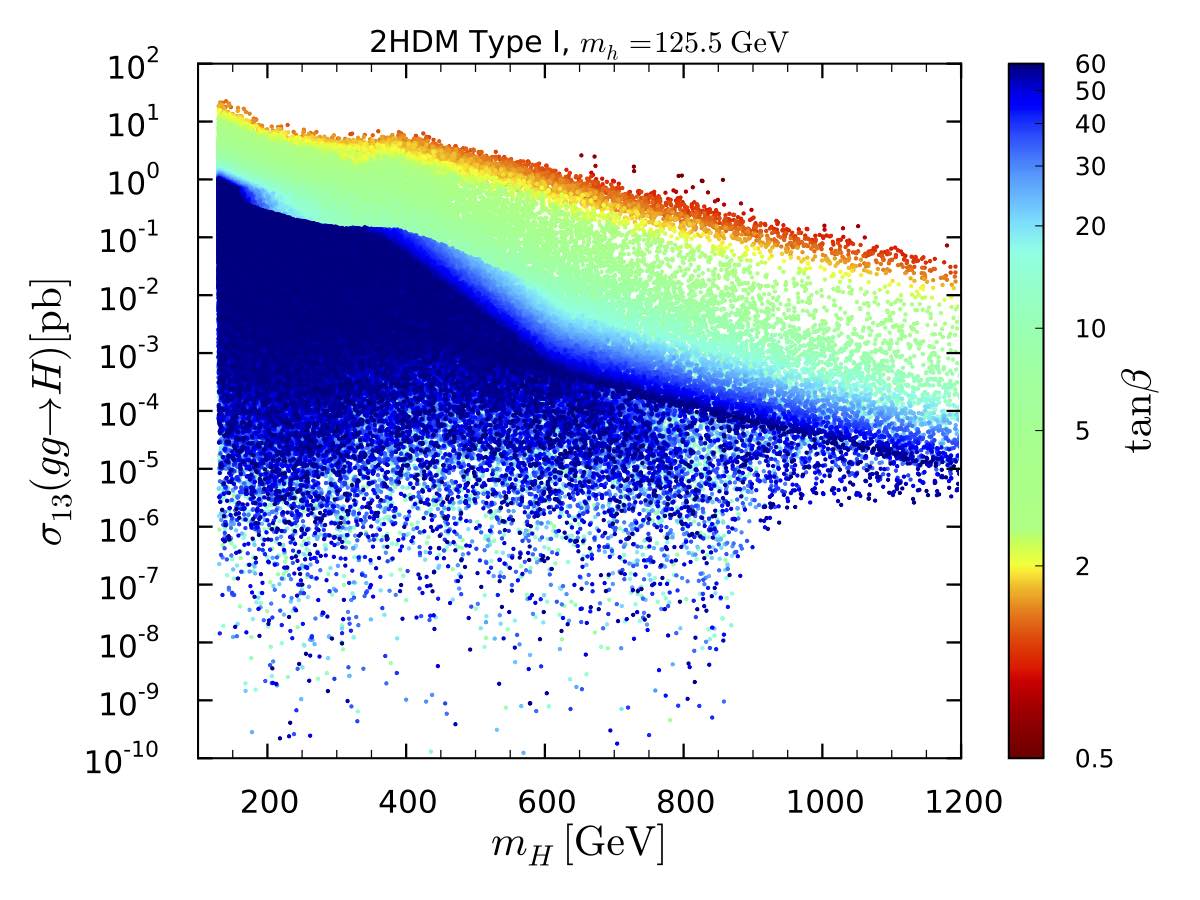}\includegraphics[width=0.51\textwidth]{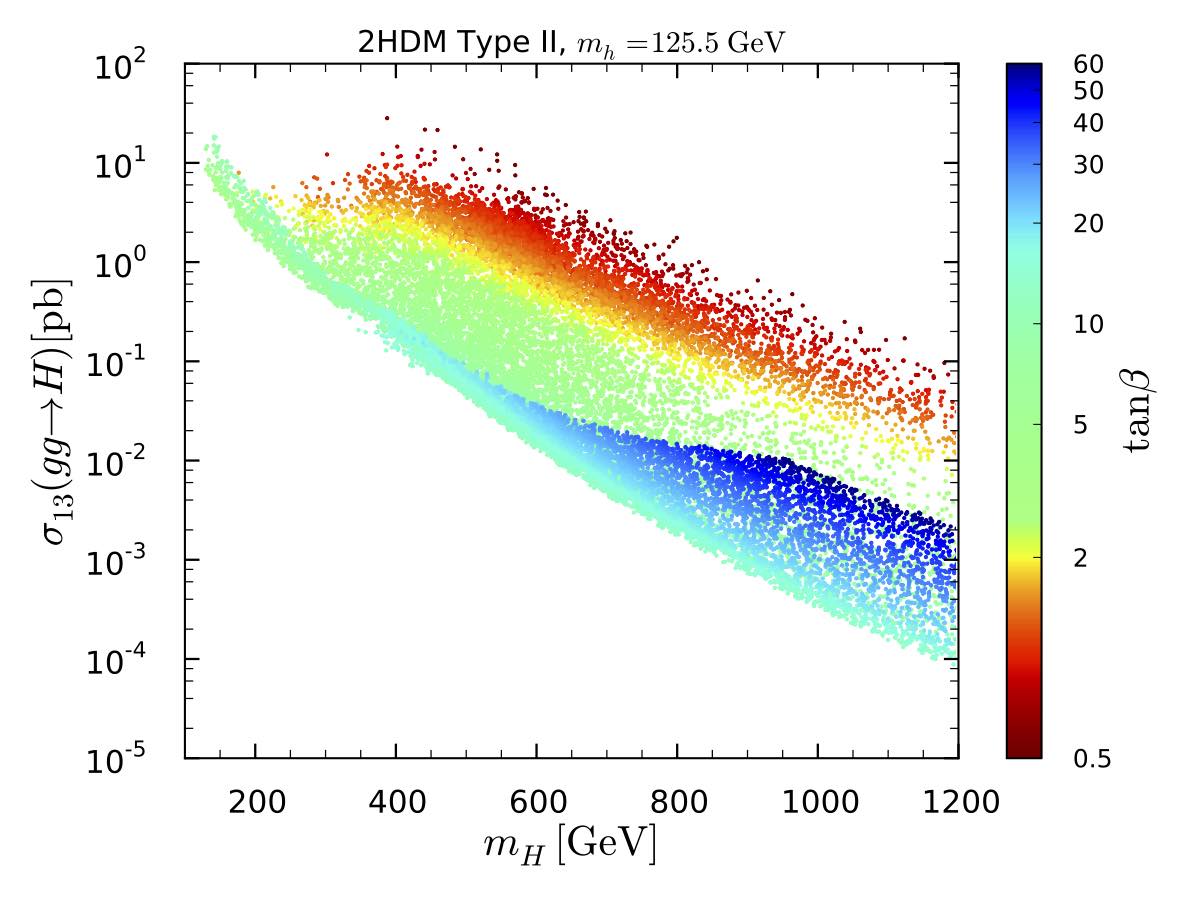}
  \caption{Cross sections in Type~I (left) and Type~II (right) for $gg\to X$ at the 13~TeV LHC as functions of $m_X$ for $X=A$ (upper row) and $X=H$ (lower row) with $\tan\beta$ color code. In all four plots, points are ordered from low to high $\tan\beta$.}
  \label{xsec13}
\end{figure}

Before considering specific decay channels of $A$ and $H$, we present in Fig.~\ref{xsec13} the gluon-fusion cross sections in Type~I and Type~II as functions of $m_A$ and $m_H$ at the 13 TeV LHC. Here, the color code shows the dependence on $\tan\beta$.\footnote{To avoid a proliferation of plots, we choose to show here only the results for gluon fusion; all corresponding results for the $b\bar{b}$ cross section can be provided upon request.}
In Type~I, the $gg\to A$ cross section is proportional to $\cot^2\beta$; this explains why it is larger (smaller) at lower (higher) $\tan\beta$. A cross section of 1 (0.1)~fb is guaranteed for $m_A$ as large as $\sim 600$ (850)~GeV.   On the other hand, the
$gg\to H$ cross section in Type~I is proportional to $(C_F^H)^2$ and can take on extremely small values for $m_H\lesssim 850$ GeV. The reason is that, in this region, the reachable values of $\cba$ are high enough such that a cancellation between the two terms of $C_F^H=(\sba-\cba/\tb)$ occurs and leads to an almost vanishing coupling. In contrast, for $m_H\gtrsim 850$ GeV, this cancellation is not possible as the values of $\cba$ are forced to be smaller as can be seen in Fig.~\ref{mH_cba_Z6_h125}. 
In Type~II, the $A$ production cross section can be very large in the very low $m_A$ region as noted in~\cite{Bernon:2014nxa} and any mass smaller than $1.1$~TeV gives a $gg\to A$ cross section larger than 1~fb. 
For $gg\to H$, a cross section $>1$~(0.1)~fb is guaranteed up to $m_H\approx 850$~GeV (1.2~TeV). 
From these considerations the prospects for discovering the additional neutral states look promising should alignment without decoupling be realized. 

\begin{figure}[h!]\centering
\includegraphics[width=0.51\textwidth]{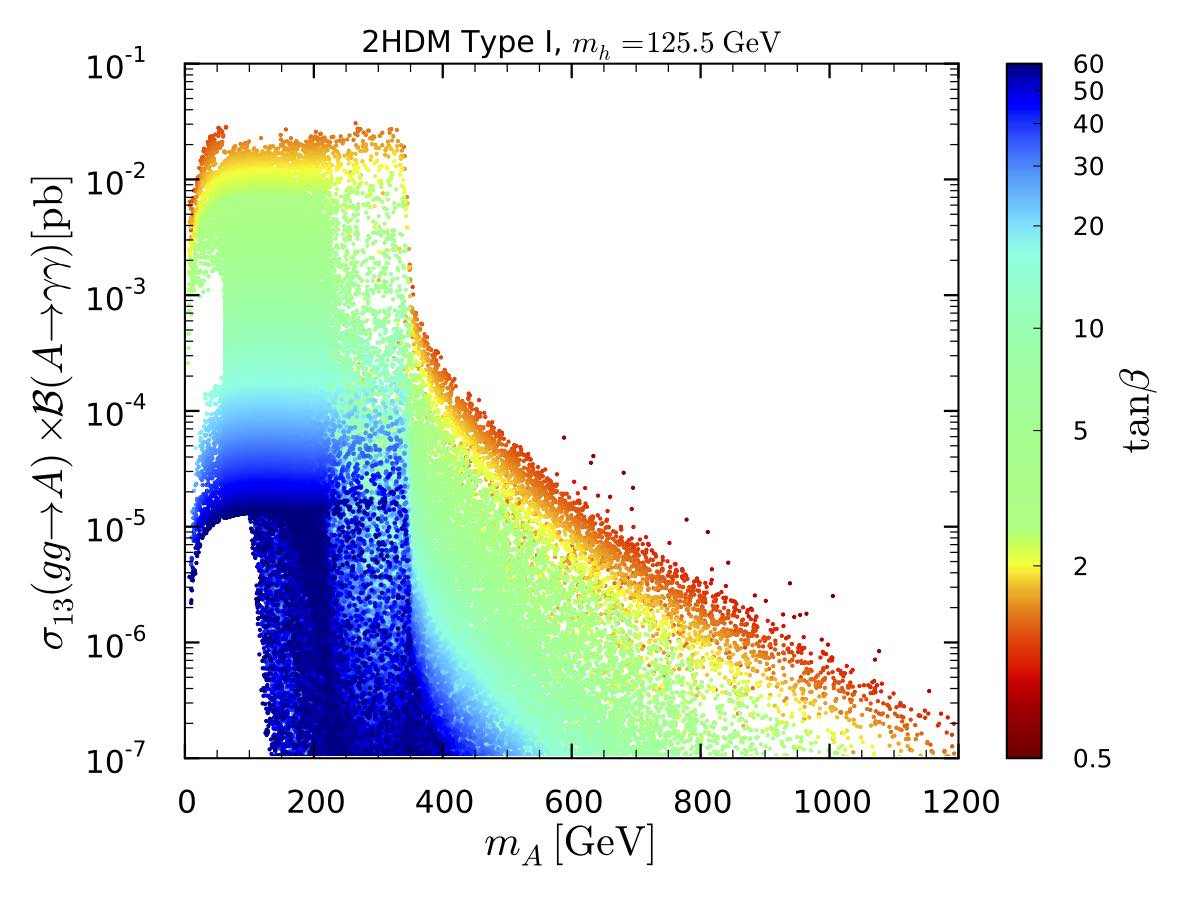}\includegraphics[width=0.51\textwidth]{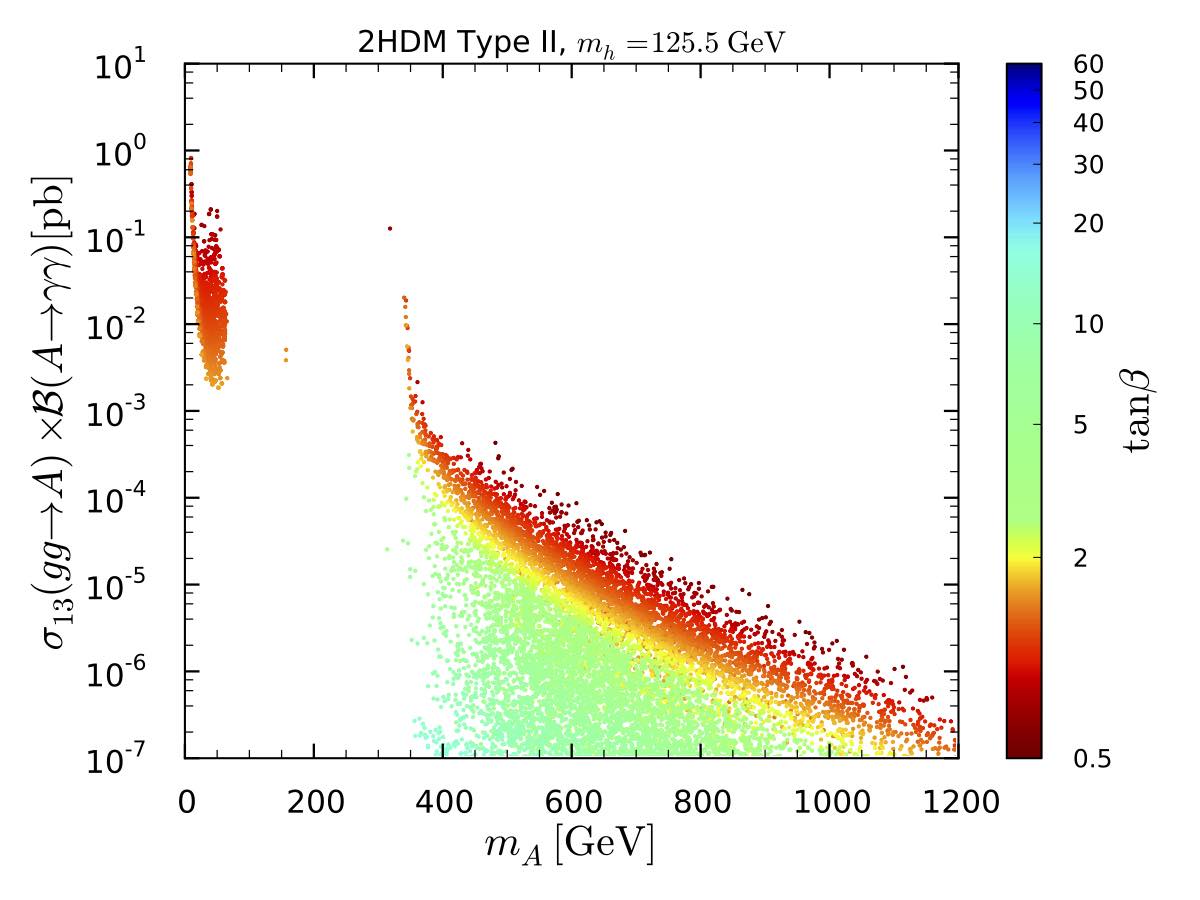}
\includegraphics[width=0.51\textwidth]{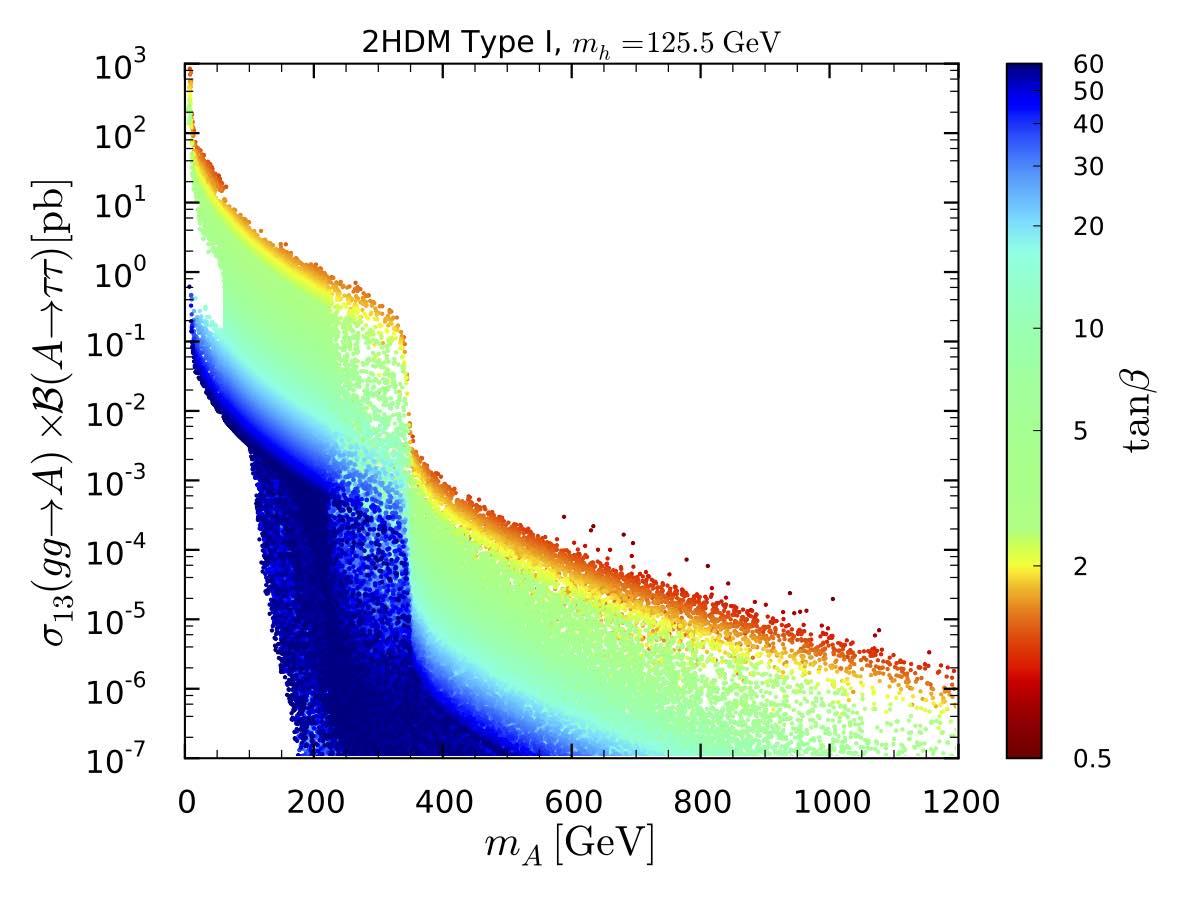}\includegraphics[width=0.51\textwidth]{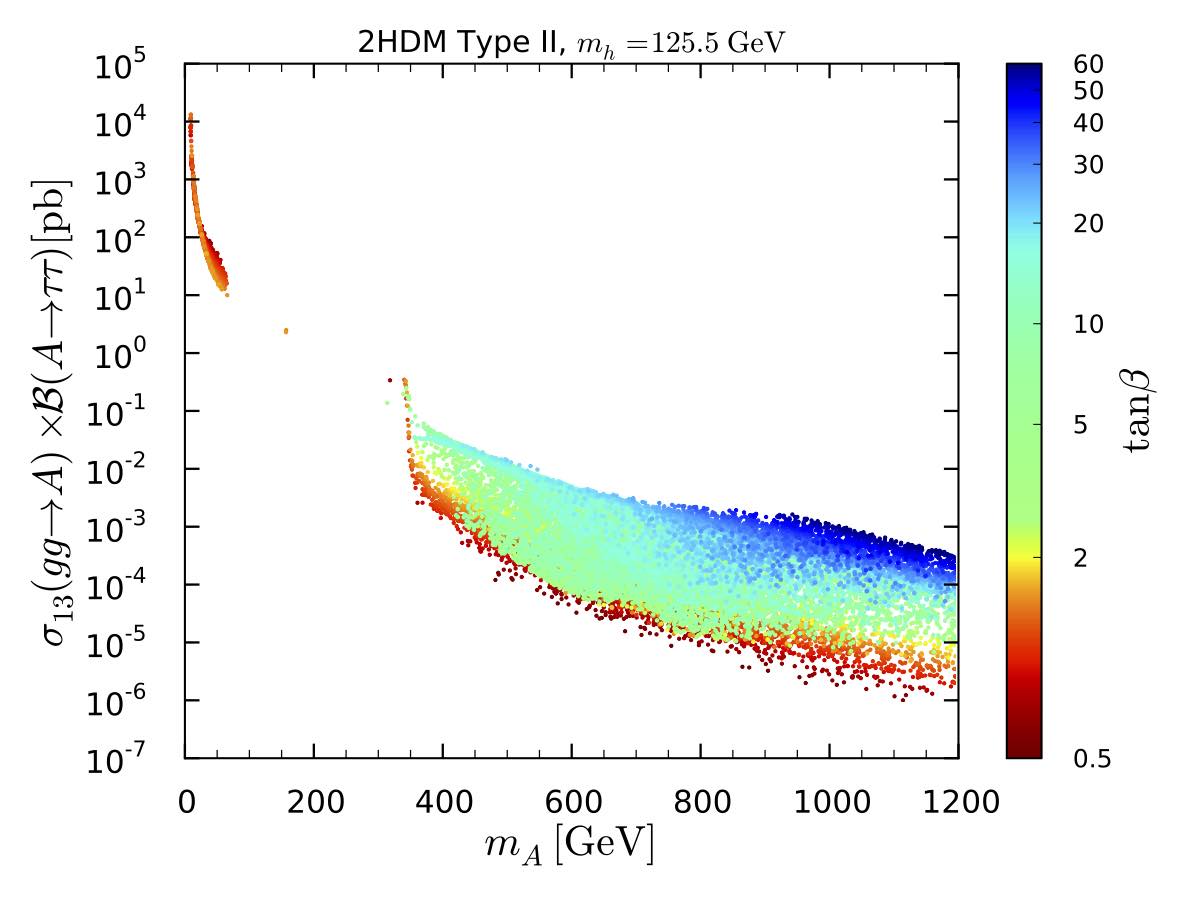}
\includegraphics[width=0.51\textwidth]{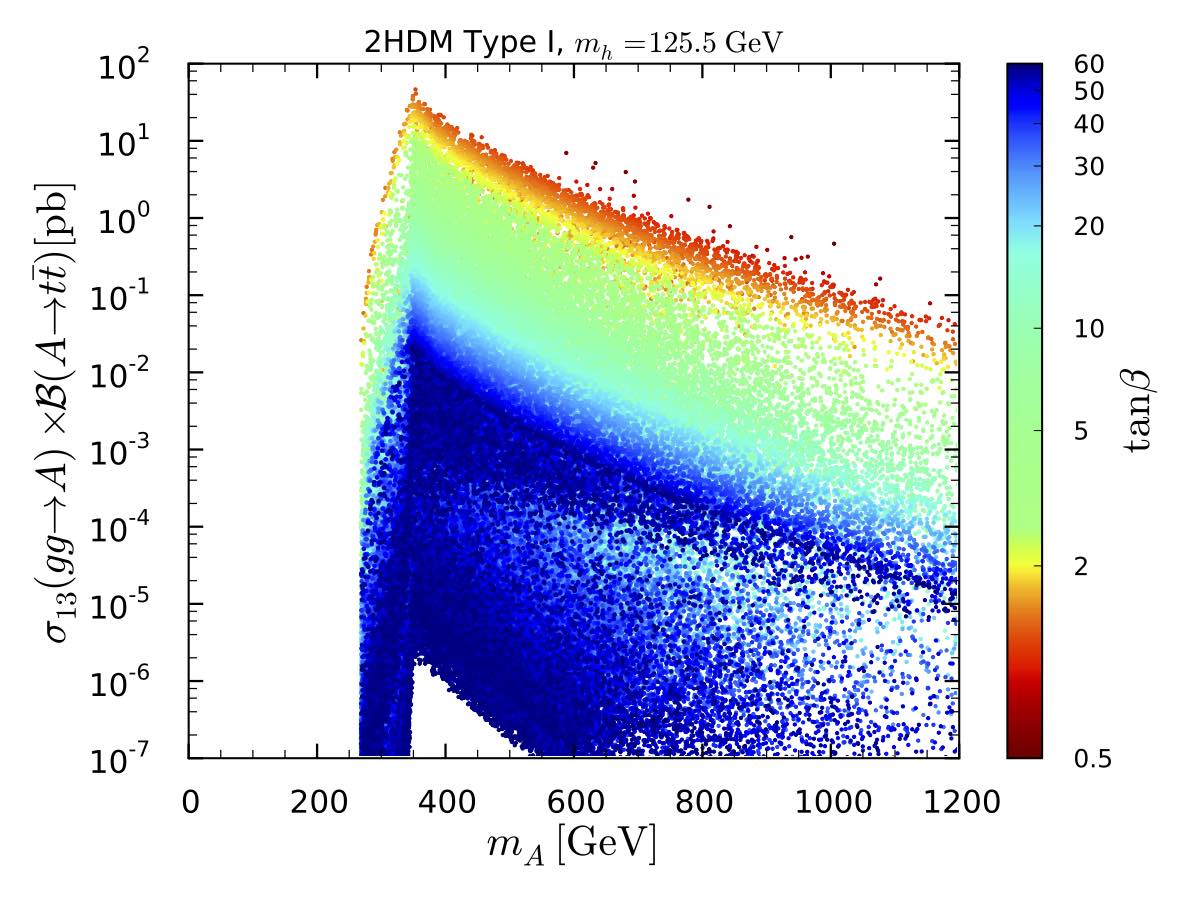}\includegraphics[width=0.51\textwidth]{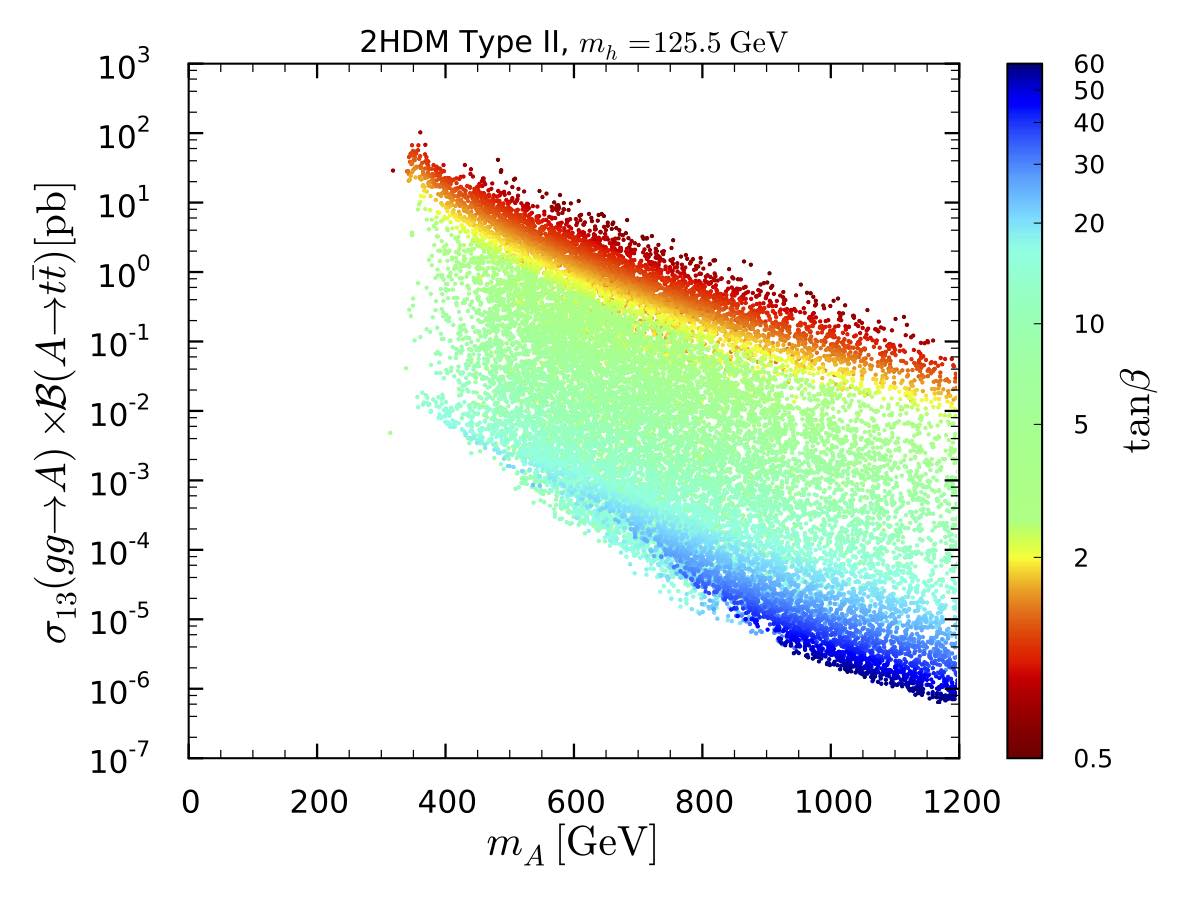}
  \caption{Cross sections times branching ratio in Type~I (left) and in Type~II (right) for $gg\to A\to Y$ at the 13~TeV LHC as functions of $m_A$ for $Y=\gamma\gamma$ (upper panels), $Y=\tau\tau$ (middle panels) and $Y=t\bar{t}$ (lower panels) with $\tan\beta$ color code. Points are ordered from low to high $\tan\beta$.}
  \label{xsecBRA13}
\end{figure}

Let us now turn to specific signatures.  Figure~\ref{xsecBRA13} presents the cross sections for $gg\to A\to Y$ for $Y=\gamma\gamma, \tau\tau, t\bar{t}$ in Types~I and II. Note that the $y$-axis is cut off at $10^{-7}$~pb.  Although much lower values of the cross section are possible, we do not show these lower values since they will certainly not be observable at the LHC.
As expected, for the $\gamma\gamma$ and $\tau\tau$ final states, the cross sections fall sharply above the $t\bar{t}$ threshold, with the noticeable exception of the $A\to \tau\tau$ decay in Type~II due to the strong constraints from LHC direct searches that exclude points with large corresponding cross section. For the $A\to\gamma\gamma$ decay, cross sections of 0.1 fb are reachable for $m_A\lesssim 470$ GeV ($m_A \lesssim 530$~GeV) in Type~I (II) but not guaranteed. The maximal cross section is $\sim 30$ fb in Type~I and  $\sim 100$ fb in Type~II (not considering the $m_A\leq m_h/2$ region). In both Types I and II, the $gg\to A\to \tau\tau$ cross section can be substantially larger. In Type~I, 0.1 fb is reachable for $m_A\lesssim 600$ GeV,  while in Type~II $m_A\lesssim 550$ GeV {\it guarantees} a cross section larger than 0.1 fb. In both cases, very large cross sections are predicted at low $m_A$. The $gg\to A\to t\bar{t}$ cross section peaks around $100$ pb in both Types~I and II and is guaranteed to be larger than 0.1 fb in Type~II for $350\lesssim m_A\lesssim 600$ GeV. These sizable cross sections therefore provide interesting probes of the extended Higgs sector in the alignment limit.

\begin{figure}[h!]\centering
\includegraphics[width=0.51\textwidth]{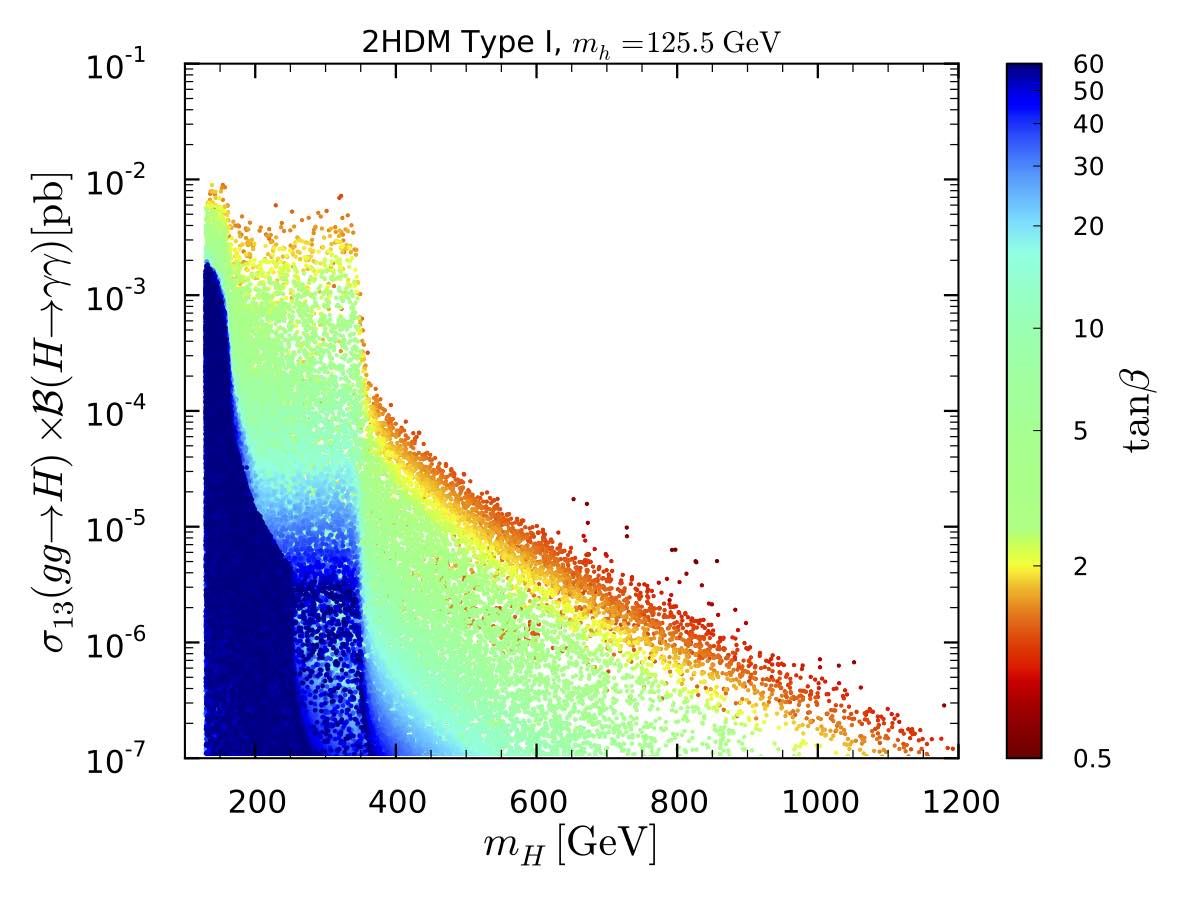}\includegraphics[width=0.51\textwidth]{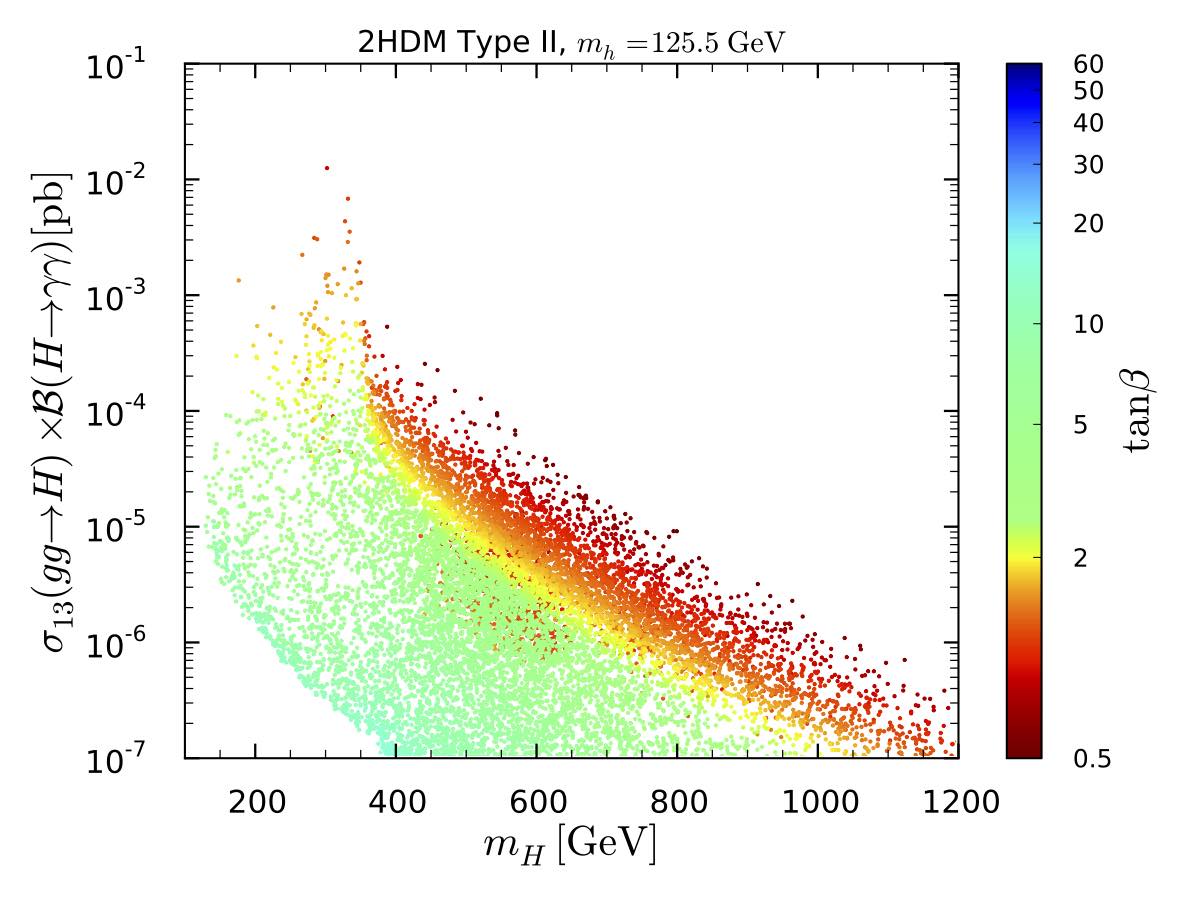}
\includegraphics[width=0.51\textwidth]{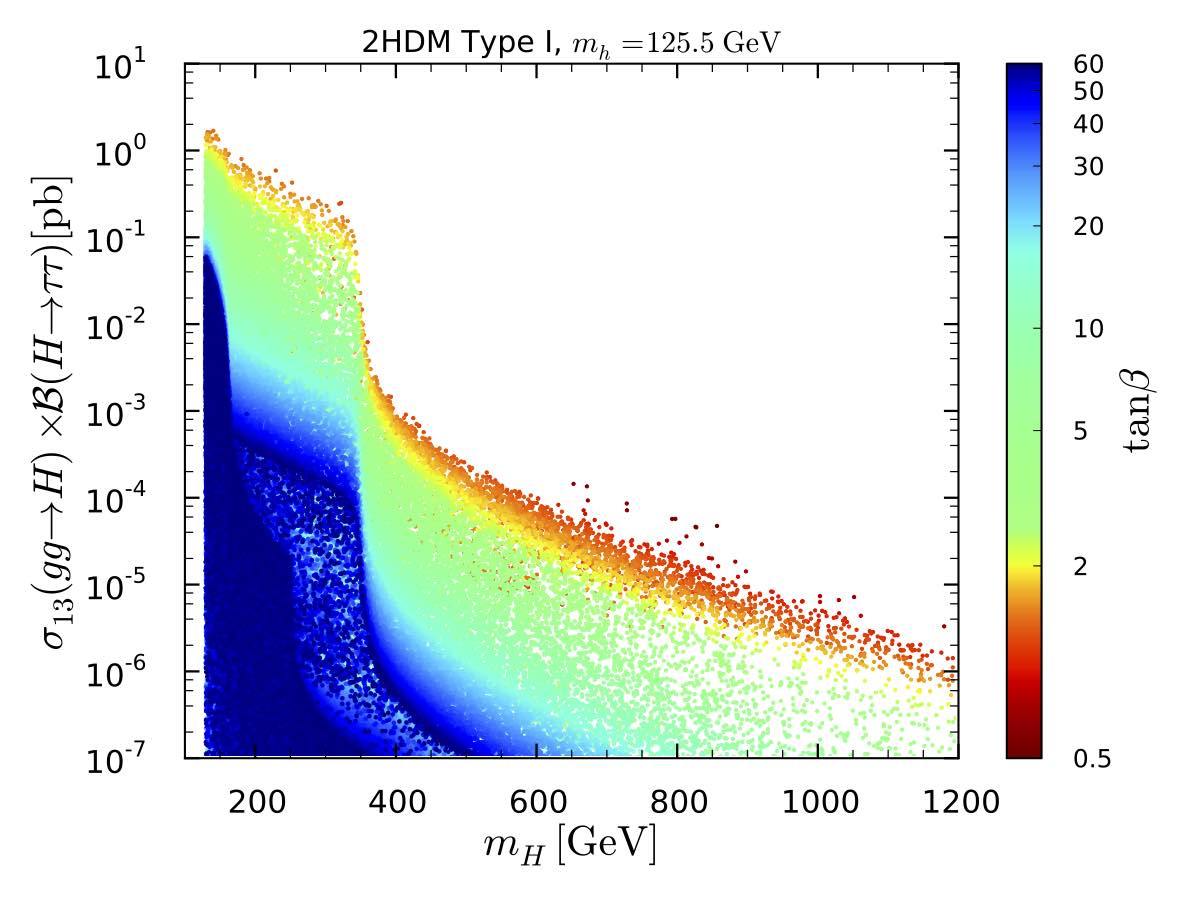}\includegraphics[width=0.51\textwidth]{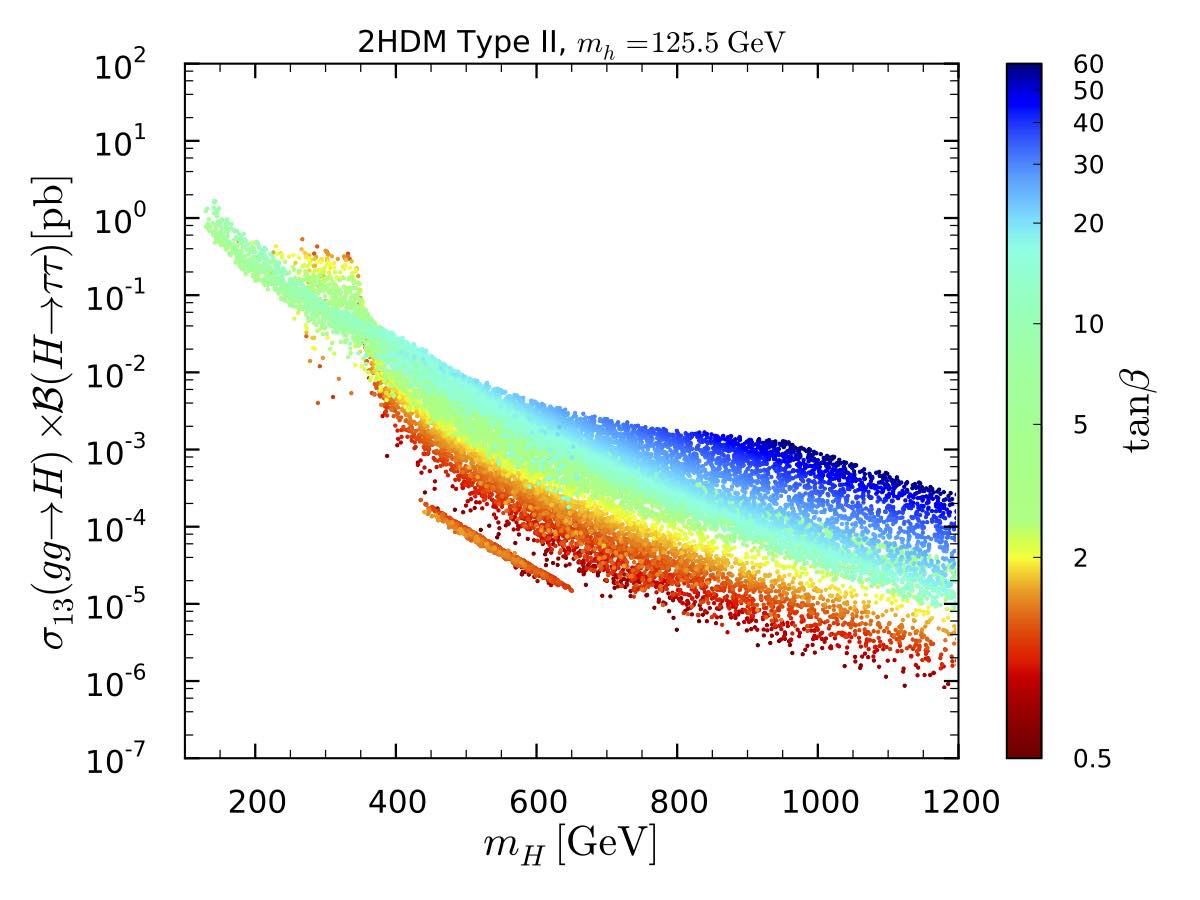}
\includegraphics[width=0.51\textwidth]{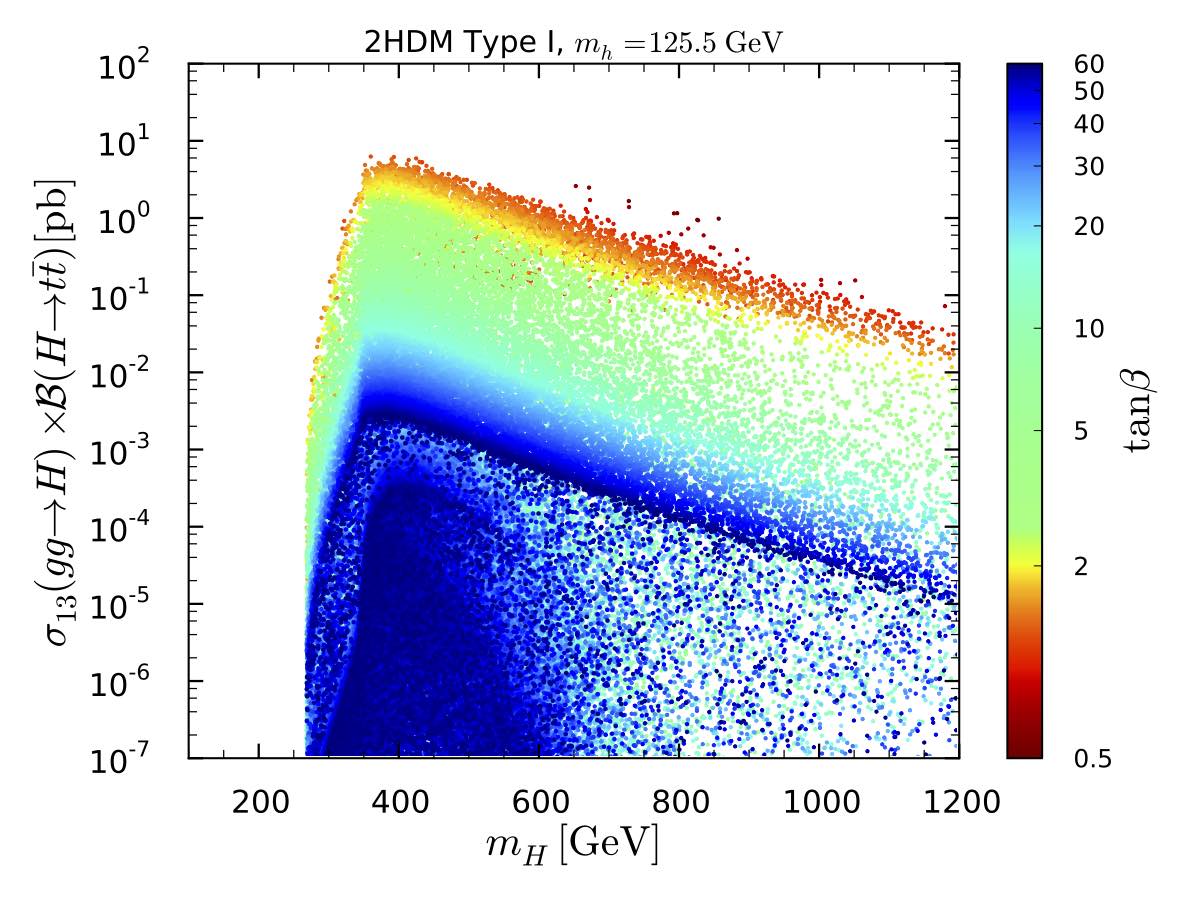}\includegraphics[width=0.51\textwidth]{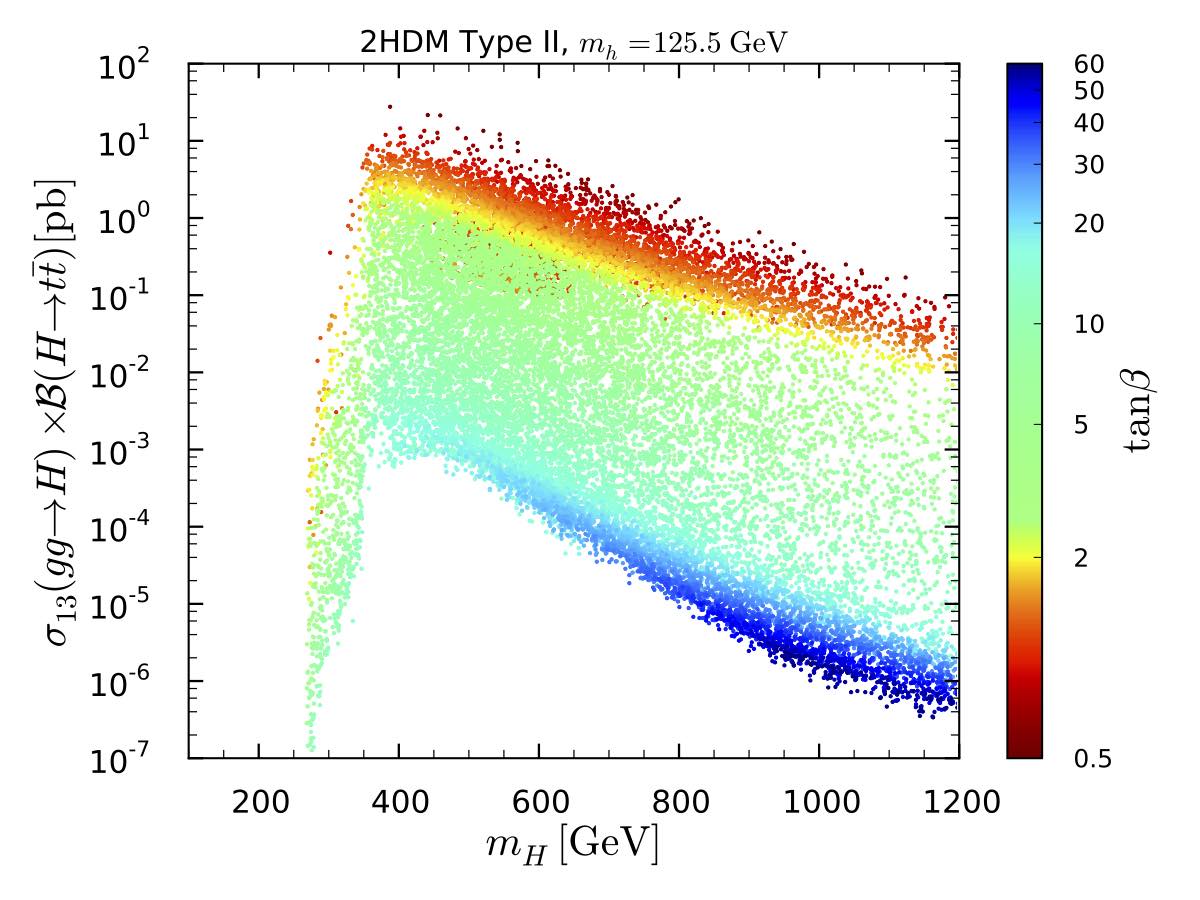}
  \caption{Cross section times branching ratio in Type~I (left) and in Type~II (right) for $gg\to H\to Y$ at the 13~TeV LHC as functions of $m_H$ for $Y=\gamma\gamma$ (upper panels), $Y=\tau\tau$ (middle panels) and $Y=t\bar{t}$ (lower panels) with $\tan\beta$ color code. Points are ordered from low to high $\tan\beta$.}
  \label{xsecBRH13}
\end{figure}

The corresponding results for the $H$ cross sections are presented in Fig.~\ref{xsecBRH13}.  Sizable values of $\sigma\times$BR are possible in both Types I and II  but heavily suppressed values are still possible for most of the cases. Only in Type~II, for $H\to\tau\tau$ (as well as for  $H\to t\bar{t}$), is the corresponding cross section guaranteed to be larger than 0.1 fb for $m_H\lesssim 460$ GeV ($m_A\approx 400$~GeV). Note that, for both Types~I and II, the cross sections for $A/H$ decays into a muon pair are related to the $A/H\to\tau\tau$ ones through $\mathcal{B}(A/H\to\mu\mu)\approx(m_\mu/m_\tau)^2\times\mathcal{B}(A/H\to\tau\tau)\approx\mathcal{B}(A/H\to\tau\tau)/280$.

\begin{figure}[t!]\centering
\includegraphics[width=0.51\textwidth]{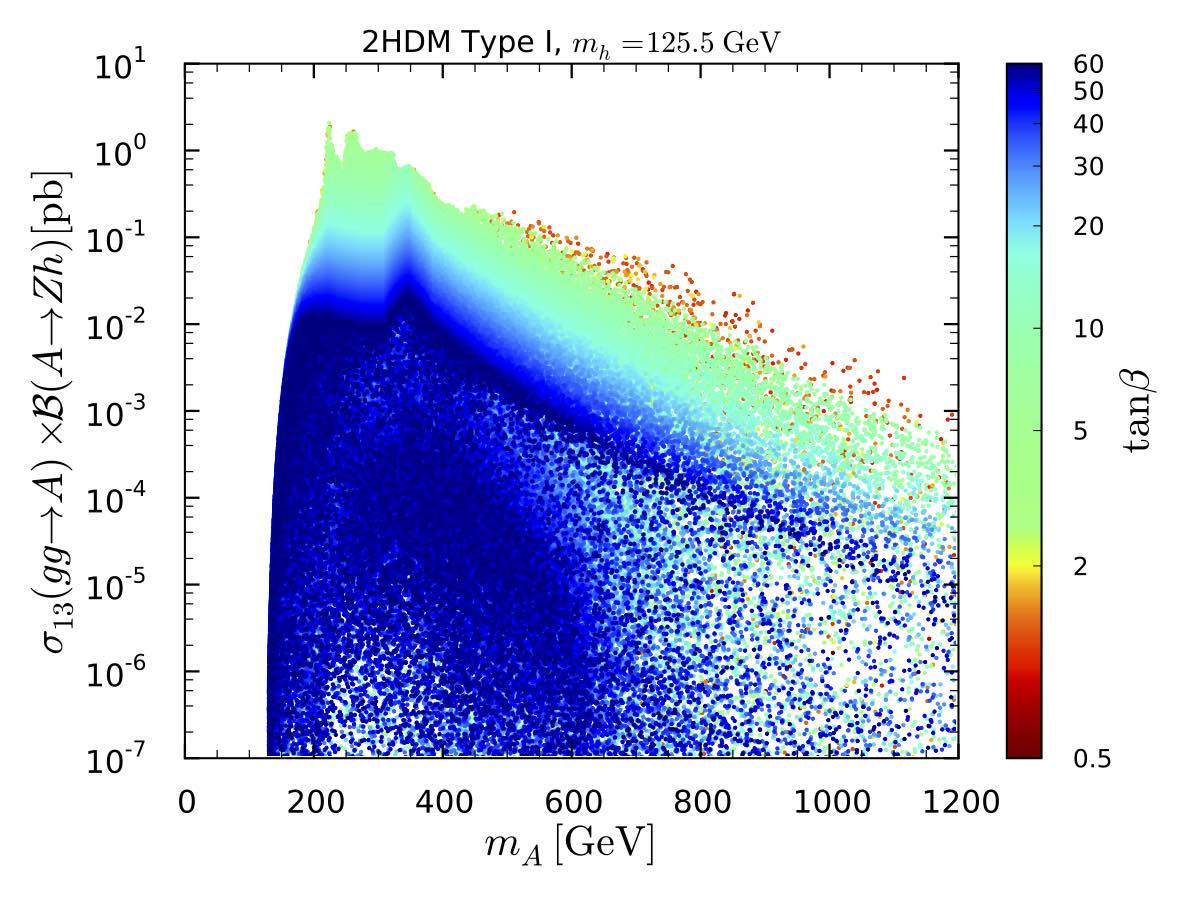}\includegraphics[width=0.51\textwidth]{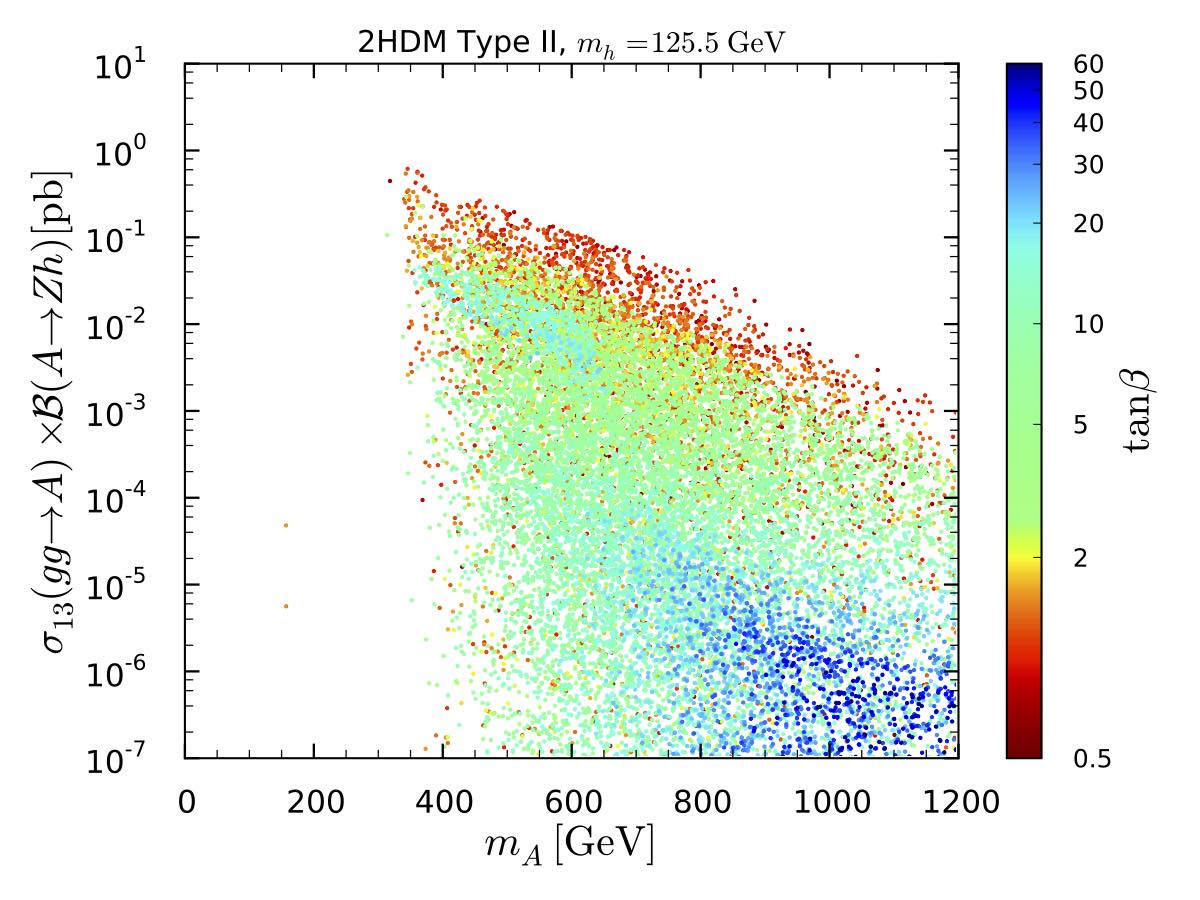}
\includegraphics[width=0.51\textwidth]{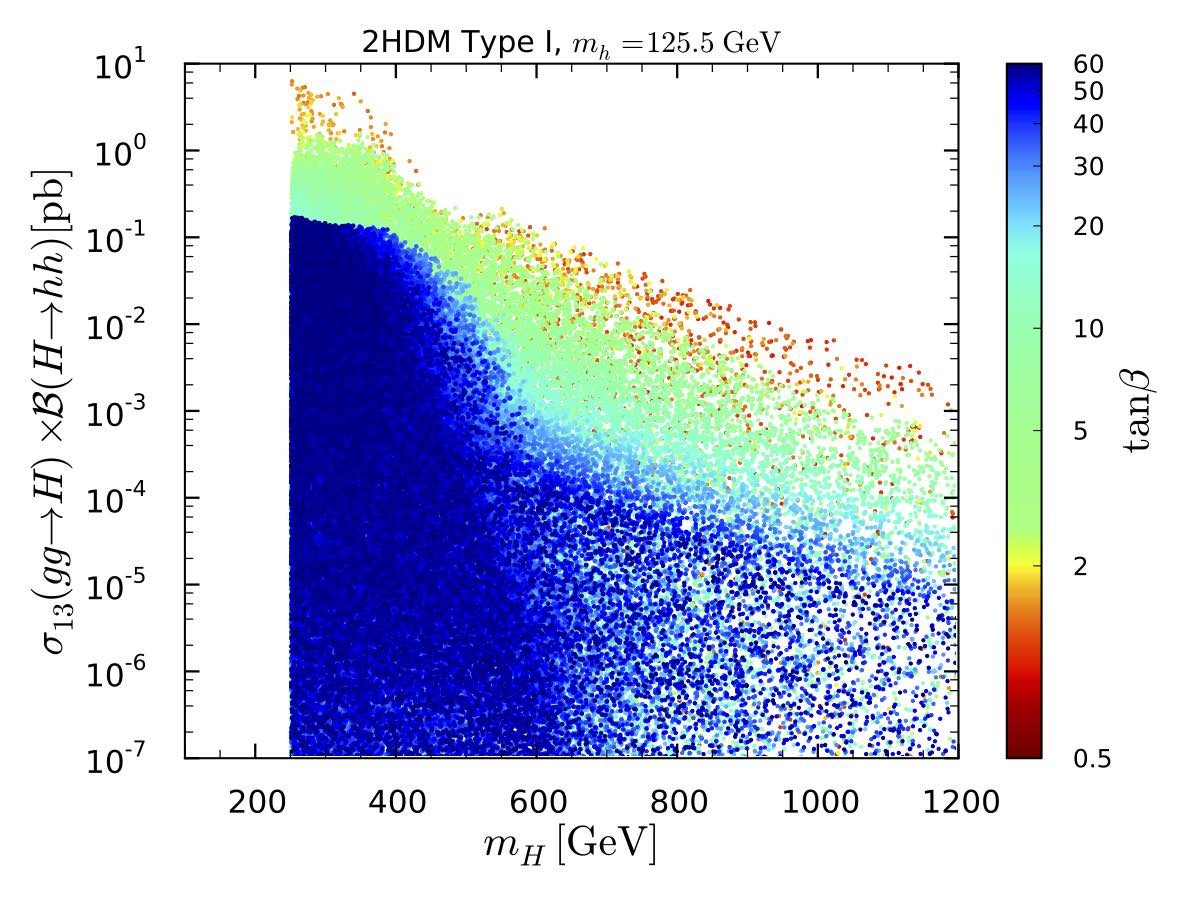}\includegraphics[width=0.51\textwidth]{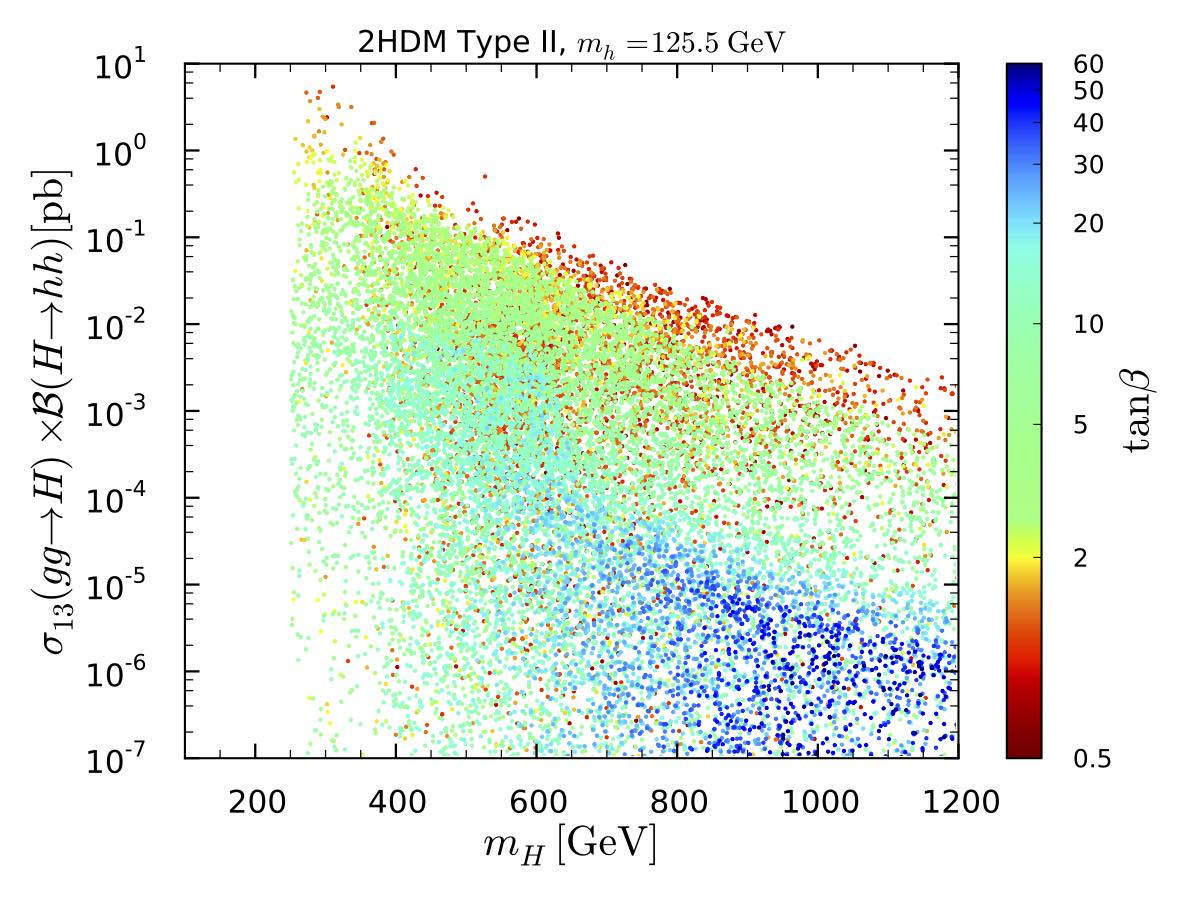}
  \caption{Cross sections times branching ratio in Type~I (left) and in Type~II (right) for $gg\to X\to Y$ at the 13~TeV LHC as functions of $m_X$ for $X,Y=A,Zh$ (upper panel) and $X,Y=H,hh$ (lower panel) with $\tan\beta$ color code. Points are ordered from low to high $\tan\beta$.}
  \label{exoticxsecBRH13}
\end{figure}

Non-standard production modes of the SM-like state, through $A\to Zh$ and $H\to hh$, are presented in Fig.~\ref{exoticxsecBRH13}. While these can be interesting discovery modes for the $A$ and/or $H$, their cross sections  
can also be extremely suppressed. 
For $gg\to A\to Zh$, the $\tan\beta$ dependence, which follows the dependence of the $gg\to A$ cross section shown in Fig.~\ref{xsec13}, explains a part of this suppression. 
Moreover, the $AZh$ coupling is proportional to $\cba^2$ which is suppressed in the alignment region. Nevertheless, the $gg\to A\to Zh$ cross section can be 
$>100$~fb for $m_A \lesssim 600$ GeV in both Types~I and~II.
The $gg\to H\to hh$ cross section, as expected, attains its maximum below the $t\bar{t}$ threshold in both Types~I and~II and can reach about $10$ pb at low $\tan\beta$. For any $m_H$, the cross section can however also be extremely suppressed. 

A comment is in order here on the possible ``feed down'' (FD) \cite{Arhrib:2013oia,Dumont:2014wha} to the production of the 125~GeV state through the decay of heavier Higgs states, which might distort the Higgs signal strengths. 
This issue was approximately addressed in section III.C of \cite{Dumont:2014wha} by imposing the ``FDOK'' conditions $\mu^{\rm FD}_{Zh}<0.3$ and $\mu^{\rm FD}_{{\rm ggF}h+bbh}<0.1$, which limit the FD contamination of $Zh$ associated production and of ggF$+bbh$ production to 30\% and to 10\% respectively, at the cross section times branching ratio level. 
Imposing these conditions here would remove the points with $\sigma_{13}(gg\to A)\times \textrm{BR}(A\to Zh)\gtrsim 0.2$~pb and $\sigma_{13}(gg\to H)\times \textrm{BR}(H\to hh)\gtrsim 2$~pb in Fig.~\ref{exoticxsecBRH13}. 
This is, however, a maximally conservative constraint for two reasons. Firstly, the amount of FD is computed without accounting for any reduced acceptance of such events into the 125~GeV signal as a result of the experimental cuts used to define the $gg\to h$, $bbh$ or $Z^*\to Zh$ channels. Secondly, it puts individual limits on specific production$\times$decay modes instead of including all signal strengths in a global fit, which is the approach followed in this paper.
Indeed, when directly adding the contribution of $gg\to A\to Zh$ to the $Zh$ signal strength in the global fit, it turns out that only cross sections of  $\sigma_{13}(gg\to A)\times \textrm{BR}(A\to Zh)\gtrsim 1.6$~pb are definitely excluded. This still assumes that the signal acceptance of the experimental analysis is the same for $gg\to A\to Zh$ as for $gg\to Z^*\to Zh$, which should however be a reasonable approximation, as the main difference would be the $Zh$ invariant-mass distribution, which is not used as a selection criterion in this case. 
The contribution of $H\to hh$ to the $h$ signal strengths is a more difficult question, as here the acceptances (in each final state considered in the experimental analyses) will certainly be different from those of single $h$ production.  
A detailed study based on event simulation would be necessary to better understand the impact of FD on the 125~GeV Higgs signal, but this is beyond the scope of this paper.  

\begin{figure}[h!]\centering
\includegraphics[width=0.51\textwidth]{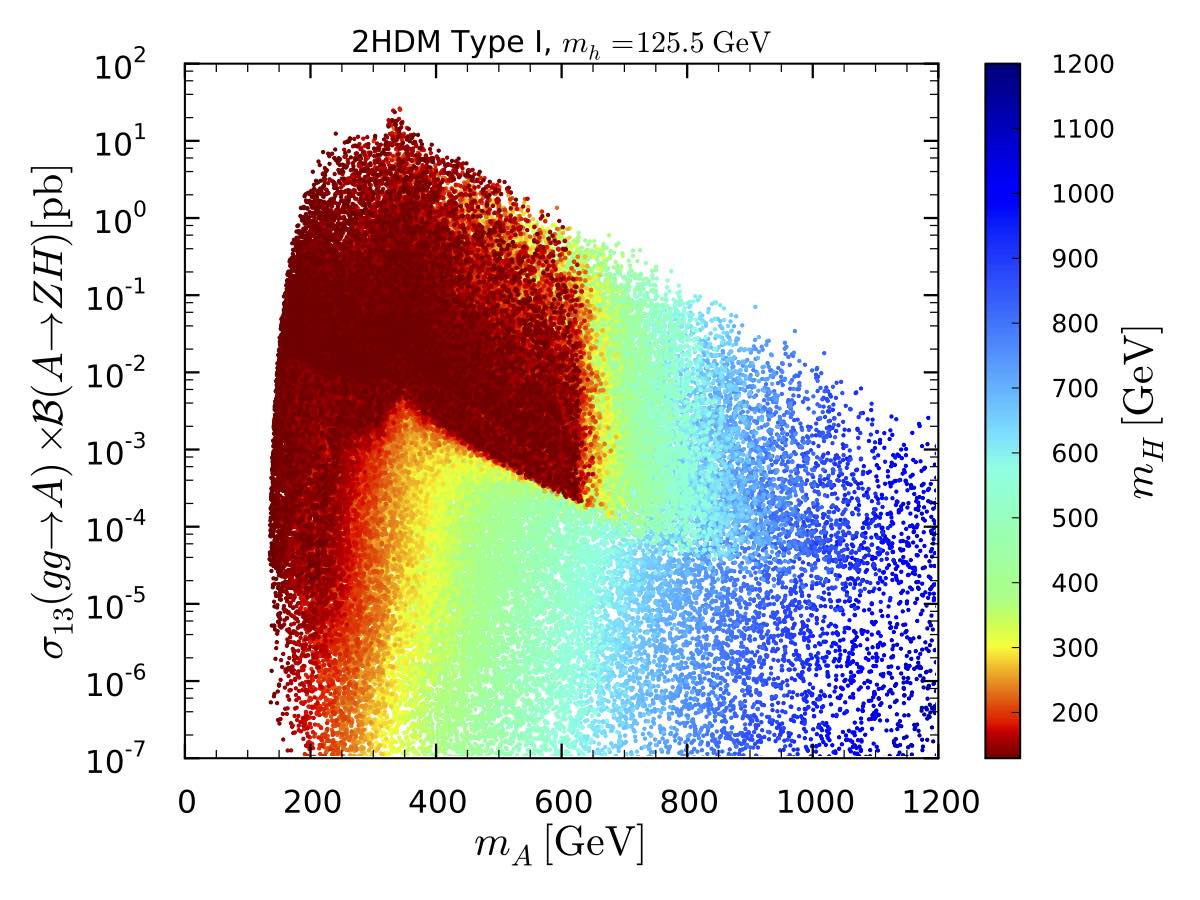}\includegraphics[width=0.51\textwidth]{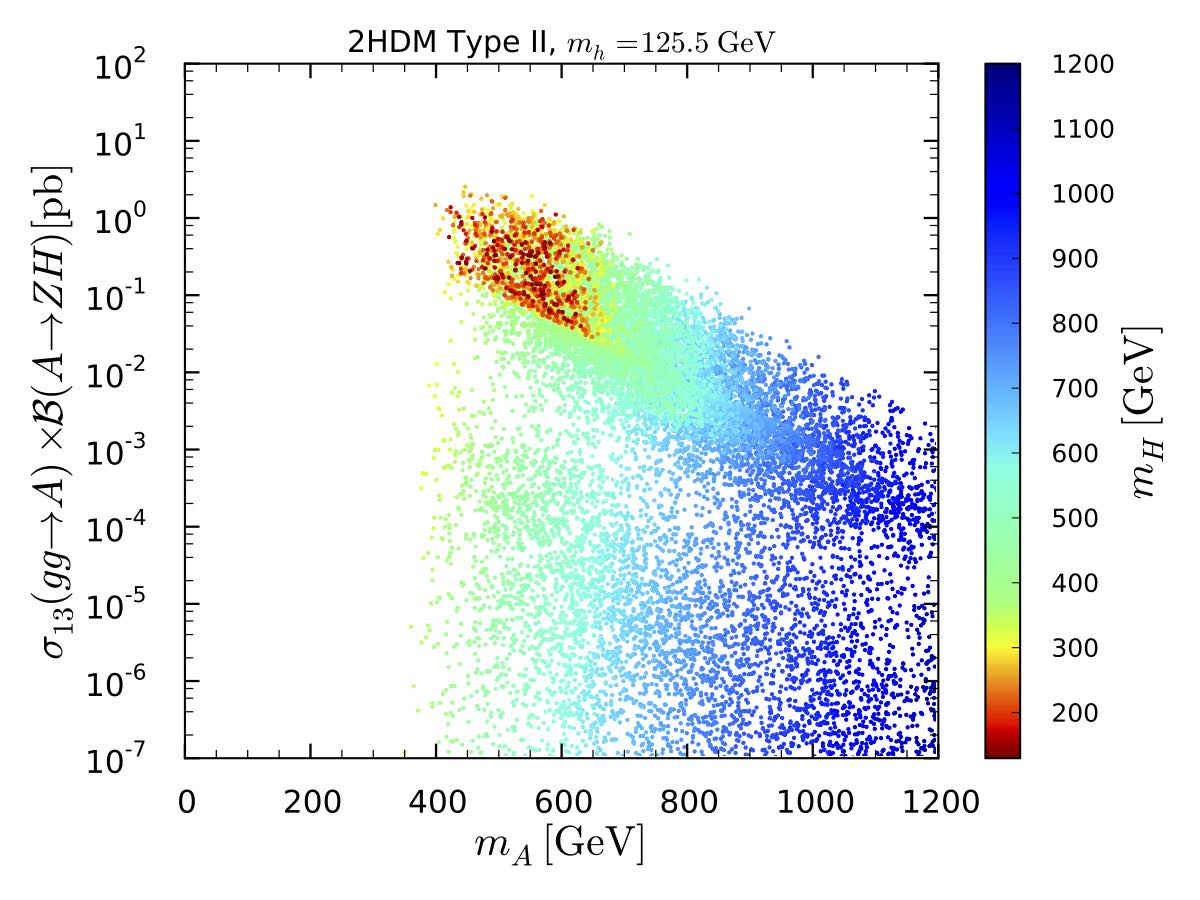}
\includegraphics[width=0.51\textwidth]{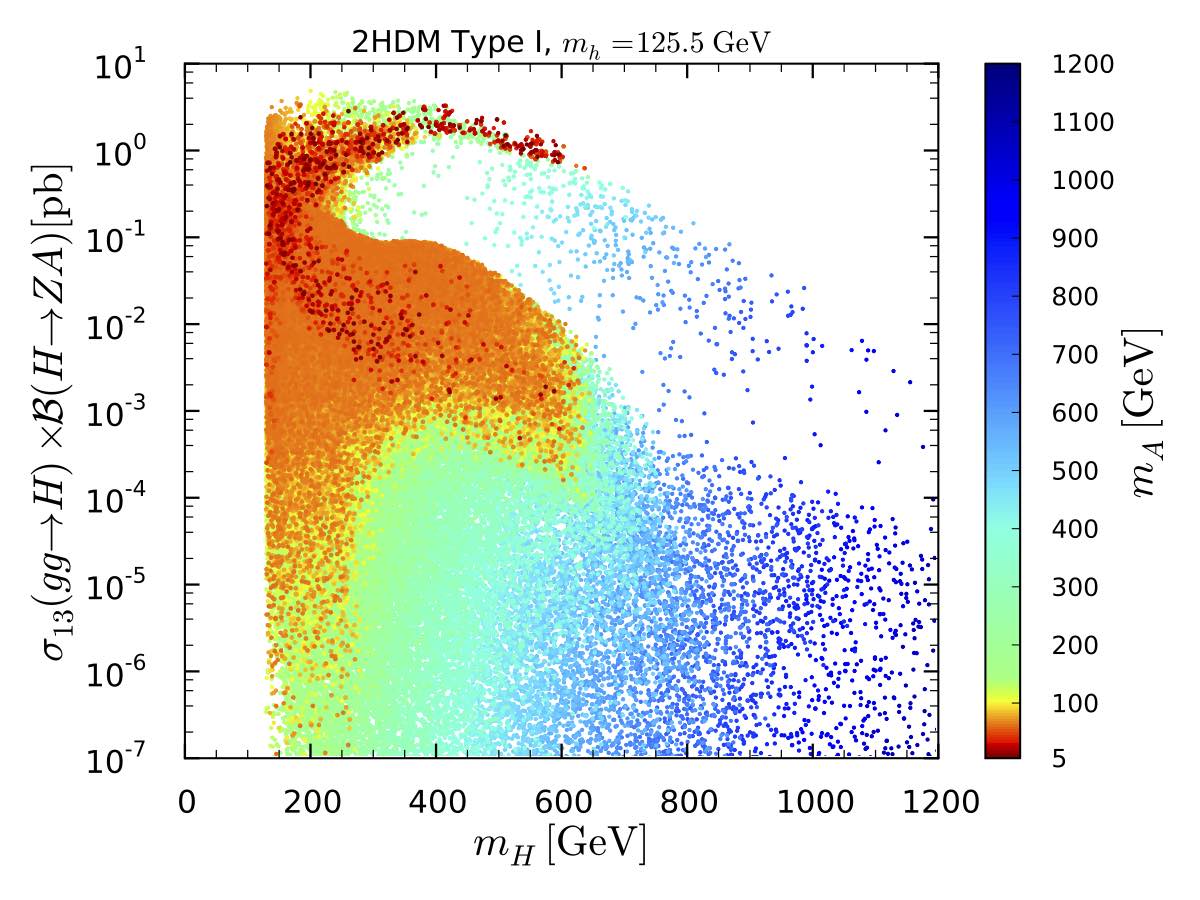}\includegraphics[width=0.51\textwidth]{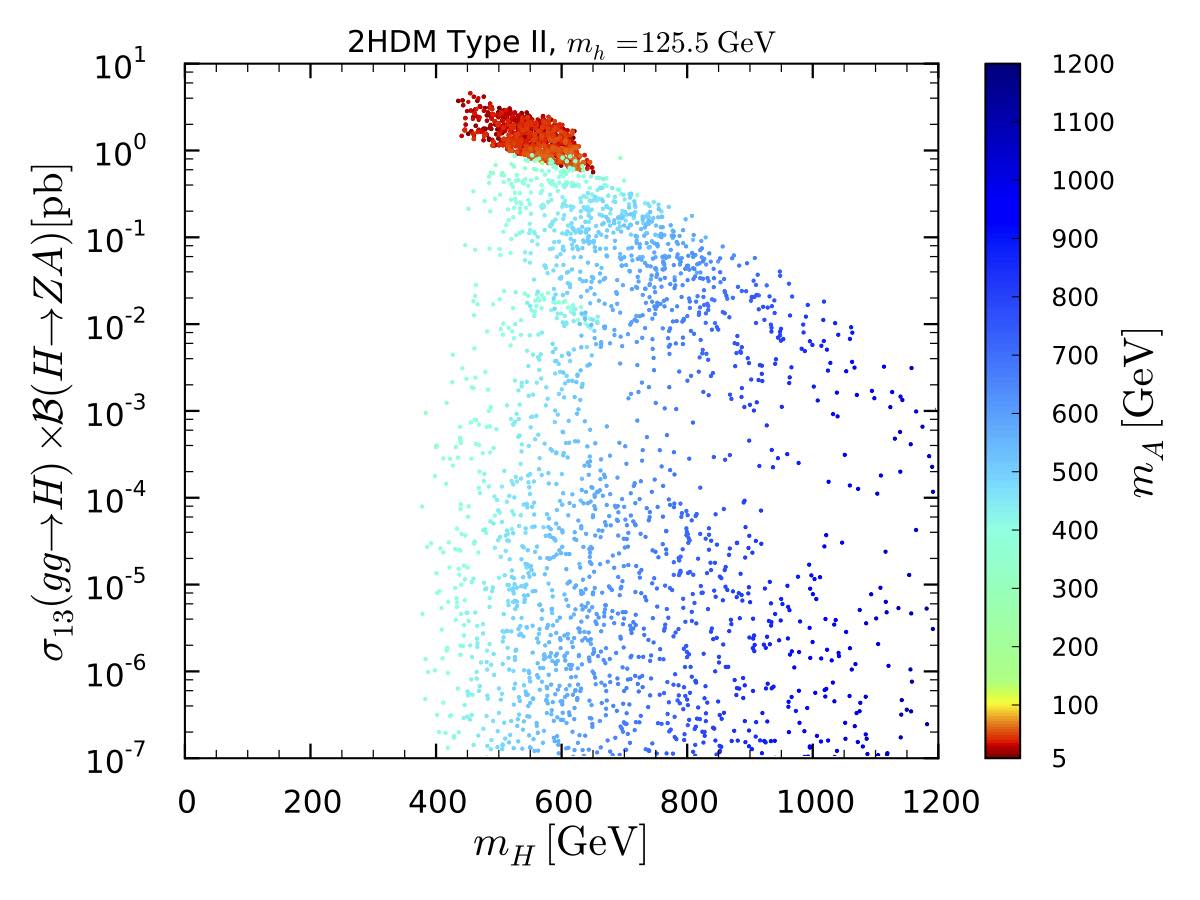}
\includegraphics[width=0.51\textwidth]{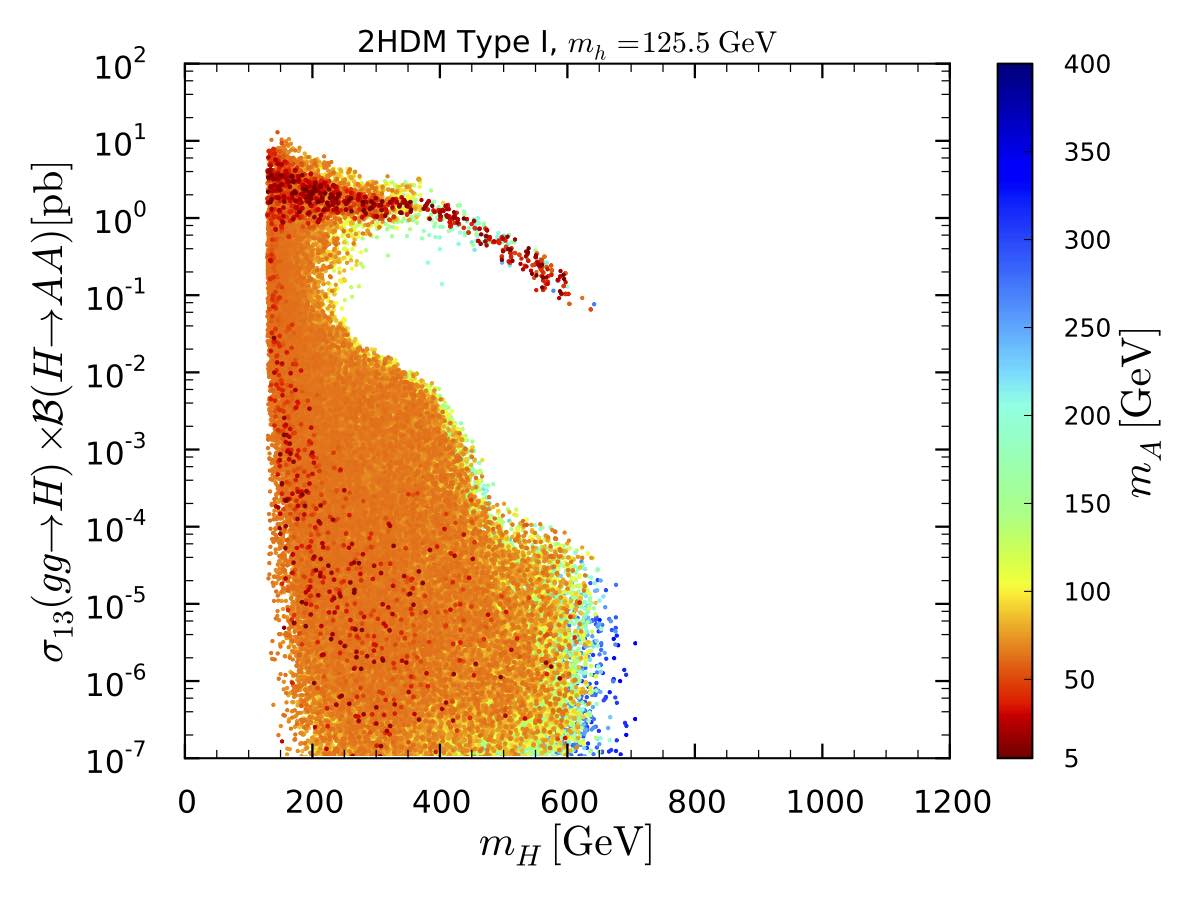}\includegraphics[width=0.51\textwidth]{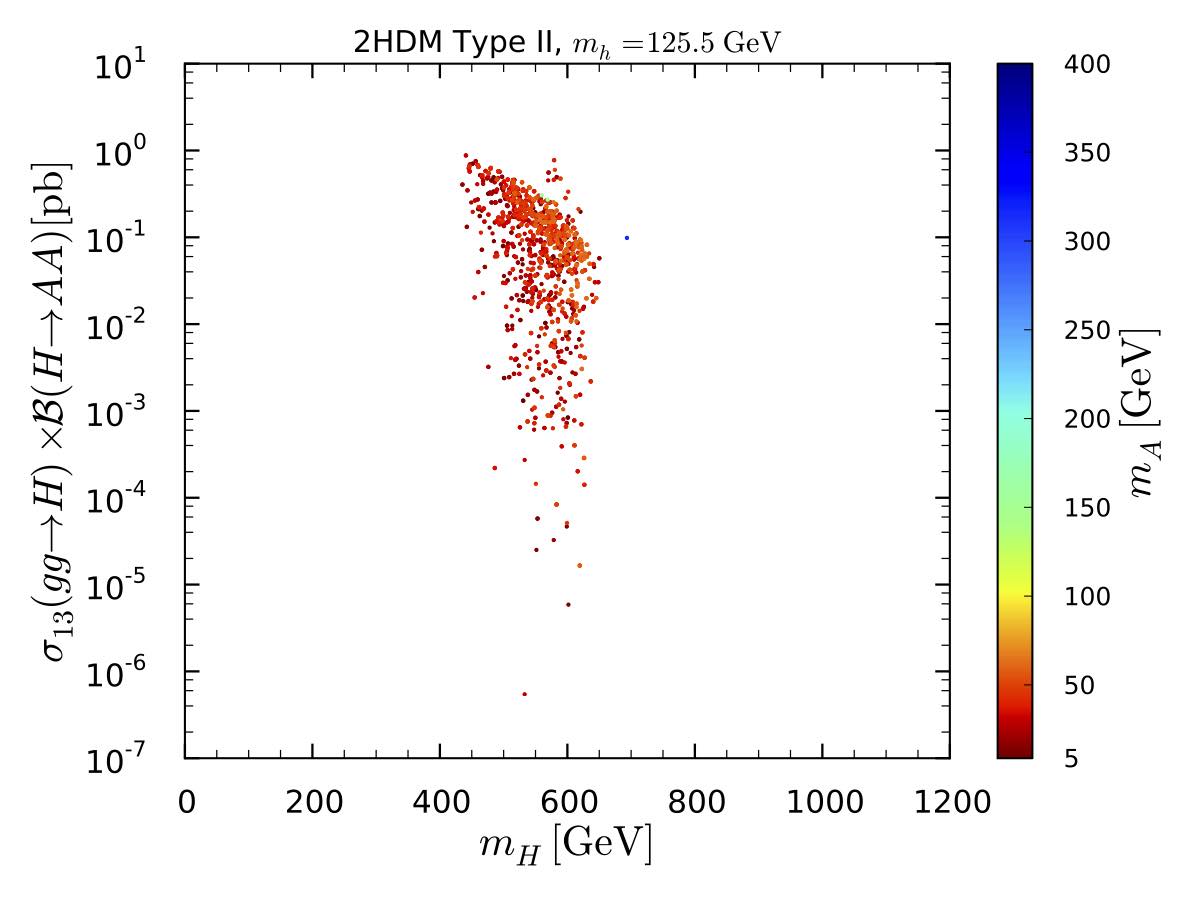}
  \caption{Cross sections times branching ratio in Type~I (left) and in Type~II (right) for Higgs-to-Higgs signatures 
  at the 13~TeV LHC, 
  in the upper panel  $gg\to A\to ZH$ with $m_H$ color code, and in the middle and lower panels for 
  $gg\to H\to ZA$ and $gg\to H\to AA$, respectively, with $m_A$ color code. 
  Points are ordered from high to low $m_A$ or $m_H$, with the exception of the $H\to ZA$, and $H\to AA$ plots in Type~II, where points are ordered from low to high $m_A$.}
  \label{exoticxsecBRH13_HA}
\end{figure}

Finally, if the mass splitting is large enough, $A\to ZH$,  $H\to ZA$, and $H\to AA$ decays offer intriguing possibilities for discovering the extra non-SM-like neutral Higgs states in the regime of approximate alignment without decoupling.  In Fig.~\ref{exoticxsecBRH13_HA},
the cross sections for $gg\to A\to ZH$, $gg\to H\to ZA$ and $gg\to H\to AA$ are exhibited.  
Large $gg\to A\to ZH$ cross sections are obtained for large $m_A-m_H$ splitting.\footnote{A large splitting $m_A-m_H\approx v$ can be motivated by the possibility of a strong first order phase transition in 2HDMs~\cite{Dorsch:2014qja}.}  
Looking back at Fig.~\ref{mA_mH_Z6} one sees that, in both Type~I and Type~II, the splitting can only be large for $m_A\lesssim 650$~GeV. This explains the preponderance of low $m_H$ points with cross sections up to 20~pb (6~pb) in Type~I (II) for $m_A\lesssim 650$~GeV. (In Type~II the $\mhpm > 480$~GeV constraint allows a large enough $m_A-m_H$ mass splitting only for $m_A\gtrsim 350$~GeV.) However, $gg\to A\to ZH$ can also be heavily suppressed; since the $AHZ$ coupling is proportional to $\sbma$, this suppression is a purely kinematical effect.

Turning to the $H \to ZA$ and $H \to AA$ signatures, in Type~I we observe a depleted
area for $m_H>300$~GeV and cross sections of the order of $0.1$~pb.
In this region, $\tan\beta=2$--$10$ and $Z_5$ is small or negative leading to
$H$ and $A$ masses for which the $H\to ZA$, $AA$ decays are kinematically forbidden
[cf.~\eq{hh2mass}]. 
In the region below, $\tan\beta>10$ and $Z_5$ can be large enough to achieve 
$m_H>m_A+m_Z$ and/or $m_H>2m_A$, but nevertheless the cross section is small because of the
$\tan\beta$ dependence of $\sigma(gg\to H)$, see Fig.~\ref{xsec13}.
The distinct branch with $gg \to H \to ZA$ and $gg \to H \to AA$ cross sections larger than about 1 pb, on the other hand, has $\tan\beta\lesssim 2$ and $\lambda_5\approx 0$.
Here, the term proportional to $\sin 2\beta$ in Eq.~\eqref{zeefive} gives a large enough $Z_5>0$
so that the $H \to ZA$ and/or $H \to AA$ decay is kinematically allowed. The small
$\tan\beta$ leads to a large $gg\to H$ production cross section, see again Fig.~\ref{xsec13}.
In Type~II, $gg \to H \to ZA$ and $gg \to H \to AA$ cross sections can also be large (even above 1 pb for $H \to ZA$) in the non-decoupling regime. However, due to the charged Higgs mass constraint these processes are allowed only for $m_H\gtrsim 430$~GeV. 
A detailed phenomenological analysis of the $A\to ZH$ and $H\to ZA$ decays at the LHC was performed in~\cite{Coleppa:2014hxa}.

Last but not least, note that due to the kinematic constraint $m_H\geq 2m_A$ and the non trivial correlation between $m_H$ and $m_A$ observed in Fig.~\ref{mA_mH_Z6}, the $H\to AA$ channel is only open for $m_H\lesssim 700$ GeV. In Type~I the branch of points with cross sections ranging from about $10^{-1}$~pb to 10~pb is mainly populated by $m_A\leq 100$~GeV points with relatively low $\tan\beta\lesssim 10$.
In Type~II, due to the $A\to\tau\tau$ and $H\to ZA$ experimental constraints, only points with low $m_A\lesssim 60$~GeV and $\tan\beta\lesssim 2$ experience $H\to AA$ decays.
All in all, this channel offers a complementary probe to the low $m_A$ region discussed in~\cite{Bernon:2014nxa}.

\clearpage
\section{Conclusions}\label{conclusions}

While the Higgs measurements at Run~1 show no deviations from the SM, conceptually there is no reason why the Higgs sector should be minimal. Indeed a non-minimal Higgs sector is theoretically very attractive and, if confirmed, would shine a new light on the mechanism of electroweak symmetry breaking dynamics.

In this paper we focused on CP-conserving 2HDMs of Type~I and Type~II, investigating the special situation that arises when one of the Higgs mass eigenstates is approximately aligned with the direction of the scalar field vacuum expectation values. In this case, the $W^\pm$ and $Z$ gauge bosons dominantly acquire their masses from only one Higgs doublet of the Higgs basis.  Moreover, the coupling of that CP-even Higgs boson to the gauge bosons tends towards the SM value, $C_V\to 1$. While this is automatically the case in the decoupling limit when the extra non-SM Higgs states are very heavy, such an alignment can also occur when the extra Higgs states are light, below about 600~GeV. 
We specifically investigated the phenomenological consequences of alignment without decoupling and contrasted them to the decoupling case. Two aspects are interesting in this respect: one being precision measurements of the couplings and signal strengths of the SM-like Higgs boson at 125~GeV, the other being the ways to discover the additional Higgs states of the 2HDM when they are light. 

In addition to an in-depth theoretical discussion, we performed a detailed numerical analysis for the case that the SM-like state observed at  $125\gev$ is the lighter of the two CP-even Higgs bosons of the 2HDM, $h$.  
In this study we allowed for 1\% deviation from unity in $C_V^h$, which corresponds to the ultimate expected LHC precision at high luminosity. The results can be summarized as follows:

\ben
\item In the alignment limit without decoupling, despite $C_V^h$ being very close to 1, the fermionic couplings of the 125~GeV Higgs can deviate substantially from the SM values. Concretely, $C_U^h$ can deviate as much as about 10\% (20\%) from unity in Type~I (Type~II), and $C_D^h$ as much as 30\% in Type~II.

\item While $C_U^h$ rather quickly approaches 1 with increasing $m_H$ and/or $\cbma\to 0$, the approach of the bottom Yukawa coupling to its SM value in the alignment limit is delayed in Type~II, with $C_D^h\approx 0.70$--$1.15$ even for $|\cbma|\sim 10^{-2}$. Large values of $C_D^h>1$ are associated with light $H,A$. Moreover, for $330 \gev \leq m_H\leq 660 \gev$ and $350 \gev \leq m_A\leq 660 \gev$, there is an allowed region with $C_D^h\approx-1\pm0.2$; this ``opposite-sign'' solution can be tested decisively at Run~2. 

\item The trilinear $hhh$ coupling can also exhibit large deviations. Large values of $C_{hhh}>1$ (up to $C_{hhh}\approx 1.7$ in Type~I and up to 
$C_{hhh}\approx 1.35$ in Type~II) can be achieved in the non-decoupling regime $m_H\lesssim 600\gev$, for $|\cba|$ of the order of $0.1$,  whereas 
for heavier $m_H$,  $C_{hhh}$ is always suppressed as compared to its SM prediction. 
The suppression can be about 50\% for $m_H$ of $\sim 1$~TeV and much larger for lighter $m_H$.  

\item For the ratios $\mu^h_X(Y)$ of the $X\to h \to Y$ signal rates relative to the SM prediction, we found 
distinct correlations of these signal strengths in both Type~I and Type~II that depend on whether the additional Higgs states are decoupled or not.  In fact, in the regime of alignment without decoupling, there are characteristic combinations of the $\mu^h_X(Y)$ signal strengths that cannot be mimicked by the decoupling limit. However, it is of course also possible that all signal strengths converge to 1 even though the additional Higgs states are very light. 

\item A decisive test of the alignment without decoupling scenario would of course be the observation of the additional Higgs states of the 2HDM in the mass range below about 600 GeV. We delineated the many possibilities for such observations. While there are no guarantees in the case of the Type~I model, in the Type~II model there is always a definite lower bound on the $gg\to A,H\to \tau\tau$ cross sections at the LHC at any given $m_A$.  For low $\tanb\sim 1$, this lower bound is still of order $0.1$ fb for $m_A\sim 500 \gev$, a level that we deem likely to be observable at the LHC during Run~2. For high $\tanb$, the lower bound is roughly two orders of magnitude higher and only falls below the $0.1$ fb level for $m_{A,H}\gsim 1.2\tev$, which is already in the decoupling region. Moreover, while in Type~I gluon-gluon fusion is always dominant for $H$ or $A$ production, in Type~II both $b\bar b$ associated production and gluon-gluon fusion modes are in principle important since either can be dominant in different regions of the parameter space.

\item Higgs-to-Higgs decays of the non-SM-like states ($A\to ZH$, $H\to ZA$, $H\to AA$) also open intriguing possibilities for testing the regime of  alignment without decoupling, with cross sections often in the range of 1--10~pb (although they can also be quite suppressed). Particularly promising are $gg\to H\to ZA$ and $gg\to H\to AA$ in Type~II for light pseudoscalars below about 100~GeV; for such a light $A$, $m_H$ can be at most $\sim 650$~GeV, 
and $\sigma\times \br$ values for these channels typically range from 10~fb to 10~pb.

\een

In short, it is possible that the observed 125~GeV Higgs boson appears SM-like due to
the alignment limit of a multi-doublet Higgs sector.  The alignment limit does not 
necessarily imply that the additional Higgs states of the model are heavy. Indeed, they can be light and non-decoupled and thus lead to exciting new effects to be probed at Run~2 of the LHC.

\section*{Acknowledgements}

J.B. and S.K. thank Jose Miguel No and Ken Mimasu for discussions on the $A\to ZH$ decay 
during the ``Physics at TeV Colliders'' workshop in Les Houches 10--19 June 2015.

This work was supported in part by the Research Executive Agency (REA) of the
European Union under the Grant Agreement PITN-GA2012-316704
(HiggsTools) and by the ``Investissements d'avenir, Labex ENIGMASS''.
H.E.H. is supported in part by U.S. Department of Energy grant DE-FG02-04ER41286.
J.F.G.\ and Y.J.\ are supported in part by the US DOE grant DE-SC-000999. 
Y.J.\  also acknowledges support by the LHC-TI fellowship  US NSF grant PHY-0969510, and 
he thanks the LPSC Grenoble for its hospitality. 
J.F.G., H.E.H. and S.K. are grateful for the hospitality and the inspiring working atmosphere  
of the Aspen Center for Physics, supported by the National Science Foundation Grant No.\ PHY-1066293, 
where this project was initiated.

\bigskip\bigskip

\appendix
\numberwithin{equation}{section}  

\section{Scalar potential quartic coefficients in the $\mathbb{Z}_2$-basis in terms of Higgs basis coefficients}
\label{AppInverse}

In \eqst{zeeone}{zeeseven}, we have provided expressions for the Higgs basis quantities $Z_i$ in terms of the quartic coefficients of the scalar potential $\lambda_i$ defined in \eq{lambdapotential}.
In this appendix, we provide the inverse of \eqst{zeeone}{zeeseven} by expressing the $\lambda_i$ in terms of the $Z_i$.   
\beqa
\lambda_1&=& Z_1 c_\beta^4+Z_2 s_\beta^4+\half Z_{345}s_{2\beta}^2-2s_{2\beta}(c^2_\beta Z_6+s^2_\beta Z_7)\,,\\
\lambda_2&=& Z_1 s_\beta^4+Z_2 c_\beta^4+\half Z_{345}s_{2\beta}^2+2s_{2\beta}(s^2_\beta Z_6+c^2_\beta Z_7)\,,\\
\lambda_i&=& Z_i+\tfrac{1}{4}s_{2\beta}^2(Z_1+Z_2-2Z_{345})+s_{2\beta}c_{2\beta}(Z_6-Z_7)\,,
\quad \text{for $i=3,4,5$}\,,
\eeqa
where $Z_{345}\equiv Z_3+Z_4+Z_5$.
However, these results do not take into account the fact that $\lambda_6=\lambda_7=0$, which yields 
two relations among the $Z_i$.  These relations were given in \eqs{ztwo}{z345} and are repeated below for the convenience of the reader.  Recall that we employ a convention where 
$0\leq\beta\leq\half\pi$.  Then, $Z_2$ and $Z_{345}$ are dependent quantities for
$\beta\neq 0$, $\tfrac{1}{4}\pi$, $\half\pi$,
\beqa 
Z_2&=& Z_1+2(Z_6+Z_7)\cot 2\beta\,,\label{app:ztwo}\\
Z_{345}&=&Z_1+2Z_6\cot 2\beta-(Z_6-Z_7)\tan 2\beta\,,\label{app:z345}
\eeqa
An alternative form of \eq{app:z345} is obtained by combining the results of \eqs{app:ztwo}{app:z345}, which yields
\beq \label{app:z345alt}
Z_{345}=Z_2-2Z_7\cot 2\beta-(Z_6-Z_7)\tan 2\beta\,.
\eeq
Taking the average of \eqs{app:z345}{app:z345alt} provides one more useful relation that can be used as the second condition for the softly-broken $\mathbb{Z}_2$ symmetry along with \eq{app:ztwo},
\beq \label{app:z345ave}
Z_{345}=\half(Z_1+Z_2)+2(Z_6-Z_7)\cot 4\beta\,.
\eeq
Using \eqs{app:ztwo}{app:z345ave} it follow that if
$\beta=0$, $\half\pi$ then $Z_6=Z_7=0$; if
$\beta=\tfrac{1}{8}\pi$, $\tfrac{3}{8}\pi$ then $Z_{345}=\half(Z_1+Z_2)$; and if
$\beta=\tfrac{1}{4}\pi$ then $Z_1=Z_2$ and $Z_6=Z_7$.   

Consequently, the expressions for the $\lambda_i$ in terms of the $Z_i$ can be written in numerous equivalent ways depending on the choice of independent quantities.  For example, if $\beta\neq \tfrac{1}{4}\pi$, then eliminating $Z_{345}$ and either $Z_1$ or $Z_2$ yields
\beqa
\lambda_1&=& \begin{cases} 
Z_1-Z_6\tan 2\beta+\half\tan^2 \beta\tan 2\beta(Z_6+Z_7)\,, & \text{if $\beta\neq\half\pi$}\,,\\
Z_2+Z_7\tan 2\beta-\half\cot^2\beta\tan 2\beta(Z_6+Z_7)\,,& \text{if $\beta\neq 0$}\,,\end{cases}\label{Z1}\\
\lambda_2&=& \begin{cases}
Z_1-Z_6\tan 2\beta+\half\cot^2 \beta\tan 2\beta(Z_6+Z_7)\,, & \text{if $\beta\neq 0$}\,,\\
Z_2+Z_7\tan 2\beta-\half\tan^2 \beta\tan 2\beta(Z_6+Z_7)\,,& \text{if $\beta\neq \half\pi$}\,,\end{cases}\label{Z2} \\
\lambda_i&=&Z_i+\half(Z_6-Z_7)\tan 2\beta\,,\quad \text{for $i=3,4,5$}\,.\label{Z345}
\eeqa
Note that the $\mathbb{Z}_2$-basis and the Higgs basis coincide if $\beta=0$ (in which case $\Phi_1=H_1$ and $\Phi_2=H_2$) and if $\beta=\half\pi$ (in which case $\Phi_1=H_2$ and $\Phi_2=H_1$).   
The two alternative forms given in \eqs{Z1}{Z2} are a consequence of the symmetry of
\eqst{app:ztwo}{app:z345ave} under the interchanges, $Z_1\longleftrightarrow Z_2$, $Z_6\longleftrightarrow Z_7$, $\beta\longleftrightarrow \half\pi-\beta$.

The exclusion of $\beta=\tfrac{1}{4}\pi$ in \eqst{Z1}{Z345} is an artifact of expressing these results in terms of both $Z_6$ and $Z_7$.  Nevertheless, there is no discontinuity, since $Z_6=Z_7$ at $\beta=\tfrac{1}{4}\pi$.  One way to avoid this inconvenience is to eliminate either $Z_6$ or $Z_7$ in favor of $Z_{345}$.  The end result~is
\beqa
\lambda_1&=&\begin{cases} Z_1(1-\half\tan^2\beta)+\half Z_{345}\tan^2\beta-\half Z_6\tan\beta(5-\tan^2\beta)\,,& \text{if $\beta\neq\half\pi$}\,,\\
Z_2(1-\half\cot^2\beta)+\half Z_{345}\cot^2\beta-\half Z_7\cot\beta(5-\cot^2\beta)
\,,& \text{if $\beta\neq 0$}\,,\end{cases}\,,\label{Z1alt}\\
\lambda_2&=&\begin{cases} Z_1(1-\half\cot^2\beta)+\half Z_{345}\cot^2\beta+\half Z_6\cot\beta(5-\cot^2\beta)\,,& \text{if $\beta\neq 0$}\,,\\
Z_2(1-\half\tan^2\beta)+\half Z_{345}\tan^2\beta+\half Z_7\tan\beta(5-\tan^2\beta)
\,,& \text{if $\beta\neq \half\pi$}\,,\end{cases}\,,\label{Z2alt}\\
\lambda_i&=&\begin{cases}Z_i+\half(Z_1-Z_{345})+Z_6\cot 2\beta\,,&
\text{for $i=3,4,5$ and $\beta\neq 0$, $\half\pi$}\,,\\
Z_i+\half(Z_2-Z_{345})-Z_7\cot 2\beta\,,&
\text{for $i=3,4,5$ and $\beta\neq 0$, $\half\pi$}\,.\end{cases}\label{Zeye}
\eeqa

Finally, one may choose to eliminate both $Z_6$ and $Z_7$, using
\eqs{app:ztwo}{app:z345ave}.  The end result is valid for $\beta\neq\tfrac{1}{8}\pi$, 
$\tfrac{1}{4}\pi$, $\tfrac{3}{8}\pi$,\footnote{Eliminating both $Z_6$ and $Z_7$ is not particularly useful in the cases of $\beta=\tfrac{1}{8}\pi$, $\tfrac{3}{8}\pi$, where $Z_1+Z_2=2Z_{345}$ and in the case of $\beta=\tfrac{1}{4}\pi$, where $Z_1=Z_2$ [cf.~\eqs{app:ztwo}{app:z345ave}].}
\beqa
\lambda_1&=&\half(Z_1+Z_2)+\frac{s_{2\beta}^2}{4c_{4\beta}}(Z_1+Z_2-2Z_{345})+\frac{1}{2c_{2\beta}}(Z_1-Z_2)\,,\\[6pt]
\lambda_2&=&\half(Z_1+Z_2)+\frac{s_{2\beta}^2}{4c_{4\beta}}(Z_1+Z_2-2Z_{345})-\frac{1}{2c_{2\beta}}(Z_1-Z_2)\,,\\[6pt]
\lambda_i&=&Z_i-\frac{s^2_{2\beta}}{2c_{4\beta}}(Z_1+Z_2-2Z_{345})\,,\quad \text{for $i=3,4,5$}\,.
\eeqa

The conditions for stability of the scalar potential [\eq{lambdapotential}] for $\lambda_6=\lambda_7=0$ were first given in \cite{Deshpande:1977rw},
\beq \label{stability}
\lambda_1>0\,,\quad\lambda_2>0\,,\quad \lambda_3>-\sqrt{\lambda_1\lambda_2}\,,\quad
\lambda_3+\lambda_4-|\lambda_5|>-\sqrt{\lambda_1\lambda_2}\,.
\eeq
Using the results of this Appendix, one can rewrite the stability conditions in terms of the~$Z_i$.  The resulting expressions are not especially illuminating, so we will not exhibit them explicitly.

In addition, we note that (under the assumption of $\lambda_6=\lambda_7=0$) the $\lambda_i$ ($i=1, 2, \ldots, 5$) can be reconstructed in principle as follows.   Assume that $\cbma$ has been deduced from precision measurements of the SM-like Higgs boson (assumed to be $h$), and $\beta$ is determined via the properties of the heavier Higgs states.  We also assume that all four Higgs masses ($m_h$, $m_H$, $m_A$ and $\mhpm$) have been measured.  Lastly, we assume that a small deviation in the signal strength for $h\to\gamma\gamma$ can be attributed to the presence of a charged 
Higgs loop,\footnote{In absence of a clear deviation from the SM in the $\gamma\gamma$ signal, one would be forced to seek out some measurable triple Higgs coupling involving no more than a single SM-like Higgs boson to avoid a suppression of the term that is sensitive to $Z_3$ or $Z_7$ [cf.~\eqst{hlhaha}{hhhphm}].}
in which case we can extract a value for $g_{hH^+H^-}$. With all this information in hand, we begin by using \eq{cbmaeq} [or equivalently, \eq{z6v}] to obtain $Z_6$.  Next, we employ \eqs{z1v}{z5v} to obtain $Z_1$ and $Z_5$, and \eqs{chhiggs}{cpodd} for the squared-mass difference, $m_{H^\pm}^2-m_A^2$ to deduce $Z_4-Z_5$, which together with the previous determination yields a value for $Z_4$.  Close to the alignment limit, we can use $g_{hH^+H^-}$ to extract $Z_3$ [cf.~\eqs{hhhphz}{HHHlim}].  We now have enough information to evaluate $Z_{345}$.  Finally, we can use \eqs{Z2}{Z345} to obtain $Z_2$ and $Z_7$.  We now have all the $Z_i$ (for $i=1,2,\ldots 7$), which can then be employed with the formulae provided in this Appendix to obtain the $\lambda_i$ ($i=1, 2, \ldots, 5$). 

\section{Trilinear Higgs self-couplings in terms of physical Higgs masses}
\label{AppTrilinear}

It is convenient to re-express the trilinear Higgs self-couplings in terms of the physical Higgs masses.
First, \eqs{mbarid}{cpodd} yield
\beq
(Z_3+Z_4-Z_5)v^2=2(\mha^2-\overline{m}^{\,2})+Z_1 v^2+2Z_6 v^2\cot 2\beta .
\eeq
Using this result along with \eqst{z1v}{z5v} and (\ref{z7v}), we end up with
{
\allowdisplaybreaks
\beqa
\!\!\!\!\!\!\!\!\!\!\!\!\bigl[(Z_3+Z_4-Z_5)\sbma+Z_7\cbma\bigr]v^2&=&\bigl[\mhl^2+2(\mha^2-\overline{m}^{\,2})\bigr]\sbma+2\cot 2\beta(\mhl^2-\overline{m}^{\,2})\cbma\,,\\
\!\!\!\!\!\!\!\!\!\!\!\!\bigl[(Z_3+Z_4-Z_5)\cbma-Z_7\sbma\bigr]v^2&=&\bigl[\mhh^2+2(\mha^2-\overline{m}^{\,2})\bigr]\cbma-2\cot 2\beta(\mhh^2-\overline{m}^{\,2})\sbma\,.
\eeqa
Noting that \eqs{chhiggs}{cpodd} yield $\mha^2-\mhpm^2=\half(Z_4-Z_5)v^2$, the above results immediately yield
\beqa
\bigl[Z_3\sbma+Z_7\cbma\bigr]v^2&=&\bigl[\mhl^2+2(\mhpm^2-\overline{m}^{\,2})\bigr]\sbma+2\cot 2\beta(\mhl^2-\overline{m}^{\,2})\cbma\,,\\
\bigl[Z_3\cbma-Z_7\sbma\bigr]v^2&=&\bigl[\mhh^2+2(\mhpm^2-\overline{m}^{\,2})\bigr]\cbma-2\cot 2\beta(\mhh^2-\overline{m}^{\,2})\sbma\,.
\eeqa
Thus, from \eqst{hlhaha}{hhhphm}, we obtain
\beqa
\!\!\!\!\!\! g\ls{\hl\ha\ha} &=&-\frac{1}{v}\biggl\{\bigl[\mhl^2+2(\mha^2-\overline{m}^{\,2})\bigr]\sbma+2\cot 2\beta(\mhl^2-\overline{m}^{\,2})\cbma\biggr\}\,,\label{haa}
 \\[4pt]
\!\!\!\!\!\! g\ls{\hh\ha\ha} &=&-\frac{1}{v}\biggl\{\bigl[\mhh^2+2(\mha^2-\overline{m}^{\,2})\bigr]\cbma-2\cot 2\beta(\mhh^2-\overline{m}^{\,2})\sbma\biggr\}\,,
 \\[4pt]
\!\!\!\!\!\! g\ls{\hl\hh\hh} &=& \phm\frac{\sbma}{v}\biggl\{2\overline{m}^{\,2}-2\mhh^2-\mhl^2+2(3\overline{m}^{\,2}-2\mhh^2-\mhl^2)(\sbma\cot 2\beta-\cbma)\cbma\biggr\}\,,
 \\[4pt]
\!\!\!\!\!\! g\ls{\hh\hl\hl} &=& -\frac{\cbma}{v}\biggl\{4\overline{m}^{\,2}-\mhh^2-2\mhl^2+2(3\overline{m}^{\,2}-\mhh^2-2\mhl^2)(\sbma\cot 2\beta-\cbma)\cbma\biggr\}\,,\label{Hhh}
 \\[4pt]
\!\!\!\!\!\! g\ls{\hl\hl\hl} &=& -\frac{3}{v}\biggl\{\mhl^2\sbma+2(\mhl^2-\overline{m}^{\,2})(\cbma\cot 2\beta+\sbma)c_{\beta-\alpha}^2\biggr\}\,,\label{hlhlhl2}\\[4pt]
\!\!\!\!\!\! g\ls{\hh\hh\hh} &=& -\frac{3}{v}\biggl\{\mhh^2\cbma-2(\mhh^2-\overline{m}^{\,2})(\sbma\cot 2\beta-\cbma)s_{\beta-\alpha}^2\biggr\}\,,\label{hhhhhh2}\\[4pt]
\!\!\!\!\!\! g\ls{\hl\hp\hm} &=& -\frac{1}{v}\biggl\{\bigl[\mhl^2+2(\mhpm^2-\overline{m}^{\,2})\bigr]\sbma+2\cot 2\beta(\mhl^2-\overline{m}^{\,2})\cbma\biggr\}\,,\label{hlhphm}
\\[4pt]
\!\!\!\!\!\!\!\!\!\!\!\! g\ls{\hh\hp\hm} &=& -\frac{1}{v}\biggr\{\bigl[\mhh^2+2(\mhpm^2-\overline{m}^{\,2})\bigr]\cbma-2\cot 2\beta(\mhh^2-\overline{m}^{\,2})\sbma\biggr\}\,.
\eeqa
}

\clearpage
\providecommand{\href}[2]{#2}\begingroup\raggedright\endgroup

\end{document}